\documentclass[a4paper,10pt]{article}
\usepackage[english]{babel}
\usepackage{amsmath,amsthm,amsfonts,amssymb}
\usepackage{cmap}
\usepackage[dvips]{graphicx}
\usepackage{wrapfig,epsfig,epsf}
\usepackage{multicol}

\newtheorem{defenition}{Defenition}

\begin{document}
\title{Transient dynamics of perturbations in astrophysical disks}

\author{ 
D.N. Razdoburdin$^{1,2}$, V.V. Zhuravlev$.^{2}$}

\date{\it\small
(1)Faculty of Physics\\
M.V.Lomonosov Moscow State University\\
(2) Sternberg Astronomical Institute\\
M.V.Lomonosov Moscow State University
\footnote {E-mail: zhuravlev@sai.msu.ru}
}
\maketitle

\label{firstpage}

\begin{abstract}

This paper reviews some aspects of one of the major unsolved problems in understanding astrophysical (in particular, accretion) disks: whether the disk interiors may be effectively viscous in spite of the absence of marnetorotational instability?
In this case a rotational homogeneous inviscid flow with a Keplerian angular velocity profile is spectrally stable, making the transient growth of perturbations a candidate mechanism for energy transfer from the regular motion to perturbations.
Transient perturbations differ qualitatively from perturbation modes and can grow substantially in shear flows due to the nonnormality of their dynamical evolution operator.
Since the eigenvectors of this operator, alias perturbation modes, are mutually nonorthogonal, they can mutually interfere, resulting in the transient growth of their linear combinations.
Physically, a growing transient perturbation is a leading spiral whose branches are shrunk as a result of the differential rotation of the flow.
This paper discusses in detail the transient growth of vortex shear harmonics in the spatially local limit as well as methods for identifying the optimal (fastest growth) perturbations.
Special attention is given to obtaining such solutions variationally, by integrating the direct and adjoint equations forward and backward in time, respectively.
The material is presented in a newcomer-friendly style.

\end{abstract}

\newpage

\tableofcontents

\section{Introduction: modal and non-modal analysis of perturbations}

A salient feature of disk accretion is that it is impossible without a dissipation mechanism of the differential rotation energy of matter. 
It is the internal friction in the disk, i.e. irreversible interaction of its adjacent rings, that leads to the transformation of the gravitational energy of the accreting matter into heat and electromagnetic radiation, which simultaneously allows the matter to flow towards the center and the angular momentum to flow outwards to the disk periphery.

A direct dissipation is possible already due to microscopic viscosity of gas (plasma), however in astrophysical conditions it turns out to be absolutely insufficient to explain the observed properties of the disks.
Essentially, the disks are too large for the characteristic accretion time, $t_\nu$,  to be explained by the microscopic viscosity.
For example, in protoplanetary disks with the typical size $L\sim 10$ a.u., where the kinematic viscosity is estimated to be $\nu_m \sim 10^7$ $\text{cm}^2$/s, the accretion time is $t_\nu = L^2 / \nu \sim 10^{13}$ years, (see Section 3.3.2 of book \cite{armitage-2009}).
Apparently, $t_\nu$ is several orders of magnitude as long as the age of the Universe.
At the same time, observations of gas-dust disks around young stars suggest that their lifetime is as short as only several million years (see, for example, review \cite{youdin-kenyon-2013}).
A similar conclusion is obtained for hot accretion disks around, in particular, black holes in close binary systems.
In this case, for much smaller scales $L\sim 10^{10}$ cm and somewhat smaller viscosity of hydrogen plasma $\nu_m \sim 10^5$ $\text{cm}^2$/s we get $t_\nu \sim 3 \times 10^7$ years, which exceeds by many orders of magnitude, for example, the duration of X-ray Nova outbursts caused by non-stationary disk accretion (see review \cite{remillard-mc-clintock-2006}). 

At the same time, it is known from statistical hydromechanics (see the discussion of the Reynolds equations in book \cite{monin-yaglom-1971}, v. 1, Ch. 3) that the presence of significant correlating fluctuations of velocity components in a flow is equivalent to the presence of a high effective viscosity that exceeds the microscopic viscosity because the mixing scale of matter in the flow is much larger than the free path length of individual particles.
In turn, the high effective viscosity enhances the angular momentum transfer towards the disk periphery thus decreasing $t_\nu$ to the observed values.
The perturbations under discussion generally can be regular: for example, the accretion can be due to tidal waves generated in the disk by the secondary companion of the binary system (see \cite{menou-2000}).
However, it is more natural to assume that these perturbations are generated by {\it turbulence} in the fluid.
The turbulence, on one hand, takes the energy from the rotational motion of matter on large scales, and on the other hand, via interaction of perturbations components with different wave numbers, cascades this energy to small scales where its direct dissipation into heat occurs  due to microscopic viscosity. 

It is important to recognize that energy transfer from regular flow to perturbations should be mediated by some {\it linear} mechanism that follows from the dynamics of small perturbations described by linearized hydrodynamic equations.
This can be rigorously proved for vortex fluid  motion using the Navier-Stokes equations (see book \cite{schmid-henningson-2001}, Section 1.4, as well as paper \cite{henningson-reddy-1994}).
Therefore the first natural step in theoretical study of turbulence generation in some (stationary) flow is to search for exponentially growing linear perturbations on a steady-state background. Such perturbations are usually referred to as {\it modes}, and the corresponding analysis is dubbed {\it modal} or spectral analysis of perturbations, since it is used to determine eigenvalues of the corresponding dynamical operator of the problem: (complex) mode frequencies. Turbulence arising from growing modes is called supercritical.
In astrophysical flows with Keplerian angular frequency, the spectral (magneto-rotational) instability and the corresponding supercritical (MHD) turbulence has been found in analytical and numerical calculations \cite{balbus-hawley-1991}, \cite{hawley-1995} and \cite{stone-1996} (see also reviews \cite{balbus-hawley-1998} and \cite{balbus-2003}) for disks with frozen seed magnetic field.
Nevertheless, the magneto-rotational instability does not operate in cold low-ionized disks.
Protoplanetary disks, accretion disks in quiescent states of cataclysmic variables, outer parts of accretion disks in active galactic nuclei provide the examples.
Thus, it would be very important to show that differential rotation alone is capable of exciting turbulence in Keplerian disks.
This property of Keplerian flows is universal, unlike the presence of a seed magnetic field together with sufficiently high degree of ionization of matter, or the existence of the flow inhomogeneities due to the vorticity jump (see, for example, review \cite{fridman-bisikalo-2008}), or the appearance of radial velocity gradients (see review \cite{kurbatov-2014}), of vertical and/or horizontal gradients of some thermodynamic values (see papers \cite{lovelace-1999} и \cite{klahr-hubbard-2014}, for example).
However, generation of turbulence in a {\it homogeneous} Keplerian flow without magnetic field remains questionable so far. 

The main difficulty here is that such a flow is spectrally stable: the specific angular momentum for the Keplerian rotation increases with radial distance from the center, therefore according to the Rayleigh criterion (\cite{rayleigh-1916} and \cite{landau-lifshitz-1987}, v. 6, paragraph 27) the growth of axially symmetric modes is impossible; in turn, non-axisymmetric modes cannot grow since the necessary Rayleigh condition on the existence of extremum of vorticity in the background flow (\cite{rayleigh-1880}, \cite{charney-1950}) is not fulfilled.
In spite of that (and as follows from laboratory experiments and numerical simulations),  turbulence arises in spectrally stable flows as well.
In this case it is called subcritical.
The plane-parallel Couette flow provides the simplest and the most prominent example (see classical monographs \cite{drazin-reid-1981}, \cite{joseph-1976}). 

In theory of hydrodynamic stability, the transition of some flow (with non-zero microscopic viscosity) to a turbulent state is usually characterized by a set of critical Reynolds numbers ${\rm Re}$ (see Section 1.3.2 of book \cite{schmid-henningson-2001}).
The smallest of them is the number ${\rm Re_E}$ such that at ${\rm Re} < {\rm Re_E}$ there are no initial perturbations, irrespective of their amplitudes, whose energy would grow at the initial time $t=0$.
${\rm Re_E}$ can be derived from the Reynolds-Orr energy equation (see Section 1.4 of book \cite{schmid-henningson-2001}).
For the Couette flow ${\rm Re_E}\sim 20$.
For ${\rm Re} > {\rm Re_E}$ initially growing perturbations at $t=0$ arise, but as long as ${\rm Re} < {\rm Re_G}$, again there are no initial perturbations with any amplitude that would not decay at $t\to \infty$.
This is the definition of the second critical number ${\rm Re_G} > {\rm Re_E}$.
Finally, at higher values ${\rm Re} > {\rm Re_G}$ there appear perturbations that can sustain their amplitude at all times, and starting from some ${\rm Re_T} > {\rm Re_G}$ the transition to a turbulent state is experimentally observed.
For the Couette flow ${\rm Re_T} \sim 360$.
The largest of the critical Reynolds numbers is ${\rm Re_L}> {\rm Re_T}$, starting from which growing modes arise, i.e. the flow becomes spectrally unstable.
For the Couette flow, as well as for a Keplerian flow of interest here, ${\rm Re_L} = \infty$.
However, the case of Keplerian flow is different in that up to the present time, the value of ${\rm Re_G}$ remains unknown, and ${\rm Re_T}$ has not been measured neither theoretically nor experimentally.

On the one hand, the general opinion emerged that for Keplerian flows ${\rm Re_G}={\rm Re_T}\to\infty$.
It is based on the indirect argument that (locally) the action of the tidal and Coriolis forces on the perturbation, which are absent in the Couette flow, strongly stabilizes the shear flow (see Fig. 9 in review \cite{balbus-hawley-1998}, on which the results from paper \cite{balbus-hawley-stone-1996} are shown).
This conclusion is supported by the local numerical simulations \cite{hawley-1999}, \cite{shen-2006} and series of laboratory experiments \cite{ji-2006}, \cite{schartman-2009} and \cite{schartman-2012}, in which the stability of a quasi-Keplerian flow was observed up to ${\rm Re} = 2 \times 10^6$.
Here we assume the quasi-Keplerian flow to be the so-called anti-cyclonic flow (see, for example, the definition in \cite{lesur-longaretti-2005}), where the specific angular momentum increases, while the angular velocity itself, in contrast, decreases towards the periphery.\footnote{In a cyclonic flow both these quantities increase with distance from the center.}

On the other hand, in a cyclonic flow the subcritical turbulence is observed at finite,  although large values ${\rm Re_T}$, see papers \cite{taylor-1936}, \cite{wendt-1933} on experiments with spectrally stable Taylor-Couette flow, as well as their analysis in astrophysical context in Zel'dovich's paper \cite{zeldovich-1981} and later in paper \cite{richard-zahn-1999}.
In addition, negative results obtained in numerical experiments mentioned above can be explained by insufficient numerical resolution, as discussed in paper \cite{longaretti-2002}.
In the subsequent paper \cite{lesur-longaretti-2005}, the dynamics of perturbations in cyclonic and anti-cyclonic flows was compared numerically.
It was concluded that the required numerical resolution in the second case is much higher than in the first case, and the current computational power is insufficient to discover turbulence in a Keplerian flow; also it is impossible to argue that the stabilizing action of the Coriolis force in this case excludes the existence of a finite value of ${\rm Re_T}<\infty$.
At last, another laboratory experiment presented in papers \cite{paoletti-lathrop-2011} and \cite{paoletti-2012} shows the appearance of subcritical turbulence  and angular momentum transfer outwards in the quasi-Keplerian flow.
The contradictory results claimed by different experimental groups show the complexity of the experiment due to inevitable arising of secondary flows induced by experimental tools.
Presently, the influence of axial boundaries on the laboratory flow is discussed (see \cite{avila-2012} and \cite{edlund-ji-2014}). 

Anyway, it can be stated that of all types of homogeneous rotating flows, quasi-Keplerian (anti-cyclonic) flows turned out to be the most stable relative to finite-amplitude perturbations.
Nevertheless, the smallness of microscopic viscosity in astrophysical conditions mentioned above simultaneously means that huge Reynolds numbers should exist in the disks: for example, if in the protoplanetary disk discussed above we take its thickness $H \sim 0.05L = 0.5$ a.u. as the natural limiting scale of the problem, which corresponds to the sound velocity in the disk at this radius $c_s \sim 0.5$ km/s, we get ${\rm Re}\approx 10^{10}$. In other astrophysical disks ${\rm Re}$ can be even higher.
Apparently, considering all negative results, for the possibility of turbulence in astrophysical Keplerian flows there are still several orders of magnitude: $10^6<{\rm Re_T}<10^{10}$.
\\
\\
Thus, a search for the critical value ${\rm Re_T}$ for Keplerian flows continues, and in the present paper we will discuss in detail the necessary condition for turbulence and/or enhanced angular momentum transfer to the disk periphery
--- the transition of energy from the regular flow to perturbations in such a flow.
As mentioned above, this transition should be mediated by a linear mechanism.
Here, since a Keplerian flow is spectrally stable, only (small) perturbations different from modes can provide such a mechanism.
The existence of such transiently growing non-modal perturbations in shear flow was suggested already in papers by Kelvin \cite{kelvin-1887} and Orr \cite{orr-1907a}, \cite{orr-1907b}.
In astrophysics, this problem was studied in stellar dynamics (see \cite{goldreich-lynden-bell-1965}, \cite{julian-toomre-1966}).
However, in the context of hydrodynamic stability, rigorous treatment of such perturbations and methods to determine them were elaborated only in the 1990s and were dubbed {\it non-modal} perturbation analysis.
To stress the inapplicability here of the traditional modal analysis, the corresponding concept of the transition to subcritical turbulence due to transient growth of perturbations was called the {\it bypass} transition.
The non-modal analysis of perturbations was formulated in papers \cite{farrell-1988}, \cite{butler-farrell-1992}, \cite{reddy-1993}, \cite{reddy-henningson-1993}, (see also reviews \cite{trefethen-1993}, \cite{schmid-2007} and book \cite{schmid-henningson-2001}).
These papers showed that mathematically the non-modal growth is due to non-orthogonality of the perturbation modes.
If modes with a physically motivated norm are non-orthogonal to each other, their linear combinations can grow in this norm, in spite of each separate mode being decaying, as in a spectrally stable flow (see Fig. \ref{modes_1} in Section \ref{sect_singular}).
In turn, the modes are non-orthogonal due to {\it non-normality} of the linear dynamical operator governing the perturbation evolution (see the introductory information about the operators in the same Section).
A non-normal operator does not commute with its adjoint operator, which is due to a non-zero velocity shear in the regular flow (see the concluding part of \ref{sect_adjoint} below for more detail).
Here the higher ${\rm Re}$, the higher the degree of non-orthogonality of modes to each other and, correspondingly, the higher {\it transient growth} is possible.
Papers mentioned above argue that the maximum possible transient growth of perturbations by a fixed time, called the optimal growth, is determined by the norm of a dynamical operator that can be obtained by calculating singular vectors of the operator (see \ref{sect_singular} for more detail).
Finally, the operator norm is tightly related to the notion of operator's pseudospectrum (see \cite{trefethen-1993} and book \cite{schmid-henningson-2001}).

Late this method was applied to astrophysical flows in papers \cite{ioannou-kakouris-2001}, \cite{yecko-2004}, \cite{mukhopadhyay-2005}, where different model were used to search for optimal perturbations demonstrating the optimal growth.
In particular, it was shown that for the Keplerian velocity profile the growth can be substantial only starting from ${\rm Re}\sim 10^6$, while in the similar setup for iso-momentum profile and the Couette flow the growth starts already at ${\rm Re}\sim 10^3$ (see the discussion in paper \cite{mukhopadhyay-2005}).
Here papers \cite{meseguer-2002} and \cite{maretzke-2014} should also be mentioned which discuss the transient dynamics in the spectrally stable Taylor-Couette flow and include both cyclonic and anti-cyclonic regimes.
Paper \cite{meseguer-2002} discovered a correlation between the experimentally obtained stability boundary in a laminar flow (see \cite{coles-1965}) and the optimal growth value; paper \cite{maretzke-2014} found that for one and the same Re number, in a quasi-Keplerian regime the transient growth is minimal.
Using the correlation from \cite{meseguer-2002}, the authors \cite{maretzke-2014} estimated ${\rm Re}_T\sim 10^5$ for the quasi-Keplerian regime.
As in the numerical experiments \cite{balbus-hawley-stone-1996}, \cite{hawley-1999} и \cite{shen-2006} mentioned above the effective ${\rm Re}$, caused by the numerical viscosity, were hardly above $\sim 10^4 \div 10^5$, it is not surprisingly that in these studies the Keplerian profile was stable against perturbations. 

In addition, presently there are a lot of astrophysical studies of the transient growth of local perturbations by the Lagrangian method, where the transformation to the reference frame co-moving to the shear is done and separate shear harmonics are considered (see Section \ref{sect_local_appr}).
It was found that in the local space limit, transiently growing vortex shear harmonics emit a wave shear harmonics of various type (depending on the account for the compressibility or some inhomogeneities in the flow)
at the moment of swing (see Section \ref{sect_local_appr}), which themselves demonstrate the non-modal growth \cite{lominadze-1988},\cite{fridman-1989}, \cite{chagelishvili-1997}, \cite{chagelishvili-2003}, \cite{tevzadze-2003}, \cite{afshordi-2005}, \cite{bodo-2005}, \cite{tevzadze-2008}, \cite{heinemann-papaloizou-2009a},\cite{heinemann-papaloizou-2009b}, \cite{tevzadze-2010}, \cite{volponi-2010}, \cite{salhi-pieri-2014}.

At last, papers \cite{umurhan-2006} and \cite{rebusco-2009} investigated the non-linear transient dynamics of three-dimensional perturbations with account for the global structure of the flow in the model of geometrically thin disk with $\alpha$-viscosity.
As in paper \cite{ioannou-kakouris-2001}, these papers discussed the possibility of exciting non-modal perturbations by weak turbulence already present in the disk and giving rise to low effective viscosity parametrized by the $\alpha$-parameter.
In Section \ref{sect_TG} we will also consider the influence of the effective viscosity on the transient growth of vortices with different scales in comparison to the disk thickness.
Thus, the transient growth of perturbations can be discussed not only in the context of the bypass transition of a laminar flow to turbulence, but as a mechanism to enhance the angular momentum transfer in the disk with pre-existing weak turbulence producing low viscosity.
In the last case this turbulence can be mathematically treated as an external stochastic perturbation in a shear flow, which transits to a quasi-stationary state with significant amplitude increase of perturbations due to non-normality of the linear operator governing their dynamics (see \cite{ioannou-kakouris-2001}). 

\begin{figure}[h!]
\includegraphics[width=1.\linewidth]{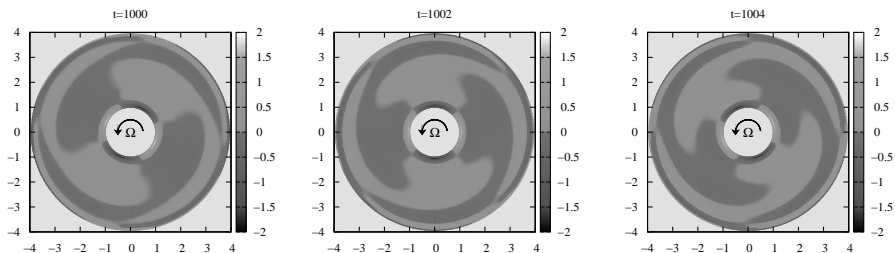}
\caption{\footnotesize{
Contours of the most unstable perturbation mode with azimuthal wave number $m=2$ in the model of a quasi-Keplerian thin torus described in \ref{sect_martix}.
Parameters of the calculation: the characteristic disk aspect ratio $\delta =0.3$, inner and outer boundaries are at $r_1=1$ and $r_2=4$, respectively, the polytropic index of matter $n=3/2$.
The mode increment and phase velocity are $\Im[\omega]\approx 0.001$ and $\Re[\omega]\approx 0.26$, respectively.
Shown is the time (in units of inverse Keplerian frequency at the inner disk edge) since the conventional moment when the mode had the unit amplitude.
The arrow shows the rotational direction of matter in the disk.
The method of calculation is described in \ref{sect_global_var}.
}}
\label{pic_modes}
\end{figure}

\begin{figure}[h!]
\includegraphics[width=1.\linewidth]{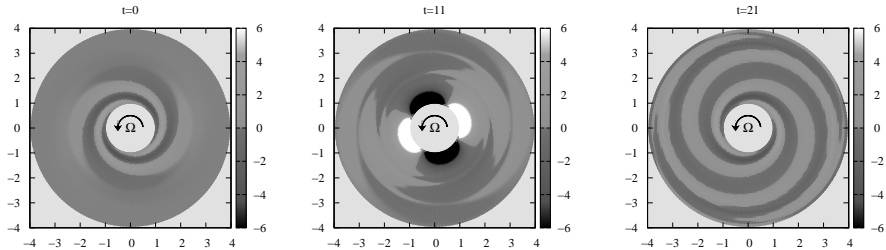}
\caption{\footnotesize{
Contours of the perturbation $m=2$ demonstrating a maximum possible transient growth of acoustic energy at time $t_{opt}=10$ counted from the beginning of the perturbation evolution in units of the inverse Keplerian frequency at the inner disk edge.
The initial perturbation has conventionally unity  amplitude, the model of the flow is the same as in Fig. {\ref{pic_modes}}.
The method of calculation is described in \ref{sect_global_var}.
}}
\label{pic_TG}
\end{figure}

The purpose of the present paper is to consider in detail the transient growth phenomenon using the simplest example of two-dimensional adiabatic perturbations in a homogeneous rotating shear flow with quasi-Keplerian angular velocity profile.
In Section 2 we present the analysis of shear vortex harmonics that are responsible for the transient growth in the spatially local treatment of the problem and discuss the mechanism of perturbation growth using them as an example.
Sections 3,4 are mainly devoted to methods of studying the non-modal perturbation growth as well as to determination of optimal perturbations attaining the maximum growth.
Two methods of obtaining the optimal growth curve are presented: a matrix and variational one.
The variational method is less applied, especially in astrophysical studies (see \cite{zhuravlev-razdoburdin-2014}), however, essentially it is more universal than the matrix one.
For example, using this method, in the present paper we calculated one of optimal transient perturbations in a geometrically thin quasi-Keplerian flow with free boundaries (Fig. \ref{pic_TG}) as well as the most unstable perturbation mode (Fig. \ref{pic_modes}), which we discuss in detail in the concluding part of Section \ref{sect_global_var}.
Comparison of Fig. \ref{pic_TG} and Fig. \ref{pic_modes} shows that these two types of perturbations are indeed qualitatively different: the transient spiral is wound up by the flow and its amplitude increases, while the modal spiral rotates as solid body and demonstrates a monotonic but very weak growth because of low instability increment.
Here the phase velocity of the modal spiral is such that its corotation radius, at which the energy is transferred from the regular flow, lies inside the flow. 

\section{Analytical treatment for two-dimensional vortices}

\subsection{Adiabatic perturbations in rotational shear flow}
\label{sect_adiabat_perts}

Consider first the dynamics of small adiabatic perturbations in perfect fluid with isentropic equation of state.
Perturbations will be described using the Euler approach, i.e. variations of physical quantities such as density $\rho$, velocity ${\bf v}$ and pressure $p$ at a given point of space at a given time in the perturbed flow relative to the unperturbed background \footnote{See monograph \cite{pringle-king-2007} concerning applications of hydrodynamics to astrophysical problems, in particular, on the application of theory of hydrodynamic perturbations.}
For simplicity we assume that there are no entropy gradients in the fluid.
Then in the right-hand side of the Euler equations it is convenient to pass from the pressure gradient to the enthalpy gradient.
Indeed, under constant entropy the enthalpy differential per unit mass is $d h = dp/\rho$ (see \cite{landau-lifshitz-1980}), and this is valid in both the background and perturbed flows.
Therefore, for Euler perturbations we get $\delta (\nabla p / \rho) = \nabla \delta h$.
Making use of this relation, we write down equations for $\delta \rho$, $\delta h$ and $\delta {\bf v}$ (see also \cite{landau-lifshitz-1987}, paragraph 26) in the form:

\begin{equation}
\label{orig_sys1}
\frac{\partial \delta {\bf v}}{\partial t} + ({\bf v}\cdot\nabla)\delta {\bf v} + (\delta {\bf v}\cdot\nabla) {\bf v} = 
-\nabla \delta h,
\end{equation}

\begin{equation}
\label{orig_sys2}
\frac{\partial \delta\rho}{\partial t} + \nabla \cdot(\rho \delta{\bf v}) + \nabla \cdot (\delta \rho {\bf v}) = 0,
\end{equation}

where we have assumed that ${\bf v}$ and $\rho$ are the velocity and density in the unperturbed (background) flow, respectively, which itself can evolve in time.
Equations (\ref{orig_sys1}) and (\ref{orig_sys2}) are linear, since perturbations are small, and all quadratic terms are omitted. 

\subsubsection{The model and basic equations}
To write down the projections of corresponding equations, let us specify the model we wish to consider to illustrate the transient dynamics.
First of all, we assume that the background flow is {\it stationary} and is purely rotational, which is well satisfied in astrophysical disks.
This means that the flow is axially symmetric, and it is convenient to use the cylindric coordinate system $(r,\varphi,z)$ in which the velocity has only azimuthal non-zero component ${\bf v} = (0,v_\varphi,0)$.
Below we will also use the angular velocity of the flow, $\Omega = v_\varphi / r$.
It is important to note that isentropicity of the  fluid  (which is a particular case of barotropicity) immediately implies that $v_\varphi$ and $\Omega$ depend only on the radial coordinate (see \cite{tassoul-1978}, paragraph 4.3).
At the same time, the density in equations (\ref{orig_sys1}, \ref{orig_sys2}) is a function of both $r$ and $z$: $\rho=\rho(r,z)$.
Most frequently, the case of geometrically thin disk, where $H(r)/r\ll 1$ and $H$ is the disk semi-thickness, takes place. 
This assumption will be useful to find how the density $\rho$ changes with height above the equatorial disk plane.
Let use the hydrostatic equilibrium condition in the background flow:
\begin{equation}
\label{vert_eq}
\frac{\partial h}{\partial z} = -\Omega^2(r) z,
\end{equation}
where the vertical gravity acceleration due to the central gravitating body around which the disk rotates stands in the right-hand side.
This acceleration is written here ignoring quadratic corrections in the small parameter $z/r$.
Integrating (\ref{vert_eq}) with the condition $h(z=H)=0$ yields the vertical enthalpy distribution:
\begin{equation}
\label{vert_enthaply}
h = \frac{1}{2} (\Omega H)^2 \left ( 1 - \frac{z^2}{H^2} \right ).
\end{equation}
Next, due to the constant entropy assumption $p\propto \rho^\gamma$, where $\gamma=1+1/n$ is the adiabatic index of matter written via the polytropic index $n$.
This means that the square of the sound velocity in the background flow is $a^2 = \gamma p/\rho$, and the density will be mainly dependent on $z$ as follows:
\begin{equation}
\label{rho_a2}
a^2 \propto \left ( 1 - \frac{z^2}{H^2} \right ),\quad \rho \propto \left ( 1 - \frac{z^2}{H^2} \right )^{n}.
\end{equation}

Finally, for simplicity we will consider only perturbations in which $\delta {\bf v}$ {\it is independent} on $z$.
Generally, this very strong assumption needs justification.
In particular, it is relevant to ask: if we take initial perturbations with such property, will it be conserved in further evolution, and if not, how rapidly will this assumption be violated?
The answer depends on the vertical disk structure.
For example, in paper \cite{okazaki-1987} it was shown that in the particular case of isothermal vertical density distribution (($n\to \infty$)), small perturbations with homogeneous velocity field in $z$ are exact solutions to equations (\ref{orig_sys1}), (\ref{orig_sys2}).
In more general case with finite $n$ this is no longer the case, however, for example, three-dimensional simulations of  barotropic toroidal flows indicate that the most unstable perturbations there weakly depend on $z$ (see paper \cite{frank-robertson-1988}).
This can be related to the fact that when the angular velocity  is independent of $z$, the Reynolds stresses, responsible for the energy transfer from the main flow to perturbations, do not depend on the vertical component of the velocity perturbation (\cite{kojima-1989}, \cite{kojima-1989b}).
At last, three-dynamical study of transient dynamics of vortices in a Keplerian flow \cite{yecko-2004} also shows that the most rapidly growing perturbations in a vertically non-stratified medium are almost independent of $z$ (see also \cite{maretzke-2014}).
Now, looking at the vertical radial and azimuthal projections of (\ref{orig_sys1}), we see that our assumption implies the independence of $\delta h$ on $z$, and therefore the right-hand side of the vertical projection of (\ref{orig_sys1}) vanishes.
Then, if we additionally assume that initial vertical velocity perturbations are absent, $\delta v_z = 0$, they will not appear later as well.
Therefore, in the perturbed flow, as well as in the background flow, the vertical hydrostatic equilibrium will take place.
It can be shown that the assumption of vertical hydrostatic equilibrium in the perturbed flow is equivalent to the assumption of the homogeneous in $z$ velocity perturbation field, i.e. one assumption is always follows from another.
At the same time, if the fluid is not isentropic and there is a radial entropy gradient in the disk, the simplifying assumptions made above are insufficient to set $\delta v_z$ to zero. 

Thus,  we came to the conclusion that we will deal with a flat velocity perturbation field, i.e. $\delta {\bf v} = \{\delta v_r, \delta v_\varphi, 0\}$, with $\delta v_r$ and $\delta v_\varphi$, like $\delta h$, being dependent on the radial and azimuthal coordinates only.
However, it is important to emphasize that this is not the case for $\delta \rho$ that enters the continuity equation (\ref{orig_sys2}).
Here it is convenient to use the relation between the pressure and density variations in isentropic fluid, $d p = a^2 d\rho$, which is the consequence of the barotropic equation of state.
Due to the universal character of this relation, small Eulerian perturbations will be related in the same way, i.e. $\delta p = a^2 \delta\rho$, where $a^2$ is the speed of sound in the background flow.
Consequently,
\begin{equation}
\label{rho_h}
\delta \rho = (\rho/a^2)\delta h,
\end{equation}
and this expression will be plugged into (\ref{orig_sys2}), after which only background quantities in equation (\ref{orig_sys2}) will depend on the radial coordinate.
When integrating equation (\ref{orig_sys2}) in its new form over $z$, we should keep in mind that 
\begin{equation}
\label{int_z}
\int\limits_{-H}^{H} \frac{\rho}{a^2} dz =  \sqrt{\pi} \frac{\Gamma(n)}{\Gamma(n+1/2)} \frac{\rho}{a^2} \Biggr |_{z=0}, \quad \int\limits_{-H}^{H} \rho dz \equiv \Sigma =  \sqrt{\pi} \frac{\Gamma(n+1)}{\Gamma(n+3/2)} \rho |_{z=0},
\end{equation}
where we have used relation (\ref{rho_a2}) and introduced the surface density $\Sigma$. 

Using in (\ref{orig_sys2}) the fundamental property of the gamma-function, $\Gamma({\bf z}+1) = {\bf z}\Gamma({\bf z})$, we can explicitly write down the system of equations (\ref{orig_sys1}), (\ref{orig_sys2}) for azimuthal complex Fourier harmonics $\delta v_r$, $\delta v_\varphi$, $\delta h \propto \exp({\rm i}m \varphi)$
\begin{equation}
\label{sys_A_1}
\frac{\partial \delta v_r}{\partial t} = -{\rm i}m\Omega\, \delta v_r + 2\Omega \delta v_\varphi - \frac{\partial \delta h}{\partial r},
\end{equation}

\begin{equation}
\label{sys_A_2}
\frac{\partial \delta v_\varphi}{\partial t} =  -\frac{\kappa^2}{2\Omega} \delta v_r  -{\rm i}m\Omega\, \delta v_\varphi -\frac{{\rm i} m}{r} \delta h, 
\end{equation}

\begin{equation}
\label{sys_A_3}
\frac{\partial \delta h}{\partial t} = -\frac{a_*^2}{r\Sigma} \frac{\partial}{\partial r} (r\Sigma \delta v_r) -\frac{{\rm i} m a_*^2}{r} \delta v_\varphi  -{\rm i}m\Omega \,\delta h,
\end{equation}
where $a_*^2 \equiv n a_{eq}^2/(n+1/2)$, and $a_{eq}$ is the background speed of sound in the equatorial disk plane.
In addition, $\kappa^2 = (2\Omega/r) d/dr (\Omega r^2)$ is the square of the epicyclic frequency, i.e. the frequency of free oscillation of the fluid in the $(r,\varphi)$ plane, which can be easily checked by writing (\ref{sys_A_1}), (\ref{sys_A_2}) for $\delta h = 0$ and substituting there the solution $\delta v_r, \delta v_\varphi \propto \exp(-{\rm i}\omega t)$.
We mention that the reducing the three-dimensional problem to the effectively two-dimensional one in a thin disk, clearly, can be performed by simple changing of the volume density by the surface density and of the polytropic index by $n+1/2$ in the original, not integrated over $z$ equations,  as was first shown in paper  \cite{churilov-shuhman-1981}. 

\subsubsection{Types of perturbations}
The system of equations (\ref{sys_A_1})-(\ref{sys_A_3}) describes the dynamics of two types of perturbations inside the disk which are possible in the two-dimensional formulation of the problem: vortices and density waves
\footnote{Density waves are also frequently referred to as inertial-acoustic waves.}
The separation between them for transient perturbations will be described below in the local framework that allows a more simple physical interpretation of the behavior of perturbations in a differentially rotating flow.
In addition, when there are free radial boundaries in the background flow (for example, in a disk with finite radial extension when at some inner and outer radii $\Sigma$ vanishes and the shear acquires super-Keplerian angular velocity gradient), the surface gravity waves arise near the boundaries (see papers \cite{blaes-glatzel-1986}, \cite{glatzel-1987a}, \cite{glatzel-1987b}).
This occurs because of the presence of a somewhat significant radial pressure gradient in the flow is equivalent to a non-zero gravitational acceleration which gives rise to waves similar to ocean waves running over the free surfaces (or radial density jumps).

\subsubsection{On the perturbation modes}
These types of perturbations were studied in detail in the 1980s by spectral method, when the system of equations (\ref{sys_A_1})-(\ref{sys_A_3}) was solved for particular temporal Fourier harmonics $\propto \exp(-{\rm i}\omega t)$ called {\it modes} (see reviews \cite{narayan-goodman-1989} and \cite{narayan-1991}).
In this analysis, the local dispersion relation gives only real values of $\omega$ in all astrophysically important cases where $\Omega(r)$ is such that the specific angular momentum $\Omega r^2$ increases with radius outwards.
This means the local stability of the disks and prohibits exponential growth of small-scale perturbations, which is also in accordance with the well-known Rayleigh criterion for the particular case of axially symmetric perturbations  (see paragraph 27 in\cite{landau-lifshitz-1987}).
Unlike this case, the global setup of the problem for axially non-symmetric modes, when the system of differential equations with respect to the radial coordinate with the corresponding boundary conditions at the inner disk radius and at infinity (or at the outer disk boundary)  is solved, yields a discrete set of $\omega$, where there can be complex frequencies as well (see, for example, \cite{papaloizou-prinle-1984}, \cite{papaloizou-prinle-1985}, \cite{papaloizou-prinle-1987}, \cite{glatzel-1987a}, \cite{glatzel-1987b}, \cite{goldreich-1986}, \cite{kojima-1986a}, \cite{kato-1987}, \cite{sekiya-miyama-1988}, etc.)
The non-zero real part of the frequency corresponds to the angular velocity of {\it solid-body} rotation of a given mode in the flow.
Generally, the solid-body azimuthal motion of constant phase of perturbations with the same azimuthal velocity $\Re[\omega]/m$ at all $r$ is the main distinctive feature of modes among other perturbations.
Here $\Re$ means the real part of the frequency $\omega$.
A non-zero imaginary part of the frequency, $\Im[\omega]$, means that the (canonical, see \cite{friedman-schutz-1978}) energy and angular momentum are exchanged between this mode and either the background flow \cite{goldreich-narayan-1985}, \cite{drury-1985}, \cite{narayan-1987}, \cite{papaloizou-prinle-1987} or the mode with (canonical) energy of the opposite sign \cite{glatzel-1987b}, \cite{glatzel-1988}, \cite{savonije-heemskerk-1990}.
In the literature, the first mechanism is also referred to as the Landau mechanism, and the second one --- as the mode coupling.
The energy exchange in both cases is resonant, i.e. always occurs in the so-called critical layer at the radius where $\omega = m\Re[\Omega]$, which is called the corotation radius.
See monograph \cite{stepanyants-fabrikant-1989} for a detailed discussion of the physics of these resonant mechanisms of mode growth (decay).
Nevertheless, in flows with almost Keplerian rotation both the mode coupling and their interaction with the background occurs extremely slow, and the corresponding increments even for substantial disk aspect ratio $H/r \sim 0.1$ is only one hundred thousandth of the characteristic Keplerian frequency \cite{zhuravlev-shakura-2007a}, \cite{zhuravlev-shakura-2007b}.
This result led to the conclusion that at least in the simplest barotropic disks the modes cannot underly any hydrodynamic activity and, in particular, cannot induce turbulence or another variant of enhanced angular momentum transfer to the flow periphery.

\subsubsection{On measurements of perturbations}
To conclude this Section, let us discuss the problem of perturbations measurements.
Indeed, in the present paper we are interested in how strongly can some perturbations grow in a given time interval.
To describe this quantitatively, it is necessary to introduce the norm of perturbations which would characterize the amplitudes of $\delta v_r, \delta v_\varphi, \delta h$ at a given time.
This should be a real and positive definite quantity.
The most natural one is the total acoustic energy of the perturbation in the disk that has the form
\begin{equation}
\label{ac_en}
E = \pi \int \Sigma \left ( |\delta v_r|^2 + |\delta v_\varphi|^2  + \frac{|\delta h|^2}{a_*^2}  \right ) r\, dr,
\end{equation}
where we have integrated over the azimuthal coordinate.

After taking derivative of (\ref{ac_en}) with respect to time and making use of (\ref{sys_A_1})-(\ref{sys_A_3}), we obtain (see also expression (8) from paper \cite{savonije-heemskerk-1990}):

\begin{equation}
\label{ac_en_der}
\frac{dE}{dt} = -2\pi\int \frac{d\Omega}{dr} r \Sigma \Re [\delta v_r \delta v_\varphi^*]\, r\, dr  - 2\pi r\Sigma \Re [\delta v_r \delta h^*]\, |_{r_1,r_2},
\end{equation}
where the symbol $*$ means complex conjugation and $r_1$ and $r_2$ are the inner and outer boundaries of the flow, respectively.
Here $r_2$ can be at infinity.
As $\Sigma\to 0$ at the flow boundaries, the second term in the right-hand side of (\ref{ac_en_der}) disappears, and we see that $E$ can change exactly in the differentially rotating body.
Without rotation or for solid-body rotation $E$ remains time constant.
It is important to note that the increase/decrease of $E$ will imply that the average in the flow amplitudes $\delta v_r, \delta v_\varphi$ and $\delta h$, also increase/decrease, since (\ref{ac_en}) contains squares of modules of these values taken with the same signs.
Note that for modes, equation (\ref{ac_en_der}) implies
\begin{equation}
\label{E_mode_dot}
\frac{dE}{dt} \propto \exp(2\Im[\omega]t),
\end{equation}
i.e. small increments obtained for quasi-Keplerian flows allows us to conclude that the total acoustic energy of modes there $E\simeq const$ on dynamic $\sim\Omega^{-1}$ and sound $\sim (\Omega H/r)^{-1}$ time scales.

Our task now is to understand how can $E$ change over the same time intervals for arbitrary perturbations.
Thus, by introducing the perturbation vector ${\bf q}(t)$ as a set of functions $\{\delta v_r(r), \, \delta v_\varphi(r), \delta h(r)\}$ taken at some time $t$, the norm of the perturbation can be chosen as
\begin{equation}
\label{ac_norm}
{||{\bf q}(t)||^2} = E(t).
\end{equation}

\subsection{Local approximation: transition to shear harmonics}
\label{sect_local_appr}

The easiest solution of the problem formulated above can be obtained in the local space approximation.
In this approximation it is assumed that the characteristic scale of perturbations, $\lambda$, is a small fraction of some fiducial radial coordinate $r_0$ around which the dynamics of perturbation is studied, $\lambda\ll r_0$.
Introduce new radial variable $x\equiv r-r_0\ll r_0$ and also new azimuthal variable $y\equiv r_0(\varphi-\Omega_0 t) \ll r_0$, where $\Omega_0\equiv \Omega(r_0)$ is the angular velocity of rotation of the new coordinate system.
Here in equations (\ref{sys_A_1})-(\ref{sys_A_3}) only leading terms in small $x$ are retained. In practice, this means that only linear in $x$ dependence should be taken into account in the angular velocity profile:
\begin{equation}
\label{Om_loc}
\Omega = \frac{d\Omega}{dx}\Biggr |_{r_0} x = -q\Omega_0 \frac{x}{r_0} \ll \Omega_0,
\end{equation}
where $q \equiv - (r/\Omega) (d \Omega/dr )|_{r=r_0}$ and $\Omega(x=0)=0$, because we are working in the frame rotating with angular velocity $\Omega_0$.
The corresponding linear background velocity is $v_y^{loc}=r_0\Omega = -q\Omega_0 x$. 

Next, in the right-hand side of equations (\ref{sys_A_1})-(\ref{sys_A_3}) we keep only terms of the order up to $\sim x/\lambda$ and drop the terms $\sim x/r_0$ and lower.
For clarity, write down also the coefficient before $\delta v_r$ in the term from (\ref{sys_A_2}) that includes $\kappa^2$:
$$
-\frac{\kappa^2}{2\Omega} = -2\Omega - r\frac{d\Omega}{dr} = 2q\Omega_0\frac{x}{r_0} + (r_0+x) \frac{q\Omega_0}{r_0} = 3q\Omega_0\frac{x}{r_0} + q\Omega_0
$$
and only the term $q\Omega_0$ is sufficient to take into account.
Next, bearing in mind that the new reference frame is not inertial, it is necessary to add the perturbed Coriolis force components $2\Omega_0 \delta v_\varphi$ to the right-hand side of (\ref{sys_A_1}) and $-2\Omega_0 \delta v_r$ to the right-hand side of (\ref{sys_A_2}).

After substituting ${\rm i}m \to \partial/\partial \varphi$ in the system (\ref{sys_A_1})-(\ref{sys_A_3}), i.e. after returning back to the arbitrary dependence of the Eulerian perturbations on $\varphi$ and by denoting the local analogs of perturbations of the velocity components as $u_x$, $u_y$ and $W$, respectively, we arrive at the following equations:
\begin{equation}
\label{sonic_sys1}
\left ( \frac{\partial}{\partial t} - q\Omega_0 x\frac{\partial}{\partial y} \right ) u_x - 2\Omega_0 u_y =
-\frac{\partial W}{\partial x},
\end{equation}
\begin{equation}
\label{sonic_sys2}
\left ( \frac{\partial}{\partial t} - q\Omega_0 x\frac{\partial}{\partial y} \right ) u_y + 
(2 - q)\Omega_0 u_x =
-\frac{\partial W}{\partial y},
\end{equation}
\begin{equation}
\label{sonic_sys3}
\left ( \frac{\partial}{\partial t} - q\Omega_0 x\frac{\partial}{\partial y} \right ) W + 
a_*^2 \left ( \frac{\partial u_x}{\partial x} + \frac{\partial u_y}{\partial y} \right ) = 0.
\end{equation}

The system of equations (\ref{sonic_sys1})-(\ref{sonic_sys3}) was first derived in paper \cite{goldreich-lynden-bell-1965}
\footnote{Even earlier, in the context of lunar dynamics, the local approach to study the motion of matter was utilized by Hill  \cite{Hill-1878}.}
(see also paper \cite{regev-umurhan-2008}), where it is described for different background flow models. 

\subsubsection{Transition to shear harmonics}
A convenient property of the system of equations (\ref{sonic_sys1})-(\ref{sonic_sys3}) is that by changing to variables corresponding to the co-moving shear reference frame, it is possible to make it homogeneous in both $x$ and $y$, which, in turn, enables us to split arbitrary perturbation into individual spatial Fourier harmonics (SFHs) with certain wave numbers $k_x$ and $k_y$.
Indeed, introduce new {\it dimensionless} variables $x^\prime = \Omega_0 x/a_*,\, y^\prime = \Omega_0 (y+q\Omega_0 xt)/a_*,\, t^\prime=\Omega_0 t$
\footnote{Due to the vertical hydrostatic equilibrium in the disk, this means that we express the length in units of its semi-thickness,  $H = a_*/\Omega_0$.}.
Such a substitution corresponds to the change of partial derivatives according to the rule
\begin{equation}
\label{derivatives}
\frac{a_*}{\Omega_0}\frac{\partial}{\partial x} = \frac{\partial}{\partial x^\prime} + q t^\prime \frac{\partial}{\partial y^\prime}, \quad \frac{a_*}{\Omega_0}\frac{\partial}{\partial y} = \frac{\partial}{\partial y^\prime}, \quad
\Omega_0^{-1}\frac{\partial}{\partial t} = \frac{\partial}{\partial t^\prime} + q x^\prime \frac{\partial}{\partial y^\prime}
\end{equation}

Making use of (\ref{derivatives}), we get the system of equations in which all coefficients depend only on $t^\prime$.
Now substitute into this system SFH written in the form
\begin{equation}
\label{SFH}
f = \hat f (k_x,k_y,t^\prime) \exp ({\rm i} k_x x^\prime + {\rm i} k_y y^\prime),
\end{equation}
where $f$ is any of unknown variables, $\hat f$ is its Fourier amplitude, $k_x$ and $k_y$ are dimensionless wave numbers along axes $x^\prime$ and $y^\prime$, respectively, expressed in units $\Omega_0/a_*$.
Changing back to variables $x$, $y$  in particular solutions (\ref{SFH}) reveals that they represent perturbations periodic in space whose phase forms a plane front with orientation depending on time for $k_y\neq 0$.
The dimensionless wave number along $x$ has the form $\tilde k_x(t) \equiv k_x+q k_y t$ and changes with time: the wave vector turns around during advection by the shear flow, which was first noted by Kelvin \cite{kelvin-1887} and Orr \cite{orr-1907a}, \cite{orr-1907b} so the SFH are often called {\it shear harmonics}.
Note from the very beginning that for $\tilde k_x < 0$ the wave vector is directed inside the disk, and in the global scale for the Fourier harmonics with wave number $m$ this corresponds to the so-called {\it leading} spirals whose arms turned to the disk rotation direction.
Inversely, the case $\tilde k_x > 0$ corresponds to the {\it trailing} spirals whose arms are turned oppositely to the disk rotation.
If at the initial time $k_x<0$ , the arms of the initially leading spiral are deformed and shortened by the flow, and then the so-called {\it swing} moment occurs, $t_s$, when the wave vector of SFH is strictly azimuthal and $\tilde k_x(t_s)=0$, after which the spiral becomes trailing, and its arms start stretching by the flow (see Fig. \ref{pic_TG}).
This process is well-known in the dynamics of stellar galactic disks (see paragraph 6.3.2 in \cite{binney-tremaine-2008}). 

Thus, for SFH we arrive at the following system of ordinary differential equations:
\begin{equation}
\label{sonic_sys1_sh}
\frac{d \hat u_x}{d t} = 2\hat u_y - {\rm i}\,\tilde k_x(t) \hat W,
\end{equation}

\begin{equation}
\label{sonic_sys2_sh}
\frac{d \hat u_y}{d t} = -(2 - q) \hat u_x - {\rm i}\, k_y \hat W,
\end{equation}

\begin{equation}
\label{sonic_sys3_sh}
\frac{d \hat W}{d t} = - {\rm i}\, ( \, \tilde k_x(t) \hat u_x + k_y \hat u_y \,),
\end{equation}
where $\hat u_x$ and $\hat u_y$ are expressed in units $a_*$ and $\hat W$  – in units $a_*^2$.
Here and below we will omit the prime for the time variable notation.

\subsubsection{Potential vorticity}
Equations (\ref{sonic_sys1_sh})-(\ref{sonic_sys3_sh}) have the important property: the quantity
\begin{equation}
\label{vorticity}
I = \tilde k_x(t) \hat u_y  - k_y \hat u_x + {\rm i} (2-q) \hat W
\end{equation}
is the invariant of motion, which can be easily verified by the direct calculation of $dI/dt$.

It turns out that $I$ (to the multiplication factor ${\rm i}$) is SFH of the Eulerian perturbation of the potential vorticity.
The potential vorticity $\mbox{\boldmath$\zeta$}$, which is by definition the vorticity itself divided by density, $\mbox{\boldmath$\zeta$} \equiv \mbox{\boldmath$\omega$}/\rho$ (see  \cite{johnson-gammie-2005a}), is conserved in all fluid elements in plane-parallel barotropic flows.
Therefore, for its Eulerian perturbation we have

\begin{equation}
\label{pot_vort}
\delta \left (\frac{d \mbox{\boldmath$\zeta$}}{dt} \right ) = \frac{d\delta \mbox{\boldmath$\zeta$}}{d t} + ( \delta {\bf v} \nabla ) \mbox{\boldmath$\zeta$}_0 = 0,
\end{equation}
where $\mbox{\boldmath$\zeta$}_0$ is the potential vorticity of the background flow.
As in both background and perturbed flows the velocity fields are plane-parallel, the vorticity has only one non-zero $z$-component, which we will consider scalar below.

Next, by definition (in a non-rotating cylindrical coordinate system), the potential vorticity in the background flow is
\begin{equation}
\label{zeta_0}
\zeta_0 = (r\Sigma)^{-1}d/dr(\Omega r^2) = \kappa^2/(2\Omega\Sigma) = (2-q)\Omega/\Sigma,
\end{equation}
and should be constant in the local space approximation in use, because the velocity shear then is constant, cf. (\ref{Om_loc}).
Therefore, the second term  in the last equality in (\ref{pot_vort}) vanishes, and we see that $\delta\zeta$ is indeed conserved.
Apparently, the first two terms in (\ref{vorticity}) arise due to perturbation of the vorticity itself, which is equal to curl of the velocity perturbation, and the third term emerges due to the non-zero density perturbation represented by the dimensionless quantity $\hat W$ (the coefficient $2-q$ here arises due to multiplication by the constant background vorticity, cf. (\ref{zeta_0})). 

\subsubsection{Inhomogeneous wave equations. Density waves and vortices}
Now differentiate equation (\ref{sonic_sys2_sh}) with respect to $t$ and take into account the relations following from other two equations (\ref{sonic_sys1_sh}), (\ref{sonic_sys3_sh}), as well as the definition (\ref{vorticity}), to obtain a new equation:
\begin{equation}
\label{u_y_double}
\frac{d^2\hat u_y}{dt^2}  +  K(t) \hat u_y = \tilde k_x(t) I ,
\end{equation}
where $K(t)\equiv \tilde k_x^2(t)+k_y^2+2(2-q)$. Apparently, (\ref{u_y_double}) represents a detached {\it wave} equation for azimuthal velocity component perturbation, $\hat u_y$, with {\it inhomogeneous} part $\sim I$ \cite{bodo-2005}. 

In a similar way, from (\ref{sonic_sys1_sh}) and (\ref{sonic_sys3_sh}) we derive two equations of the same type:
\begin{equation}
\label{u_x_double}
\frac{d^2\hat u_x}{dt^2} + K(t) \hat u_x +2{\rm i}q k_y \hat W = -k_y I,
\end{equation}

\begin{equation}
\label{W_double}
\frac{d^2\hat W}{dt^2} + K(t) \hat W + 2{\rm i} q k_y \hat u_x = -2{\rm i} I,
\end{equation}
which can be separated by changing variables $\hat u_{\pm} = (\hat u_x\pm \hat W)/2$ \cite{heinemann-papaloizou-2009a}.

Consider in more detail, for example, equation (\ref{u_y_double}).
Its general solution is the sum of the general solution of the corresponding homogeneous equation and a partial solution of the inhomogeneous equation.
First consider both these solutions in the solid-body rotation limit, i.e. without the shear, $q=0$.
Then all coefficients in (\ref{u_y_double}) turn constant and 
\begin{itemize}
 \item 
the homogeneous equation has partial fundamental solutions $\hat u_y^{(dw)}\propto \exp(\pm{\rm i}\omega t)$ with frequency $\omega=\sqrt{K}$, corresponding to the density waves propagating in the opposite directions, 

\item
the partial solution with non-zero right-hand part can be taken as the constant $\hat u_y^{v} = (k_x/K)\, I$.
In other words, $u_y^{(v)}$ corresponds to the zero frequency $\omega=0$ and represents a static perturbation.
This perturbation, apparently, has a non-zero vorticity and corresponds to a vortex (it is possible to show that divergence of the velocity perturbation for this solution vanishes, by taking  the similar solution for $\hat u_x$ from equation (\ref{u_x_double}), $\hat u_x^{v}$, and checking that the combination $k_x\hat u_x^{v} + k_y \hat u_y^{v}=0$). 

\end{itemize}

\subsubsection{Amplification of the density waves}
With account for the non-zero shear, the density wave frequency becomes a function of time.
For example, for leading/trailing spirals this frequency gradually decreases/increases with the simultaneous wavelength increase/decrease, which, in turn, in the absence of viscosity leads to a monotonic decrease/increase in the energy and amplitude of the density waves.
Such a growth of the density waves amplitude was studied in \cite{chagelishvili-1994} and \cite{chagelishvili-1997}.
The reason can be understood from the fact that due to the axial symmetry of the background flow, the canonical angular momentum of the wave, $J_c$, should be conserved (see \cite{friedman-schutz-1978}).
From here we obtain that, following equation (52) from paper \cite{friedman-schutz-1978}, the canonical energy, $E_c \sim \omega J_c$, {\it linearly} increases starting from some sufficiently long time, since  $\omega =\sqrt{K}$ (see above).
The conservation of $J_c$ for the local perturbation considered here is discussed in paragraph 3.2 of paper \cite{heinemann-papaloizou-2009a}.
Unlike $J_c$, the canonical energy itself in this case is not conserved any more, since the time-variable frequency makes the problem inhomogeneous in time.
This growth (or decrease) of the energy, despite that the wave frequency $\omega$ is present here, is already essentially non-modal, since $\omega$ is a function of time, which, in turn, is connected exactly to the deformation of SFH by the shear flow.

In the present paper, however, we will be more interested in the 'classical' variant of the non-modal growth, which is called 'transient' in the literature.
In the simplest model considered here it is represented by the vortex solution which for $q\neq 0$ becomes dynamical and, oppositely to the waves, is aperiodic. 

\subsubsection{The vortex existence criterion}
Before discussing in detail the behavior of the vortex solution, let us analyze the justification of the decoupling of perturbations in waves and vortices made above in the presence of a shear.
Indeed, immediately after $\tilde k_x$ becoming variable, the solution $\hat u_y^{v}$ does not exactly satisfy equation (\ref{u_y_double}) any more, since a non-zero second derivative of $\hat u_y^{v}$ appears.
Moreover, in the limit $\tilde k_x \to 0$ equation (\ref{u_y_double}) becomes homogeneous, and its solution describes density waves only.
The region, in which $\tilde k_x \to 0$, corresponds to the swing of SFH, and thus we see that the vortex solution becomes poorly defined over there: the vortex must share wave properties.
This means that we cannot neglect the second time derivative in equation (\ref{u_y_double}) any more for the slowly evolving solutions.
In other words, $\hat u_y^{v}$ cannot be considered, even approximately, as a solution of equation (\ref{u_y_double}).
Let us discuss in more detail the criterion of decoupling of density waves and vortices in a shear flow.

In order to do this, use the fact that the vortex dynamics is possible only in the subsonic flows (see \cite{landau-lifshitz-1987}, end of paragraph 10).
In the considered case of an infinite flow this means that the difference in the fluid velocity on the characteristic scale of the problem must be smaller than the sound velocity.
The characteristic spatial scale is determined by the instant spatial period of SFH in the radial direction, $\lambda_{x} \sim H |\tilde k_x|^{-1}$.
As the infinitesimal perturbations are considered here, its is sufficient to apply the condition of the vortex dynamics for the background flow, and then the velocity difference is given simply by the change in the flow azimuthal velocity, i.e. for the flow with constant shear we get 
\begin{equation}
\label{decoupled}
\lambda_x q\Omega_0 / a_* = \frac{q}{|\tilde k_x|} \ll 1,
\end{equation}
Thus, the spatial radial period of the vortex harmonics must be smaller than the disk thickness.
It is important to note that the condition (\ref{decoupled}) does not directly contain the azimuthal wave number $k_y$, and hence perturbations can be vortex even if their azimuthal spatial scale exceeds the disk thickness.
In this connection it is the most important to consider the case of initially leading spirals, i.e. SFH with $k_x<0$. For such spirals, the swing occurs at 
\begin{equation}
\label{t_s}
t_s = -k_x/(qk_y) > 0,
\end{equation}
i.e. when $\tilde k_x = 0$.
Clearly, if the initial spiral was vortex-like, and therefore $k_x \gg 1$, and its evolution was initially described by the approximate solution $\hat u_y^{v}$, then in some time interval around $t_s$  the vortex approximation is not valid, and the complete equation (\ref{u_y_double}) should be integrated.
Let us call this time interval 'the swing interval' and obtain the condition under which its duration will be much shorter than the characteristic time of evolution of SFH determined by the time of the spiral unwinding, $t_s$ (see paper \cite{zhuravlev-razdoburdin-2014}).  

The time moments at which the vortex approximation breaks down can be estimated from the limiting case of equality in the condition (\ref{decoupled}):
\begin{equation}
\label{swing_int}
t_{s_1,s_2} = t_s \left ( 1 \pm \frac{q}{k_x} \right ),
\end{equation}
from where we see that the swing interval is much shorter than the evolution time of the entire vortex spiral, $t_{s_2}-t_{s_1} \ll t_s$, once 
\begin{equation}
\label{swing_condition}
|k_x|\gg 2q, 
\end{equation}
which does not contain $k_y$.
The condition (\ref{swing_condition}) implies that to study the vortex dynamics, we can use the solution $\hat u_y^{v}$ each time when at the initial moment the spiral is sufficiently strongly wound irrespective of the value of $k_y$, i.e. in both truly short-wave limit $k_y\gg 1$ and long-wave limit $k_y\ll 1$.
In the last case, the vortices will be referred to as 'large-scale'.
Here we exclude the case $k_y\sim 1$, since as was shown numerically in \cite{chagelishvili-1997}, \cite{bodo-2005} and analytically studied in the WKB approximation in paper \cite{heinemann-papaloizou-2009a}, in this case during the swing the vortex additionally generate a pair of density waves corresponding to trailing spirals and propagating inside and outside the disk.
This process is asymmetric, since only density wave generation is possible by vortices, and not vice versa.
In paper \cite{heinemann-papaloizou-2009a} analytical expressions for the amplitude and phase of the generated wave were obtained.
It was shown that its amplitude is proportional, at first, to the vortex vorticity $I$, and at second, to the combination $\epsilon^{-1/2}\exp(-4\pi/\epsilon)$ (see formula (53) in \cite{heinemann-papaloizou-2009a}).
Here $\epsilon$ is the small WKB parameter
\begin{equation}
\label{epsilon}
\epsilon = \frac{q k_y}{k_y^2 + \kappa^2/\Omega_0^2},
\end{equation}
were, we remind, $\kappa^2/ \Omega_0^2 = 2(2-q)$.
Expression (\ref{epsilon}) implies that the excitation of density waves is exponentially suppressed in both short-wave and long-wave limits and is significant only for $k_y\sim 1$ (here we specify that we will not consider the extreme cases where $q\ll 1$, and therefore $\epsilon\ll 1$ even for $k_y\sim 1$, as well as when $q\to 2$, and hence $\epsilon \gtrsim 1$ even for $k_y\ll 1$).

Thus, the vortex solution of equation (\ref{u_y_double}) exists when the condition (\ref{swing_condition}) holds together with the requirement $k_y\ll 1$ or $k_y\ll 1$, which excludes the density wave generation with non-zero vorticity during the swing of a vortex SFH.
At the same time, these restrictions provide the criterion to separate waves and vortices in the perturbed flow.
Indeed, under such constraints the density waves with zero vorticity propagate in the flow independently of vortices and represent the high-frequency branch of solutions of equation (\ref{u_y_double}) with zero right-hand side.
Similarly, for example, sound and wind exist independently in the Earth atmosphere. 

\subsubsection{Vortex solution}
Below we will only consider the evolution of vortex SFH in a shear flow.
To conclude Section \ref{sect_local_appr}, obtain also vortex solutions for $\hat u_x$ and $\hat W$.
This can be done most easily by neglecting second time derivatives of $\hat u_x$ and $\hat W$ in equations (\ref{u_x_double}) and (\ref{W_double}), as has been done with equation (\ref{u_y_double}) to obtain $u_y^{v}$.
Thus, we will have for all three quantities:
\begin{equation}
\label{vort_u_x}
\hat u_x^{v} = - \frac{K + 4q}{K^2 + 4 q^2 k_y^2} \, k_y I,
\end{equation}
\begin{equation}
\label{vort_u_y} 
\hat u_y^{v} =  \frac{\tilde k_x}{K} \, I,
\end{equation}
\begin{equation}
\label{vort_W}
\hat W^{v} = 2{\rm i} \, \frac{q k_y^2 - K}{K^2 + 4 q^2 k_y^2} \, I,
\end{equation}

It is important to note that the existence of aperiodic vortex solution in the form (\ref{vort_u_x})-(\ref{vort_W}) is possible because of the main simplifying assumption on the local constant velocity shear which provides the existence of time invariant $I$.
This enables us to reduce the system of three homogeneous 1st-order equations (\ref{sonic_sys1_sh})-(\ref{sonic_sys3_sh}) to one inhomogeneous 2d-order equation (\ref{u_y_double}) (other dynamical variables can be obtained from the known solution $\hat u_y(t)$, which gives two independent wave solutions (the general solution of the corresponding homogeneous equation) and one aperiodic vortex solution (the partial solution (\ref{u_y_double})).
However, with the account for the gradient of velocity shear in the flow the invariant $I$ disappears, and the reduction of the system of equations (\ref{sonic_sys1_sh})-(\ref{sonic_sys3_sh}) becomes impossible, and from this system we will need to obtain directly three independent solutions, two of which, as before, will correspond to the density waves, and the third solution will describe the vortex wave called the Rossby wave (see the discussion in paragraph 4 of paper \cite{bodo-2005})
\footnote{See book \cite{brekhovskikh-goncharov-1985}, paragraph 43, for the discussion of Rossby waves arising due to the gradient of the velocity shear (the gradient of vorticity) in an incompressible rotating flow.}. 

\subsection{Vortex amplification factor}
\label{sect_TG}
To measure the growth of local perturbations, the average density of their acoustic energy can be taken as the local analog of norm (\ref{ac_en}):
\begin{equation}
\label{loc_ac_en}
E = \frac{1}{2 \bar S}\int\limits_S \left ( (\Re [u_x])^2 + (\Re [u_y])^2 + \frac{( \Re [W])^2}{a_*^2} \right ) \,  {\rm d} x {\rm d}y.
\end{equation}
where $\bar S$ is the area of integration region $S$.
 
After substituting dimensionless SFH (\ref{SFH}) into (\ref{loc_ac_en}) and integrating over their spatial period we obtain the local variant of norm (\ref{ac_norm}):
\begin{equation}
\label{loc_norm}
{||{\bf q}||^2} = \frac{1}{2} \left ( |\hat u_x|^2 + |\hat u_y|^2 + |\hat W|^2  \right ).
\end{equation}

Making use of the vortex solution for SFH (\ref{vort_u_x})-(\ref{vort_W}), we get the norm in the following form:
\begin{equation}
\label{SFH_E_compr}
||{\bf q}||^2 = \left [ \frac{\tilde k_x^2}{K^2} + \frac{4 + k_y^2}{K^2 + 4q^2 k_y^2} \right ] I^2.
\end{equation}

Below we shall utilize the growth factor as the main quantity characterizing the perturbation dynamics:
\begin{equation}
\label{g_t}
g(t) \equiv \frac{||{\bf q(t)}||^2}{||{\bf q(0)}||^2},
\end{equation}
which is, in other words, the norm of perturbation with respect to its initial value. 

\begin{itemize}
\item {\bf Short-wave perturbations.}
For $k_y\gg 1$ we can in any case omit the factor 4 in (\ref{SFH_E_compr}) in the nominator of the second term, the term $4q^2k_y^2$ in the denominator of the second term, as well as the term $2(2-q)=\kappa^2/\Omega_0^2$ in the quantity $K$. Then 
\begin{equation}
\label{g_incompr}
g \approx \frac{k_x^2 + k_y^2}{\tilde k_x^2 + k_y^2},
\end{equation}
which is the result obtained in \cite{lominadze-1988} (see also formula 4 in \cite{afshordi-2005}).
Expression (\ref{g_incompr}) shows that SFH initially taken as a leading spiral with $k_x<0$ increases in amplitude until the time (\ref{t_s}), and at the swing moment, when $\tilde k_x=0$, reaches maximum in the norm and then decays.
The energy transfer from the background flow to perturbations is described in detail in terms of fluid particles in \cite{chagelishvili-1996} (see Fig. 2 therein).
Similar to the well-known lift-up effect (see book \cite{schmid-henningson-2001}, paragraph 2.3.3 for more detail), it is based on 'pickup' of fluid particles by the main flow as they move into the region with different shear velocity.
However, it also has an important additional ingredient being interaction of particles with each other at the planes of pressure extrema, ending up with the growth of their velocity respectively to the background flow even in situation 
when lift-up effect does not work. 

\subsubsection{On the transient growth mechanism}
Here we will provide additional consideration clarifying the transient growth mechanism.
As mentioned in the Introduction and discussed in Section \ref{sect_local_appr}, a differentially rotating flow shortens the length of the leading spiral arms of a transiently growing vortex until the swing moment (see Fig. \ref{pic_TG}).
Due to the barotropicity of the perturbed flow, the velocity circulation along a fluid contour coinciding with the spiral arm boundary must be constant.
Consequently, the contour shortening must lead to the compensating increase in gas velocity along the spiral's boundary.
Consider this suggestion more rigorously in the local space limit (see the scheme in Fig. \ref{contour}).
Let us calculate the velocity circulation for the most simple fluid contour.
Without perturbations, this is naturally a parallelogram with one pair of sides (call them the base of the parallelogram) go along the background stream lines, i.e. parallel to the $y$ axis and symmetrical on both sides from the level $x=0$.
The condition that these sides move synchronously with the fluid automatically implies that the entire contour is co-moving with the background flow, since the velocity in the flow is linear in $x$.
Now let us pass to the reference frame co-moving with the shear, in which equations (\ref{sonic_sys1_sh})-(\ref{sonic_sys3_sh}) were written: in this frame, the background velocity together with the velocity circulation along the given contour are zero.
Next, with account for small perturbations, the velocity circulation must change, strictly speaking, for two reasons: at first, the velocity perturbation arises, ${\bf u}$ (as determined in the shear reference frame), and at second, even the contour taken at the time $t=0$ as a parallelogram starts being deformed due to additional shifts caused by perturbations.
In the second case, however, for small perturbations considered here, only the contribution due to the corresponding change in the background velocity circulation will be important.
But this addition is absent, since in the shear reference frame the background velocity is zero at all points.
Thus, all we need to do is to calculate the circulation ${\bf u}$ along a contour co-moving with the background flow.
At the time $t=0$ we take it such that the parallelogram sides coincide with the SFH front lines separated by the phase $\pi$ (see Fig. \ref{contour}, where the initial front direction is denoted by the wave vector ${\bf k_0}$).
As in the shear frame SFH, by definition, has constant space phase front lines, it is clear that at times $t>0$ they remain coinciding with the contour's sides.
Now note that we consider the case $k_y\gg 1$, therefore $\hat W^{v}\to 0$, and from (\ref{sonic_sys3_sh}) we derive the orthogonality condition ${\bf u}\perp{\bf k}$.
Consequently, the velocity perturbation is directed along the parallelogram's sides and always points to their going around.
As for the parallelogram's bases, their contribution to the circulation will be mutually canceled, because along them the projection of the velocity ${\bf u}$ does not change, while the going around direction becomes opposite.
With account for the above considerations, the perturbed flow circulation in the co-moving shear frame for the left contour in Fig. \ref{contour} reads:
$$
\mathcal{C}|_{t=0} = 2  \Delta y \left ( 1 + \frac{k_y^2}{k_x^2} \right )^{1/2} |{\bf u}|_{t=0}.
$$
For the right contour in Fig. \ref{contour} taken at the spiral swing moment, we similarly find:
$$
\mathcal{C}|_{t=t_s} = 2  \Delta x \, |{\bf u}|_{t=t_s}. 
$$

By equating these two expressions, we see that the circulation conservation law yields for the vortex SFH with $k_y\gg 1$:

\begin{equation}
\label{contour_expr}
g(t_s) = \frac{|{\bf u}(t_s)|^2}{|{\bf u}(0)|^2} = \frac{k_x^2+k_y^2}{k_y^2}, 
\end{equation}

This coincides with the result following from (\ref{g_incompr}) for the spiral swing time. 

\begin{figure}[h!]
\includegraphics[width=1.\linewidth]{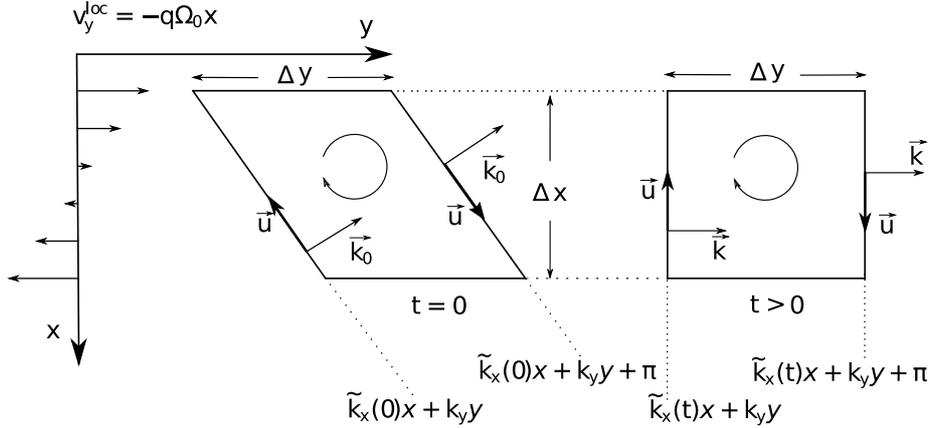}
\caption{\footnotesize{
The illustration of physical reasons for the transient growth of two-dimensional vortices in the local space limit (see Section \ref{sect_local_appr}).
The case of short-wave ($k_y\gg 1$) vortex  SFH with $k_x<0$ is taken.
A liquid contour co-moving with the background flow at two instants is shown: at the initial time $t=0$ and at the time of the SFH swing when $\tilde k_x=0$.
See text (Section \ref{sect_TG}.1) for the explanation why it is possible to ignore deformation of the contour by perturbations.
At $t=0$ the contour has the form of a parallelogram with one pair of sides along the $y$-axis symmetrically relative to $x=0$ and another pair along two SFH fronts, with the phase difference between them $\pi$.
${\bf u}$ is the velocity perturbation vector, ${\bf k_0}$ and ${\bf k}$ show the SFH wave vector at different time moments.
$\Delta x$ and $\Delta y$ are the parallelogram's height and base, respectively.
}}
\label{contour}
\end{figure}

Thus, we have been convinced that the transient growth of a vortex is in fact due to its perimeter (its 'size') shortening by the background shear flow with constant velocity circulation, $\mathcal{C}=const$, along this perimeter.
It is important to note that $\mathcal{C}$, as well as the corresponding vorticity flux, is the measure of the vortex rotation.
Therefore, it is appropriate to compare it with a body compressing with angular momentum conservation, since in that case the body's angular velocity increases inversely with the moment of inertia, $\omega_{rot} \propto I^{-1}_{rot}$, and the rotation energy $E_{rot}=1/2 I_{rot} \omega^2_{rot} \propto I^{-1}_{rot}$ increases with time.
In our case, the background flow does work on shortening the vortex size and thus transfers it the kinetic energy. 

Finally, note also that as the differential rotation is purely shear, i.e. occurs with zero divergence of the background flow, the area subtended by the contour considered above must keep constant.
Indeed, the area of the parallelogram is the production of its base (which is constant since the flow in homogeneous in $y$) by its height (which is constant since there is no radial background velocity).
Therefore, due to the constant $\mathcal{C}$ and hence the vorticity perturbation flux through the contour, the vorticity perturbation itself is constant.
The same conclusion was obtained in Section \ref{sect_local_appr} from the discussion of the invariant (\ref{vorticity}). 

\subsubsection{Estimation of the optimal growth}
Knowing the physical mechanism of the transient vortex growth, let us return to expression (\ref{g_incompr}) for their growth factor in the case of short azimuthal wavelength.
Clearly, the growth factor of an individual SFH is a function of three arguments, $g = g(k_x,k_y,t)$.
However, it is possible to consider a more general characteristic of the transient dynamics which is called the optimal growth of perturbations $G$. 
By definition,
\begin{equation}
\label{G_t}
G \equiv \max\limits_{\forall\, k_x} \{g\}.
\end{equation}

Formula (\ref{G_t}) gives the maximum possible amplification among all vortices with given $k_y$ which can occur in a time interval $t$.
Note that below we will also employ an analogue of (\ref{G_t}) used for the global space problem described by the system of equations (\ref{sys_A_1})-(\ref{sys_A_3}) (see formula (\ref{G_global})), when
the value $G$ will be determined for all perturbations with fixed azimuthal wave number $m$.

There are rigorous mathematical algorithms to search for the optimal growth, which we will discuss in the next Section.
Here, for analytical estimates in the local space limit it will be sufficient to recognize that since the growth factor $g(k_x,k_y,t)$ of a certain SFH has maximum at $\tilde k_x=0$, it is reasonable to suppose that $G$ can be estimated as 
\begin{equation}
\label{G_g}
G \approx g(k_x = -k_y q t),
\end{equation}
in other words, to adopt that of all SFH with given $k_y$, the harmonics that swings at time $t$, reaches maximum possible growth by this time. 

Making use of definition ((\ref{G_g}), from (\ref{g_incompr}) we obtain the simple expression:
\begin{equation}
\label{G_high_ky}
G_1 \approx {(qt )}^2,
\end{equation}
which can be also found in paper \cite{afshordi-2005} (see formula (5) therein).
Note that in that paper corrections to $G_1$  due to non-zero vertical projection of the wave vector and finite value of $k_y$ were also obtained.
As we see, in a sufficiently long time it is possible to reach arbitrarily large amplitude growth of small-scale vortices $k_y\gg 1$.
This growth, however, is power-law and not exponential, as it would be expected in a modal instability of the flow. 

\item {\bf Long-wave perturbations.}
Now turn to another limiting case where $k_y\ll 1$ and the azimuthal space period of SFH is much larger than the disk thickness (see paper \cite{zhuravlev-razdoburdin-2014}).
In this case, in the second term in (\ref{SFH_E_compr}) we omit $k_y^2$ in the nominator and $4q^2k_y^2$ in the denominator, and also assume that $K = \tilde k_x^2 + \kappa^2/\Omega_0^2$. Here, by the condition (\ref{swing_condition}), we see that $||{\bf q(0)}||^2 \approx k_x^{-2}$. 

Then, for the SFH growth factor we obtain
\begin{equation}
\label{g_compr}
g \approx k_x^2 \, \frac{\tilde k_x^2 + 4}{(\tilde k_x^2 + \kappa^2/\Omega_0^2)^2},
\end{equation}

This quantity increases for $\tilde k_x$ decreasing with time, i.e., similar to the short-wave vortices, the transient growth occurs for $k_x<0$.
Note that now the maximum $g$, attained during the spiral swing, is proportional to the square of the value $k_x$ itself, but not to the square of the ratio $k_x/k_y$, as in the case of the short wavelength vortices (cf. (\ref{g_incompr})).
In addition, another important difference is that now $g$ depends on the epicyclic frequency as $\kappa^{-4}$.
Such a strong dependence can be important in disks with super-Keplerian angular velocity gradient: in thin disks this can occur in the inner regions of relativistic disks, where $\kappa\to 0$ when approaching their inner boundary. 

Following the definition (\ref{G_g}), we obtain from (\ref{g_compr}) the corresponding optimal growth factor:
\begin{equation}
\label{G_low_ky}
G_2 \approx \frac{4\Omega_0^4}{\kappa^4} k_y^2 {(qt )}^2.
\end{equation}
Note that both (\ref{G_high_ky}) and (\ref{G_low_ky}) are valid only for sufficiently large timespans because in order to obtain this expression we used the condition $k_x = -q k_y t$, but at the same time the condition $k_x\gg 1$ must hold, as required by (\ref{swing_condition}).
Formula (\ref{G_low_ky}) shows that for rotation profiles weakly different form the Keplerian one, when $\kappa\sim\Omega_0$, for equal time intervals $G_2\ll G_1$, because the azimuthal wave number now explicitly entering the optimal growth factor is small, $k_y\ll 1$
\footnote{
In Section \ref{sect_global_var} below we calculate $G$ in the global problem (see Fig. \ref{fig_10}), which implies that as $m\to 1$, the difference in the transient growth rate between vortices with azimuthal wavelength shorter and longer than the disk thickness is significantly smaller.}
Therefore, in the local space limit considered here, small-scale vortices take energy from the flow more efficiently than large-scale ones.
However, it is interesting to learn which of them can display the highest growth over the entire time interval.
In an inviscid flow $G_{1,2}\to\infty$ mostly due to small-scale SFH, as we just noted.
Nevertheless, a shear flow can have noticeable effective viscosity due to, for example, some weak turbulence.
Then the dependence $G(t)$ turns out to have the global maximum $G_{max}$ corresponding to the maximum possible non-modal growth of perturbations irrespective of the time intervals we have considered so far.
Physically, the decrease of $G(t)$ after some long time is related to the fact that more tightly wound spirals have larger swing times $t_s$.
This in turn means the smaller radial scale of perturbations and hence the smaller dissipation time of perturbations due to viscosity.
Ultimately, the leading transient spirals start faster decaying than growing due to unwinding by the flow.
It is the value $G_{max}$ for cases $k_y\gg 1$ and $k_y\ll 1$ that we would like to compare below. 

\end{itemize}

\subsubsection{Account for the viscosity}
The effect of viscosity on the maximum possible transient growth of vortices can be estimated as follows.
For sufficiently long time intervals $qt\gg 1$ we have $k_x\gg k_y$ for any of the two limits of $k_y$ we consider.
Therefore, in a shearless flow the spiral would decay in the characteristic viscous time $\Delta t_\nu \sim \lambda_x^2 / \nu$, where $\nu$ is the kinematic viscosity coefficient.
Using the standard viscosity parametrization by the Shakura-Sunyaev $\alpha$-parameter, $\nu = \alpha a_* H$, we get that $\Delta t_\nu \sim (\Omega_0^{-1}\alpha k_x^2)^{-1}$ rapidly decreases with increasing $|k_x|$.
At the same time, the larger $|k_x|$, the longer is the transient growth time of the spiral, $\Delta t_{tg} \sim |k_x/(q k_y)|$.
Simultaneously with arising of a shear in the flow, the spiral starts unwinding, and therefore the viscous dissipation is delayed. Thus, the equality of these characteristic times, $\Delta t_{tg} = \Delta t_\nu$, gives the lower limit on the duration of the transient growth of vortices  in a viscous flow.
 Using it we obtain:

\begin{equation}
\label{T_max}
\max (\Delta t_{tg}) \gtrsim \alpha^{-1/3} (q k_y)^{-2/3}
\end{equation}
It can be verified that expression ((\ref{T_max}) reproduces the estimate made in paper \cite{afshordi-2005} (see formula (81) therein). 

The upper limit on the optimal growth time (\ref{T_max}) , $G_{max}\equiv G(\max(\Delta t_{tg}))$, is given by its inviscid  value taken for $G_1$ or $G_2$. We then obtain that for $k_y\gg 1$

\begin{equation}
\label{G_1_max}
{(G_{max})}_1 \approx \alpha^{-2/3} q^{2/3} k_y^{-4/3},
\end{equation}

(see also formula (83) in paper \cite{afshordi-2005}).
At the same time, for $k_y\ll 1$ we have
\begin{equation}
\label{G_max_2}
{(G_{max})}_2 \approx \frac{4\Omega_0^4}{\kappa^4} \alpha^{-2/3} q^{2/3} k_y^{2/3}.
\end{equation}

\begin{figure}[h!]
\includegraphics[width=0.9\linewidth]{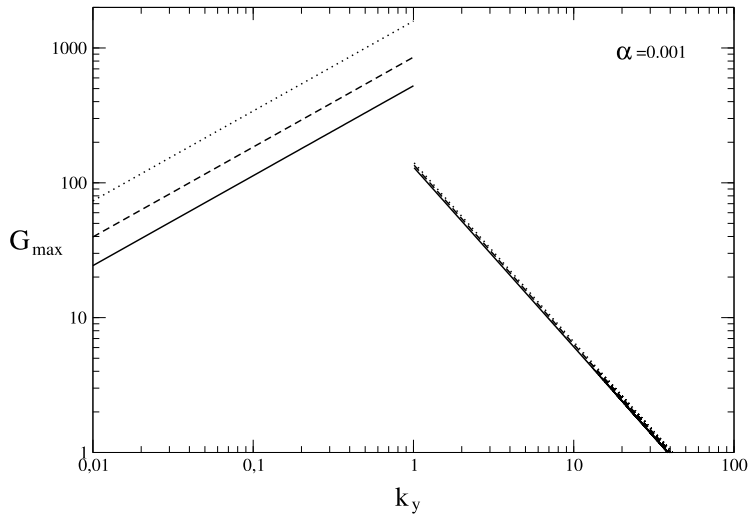}
\caption{
\footnotesize{
Estimate of the maximum possible transient growth of acoustic energy in a disk with efficient viscosity $\alpha=0.001$.
The solid, dashed and dotted lines correspond to $q=1.5,1.6$ and $1.7$, respectively.
Three uppermost right and left curves are obtained using formula (\ref{G_1_max}) and (\ref{G_max_2}), respectively. 
}}
\label{G_max}
\end{figure}

This result is shown in Fig. \ref{G_max} for some small $\alpha$ and several shears $q$: Keplerian and super-Keplerian.
We see that even for the Keplerian shear, when $\kappa=\Omega_0$, for $k_y$ different from 1, $(G_{max})_2 \gtrsim (G_{max})_1$.
This occurs because the large-scale vortices are much less dissipative which more than compensate their low growth rate compared to the small-scale vortices.
Note also that despite ${(G_{max})}_2$ decreasing with decreasing $k_y$, this occurs at lower rate compared to the case of ${(G_{max})}_1$ decreasing with increasing $k_y$.
As a result, the integral transient growth of large-scale vortices at all $k_y$ increases in comparison with small-scale ones.
Even more significant advantage of large-scale vortices appears for super-Keplerian shears, when $q>3/2$, due to ${(G_{max})}_2\propto\kappa^{-4}$ (see the comment after formula (\ref{g_compr})).
Clearly, the deviation from $q=3/2$ by several per cents would increase the transient growth rate of perturbations by a factor of a few. 

As  discussed in paper \cite{zhuravlev-razdoburdin-2014}, the estimate (\ref{G_max_2}) is in reasonable agreement with exact calculations of the optimal growth rate in thin disks in the global space limit for low azimuthal wave numbers $m$.
Thus, large-scale vortices are also able to provide additional transportation of the angular momentum to the periphery of a disk with pre-existing weak turbulence. 

\vspace{0.5cm}

In Section 3 we provide a rigorous mathematical justification of algorithms to search for the most rapidly growing perturbations in shear flows.
Such perturbations will be called optimal, and the corresponding amplification, as we already mentioned, will be referred to as the optimal growth $G$.
The solutions presented in the Introduction and shown in Fig. \ref{pic_modes} and \ref{pic_TG} were obtained using one of these algorithms.
We will also provide another example of calculation of $G$ by solving the general system of equations (\ref{sys_A_1})-(\ref{sys_A_3}) in a geometrically thin disk (see Fig. \ref{fig_10} below).
When discussing mathematical aspects of the non-modal dynamics of perturbations in shear flows, already in the introductory part to the next Section we will see that the transient growth phenomenon can be treated as a consequence of {\it non-orthogonality} of perturbation modes, which will be evident, in particular, from consideration of simple analogs presented in Fig. \ref{modes_1} and \ref{modes}. 

\section{Search for optimal perturbations}

\subsection{Definition and properties of singular vectors}

\label{sect_singular}
General solutions of the initial value problem of the small perturbation evolution described by general equations (\ref{orig_sys1})-(\ref{orig_sys2}) supplemented by an appropriate boundary conditions can be conveniently studied using abstract concepts of the functional space of the so-called state vectors of the system, as well as the notion of linear operators acting on these vectors.
In Section \ref{sect_adiabat_perts}, in application to the system of equations (\ref{sys_A_1})-(\ref{sys_A_3}), we have already introduced the particular case of the state vector as a set of azimuthal Fourier harmonics of Eulerian perturbations ${\bf q}(t)\equiv \{\delta v_r(r), \, \delta v_\varphi(r),\, \delta h(r)\}$ taken at some fixed instant $t$.
In this Section we shall assume the initial general case where ${\bf q}(t) \equiv \{\delta {\bf v}({\bf r}), \, \delta h({\bf r}),\, \delta \rho({\bf r})\}$.
Consider some properties of a dynamical operator $\mathbf{Z}$ acting in the Banach space of vectors ${\bf q}$ and corresponding to the system (\ref{orig_sys1})-(\ref{orig_sys2}).
This operator transforms the initial perturbation vector $\mathbf{q}(0)$ to the consecutive vector $\mathbf{q}(t)$, i.e. in the operator form the system of equations can be written as
\begin{equation}
\label{Z_eq}
\mathbf{q}(t)=\mathbf{Z}\mathbf{q}(0).
\end{equation} 
All functions entering $\mathbf{q}(t)$ will be assumed to be infinitely smooth and to have  a uniformly bounded derivative in the domain of definition.
The last condition follows from physical considerations: in realistic gas flows, there cannot be perturbations with arbitrarily small wavelength.
In addition, due to linearity of the problem, at the initial time all vectors $\mathbf{q}(0)$ will be assumed to have the unit norm. 

In this section we will show that the general assumptions given above imply important properties of the operator $\mathbf{Z}$.
For example, we will show that the  norm of all initial vectors $\mathbf{q}(0)$ can grow by the time $t$ only by less than some strict factor.
We will present two methods to calculate this perturbation growth limit.
In addition, we will show that in the space of initial conditions there is an orthonormal basis which can be found by solving the  eigenvalue problem for some operator {\it different} from $\mathbf{Z}$.  

\subsubsection{Continuity of a dynamical operator}
The continuity is the first important property of the operator $\mathbf{Z}$.
To see this, write down the operator in the integral form:
\begin{equation}
\label{U_view}
\mathbf{Zq}(0)=\mathbf{q}(t)=\mathbf{q}(0)+\int\limits_0^t \left(\mathbf{M}_1(s)\mathbf{q}(s)+\mathbf{M}_2(s)\frac{\partial \mathbf{q}(s)}{\partial \mathbf{r}}\right)ds,
\end{equation}
Here we have introduced matrices $\mathbf{M}_1(s)$ and $\mathbf{M}_2(s)$ composed from coefficients in the dynamical equations before the corresponding spatial derivative of $\mathbf{q}$.
The explicit form of these matrices can be obtained from equations (\ref{orig_sys1}) and (\ref{orig_sys2}).
The number of rows and columns in the matrices is equal to the number of quantities forming the state vector $\mathbf{q}$ and the number of spatial variables, respectively.
As the quantities describing the background flow are bounded and continuous, all elements of the matrices $\mathbf{M}_1(s)$ and $\mathbf{M}_2(s)$ are also bounded and continuous in the domain of definition of the operator $\mathbf{Z}$.
In addition we note that if viscous forces are included, one more term appears in equation (\ref{U_view}) corresponding to the second time derivative of $\mathbf{q}$.
This case can be treated analogously. 

Apparently, the operator (\ref{U_view}) is a superposition of continuous mappings (see \cite{vilenkin-1972}, Ch. 1), and hence is itself a continuous mapping (see \cite{kolmogorov-fomin-1975}, Ch. 2).
This  implies that it is bounded on bounded sets (see \cite{kolmogorov-fomin-1975}, Ch. 4). 

Thus, the mapping defined by (\ref{U_view}) is continuous and bounded, which implies that vectors $\mathbf{q}$ are {\it uniformly bounded} at any time $t$.
Now let us use this property.

\subsubsection{Completely continuous dynamical operator}
The next property of the operator $\mathbf{Z}$ is its complete continuity. Let us remind the definition of this notion. 
\begin{defenition}[completely continuous operator]
Operator $\mathbf{Z}$ mapping a Banach space $\mathbf{E}$ into itself is called completely continuous if it takes any bounded set to a relatively compact set (\cite{kolmogorov-fomin-1975}, Ch. 4). 
\end{defenition}

Thus, to prove the complete continuity of the operator (\ref{U_view}) it is sufficient to prove the relative compactness of its values, since the boundedness of its domain was postulated by assuming that all $\mathbf{q}(0)$ have unit norm.
Let us use the Arzel\`a-Ascoli theorem.
According to this theorem, a sequence of continuous functions defined on a closed and bounded interval is relatively compact if and only if this sequence is uniformly bounded and equicontinuous (\cite{kolmogorov-fomin-1975}, Ch. 4). 

The uniform boundedness was shown above, and for a sequence of differentiable functions to be equicontinous, it is sufficient that their derivatives be uniformly bounded (\cite{ilin-1987}, Ch. 2), which was initially postulated.
Thus, we see that the set of values of the operator ${\bf Z}$ is relatively compact and hence it is completely continuous. 

Now, if we introduce an inner product (in physical problems, as a rule, it is introduced such that the norm of a vector coincides with the energy of perturbation, as was done, for example, in equation (\ref{ac_norm})), it is possible to define the adjoint operator $\mathbf{Z^{\dag}}$ using the Lagrange identity for arbitrary vectors $\mathbf{f}, \mathbf{g}$ (see, for example, \cite{vilenkin-1972}, Ch. 1, for more details on the adjoint operators):

\begin{equation}
\left(\mathbf{Zf},\mathbf{g}\right)=\left(\mathbf{f},\mathbf{Z}^{\dag}\mathbf{g}\right)
\end{equation}

Here if the operator $\mathbf{Z}$ is completely continuous, so will be the adjoint operator $\mathbf{Z}^{\dag}$ and self-adjoint composite operators $\mathbf{ZZ}^{\dag}$ and $\mathbf{Z}^{\dag}\mathbf{Z}$ as well (\cite{kolmogorov-fomin-1975}, Ch. 4). 

\subsubsection{Linear operators: from the particular to the general}
There can be different linear operators depending on their properties.
Let us list those of them that we will need below, from the more particular to the more general case.
Start from positive definite operators, for which the inner product $(\mathbf{Zq},\mathbf{q})>0$ for any vector $\mathbf{q}$.
By definition, eigenvalues of a positive definite operator are positive.
Indeed, by multiplying the equation $\mathbf{Zq}=\lambda\mathbf{q}$ through $\mathbf{q}$, we see that its left-hand side is positive, and the right-hand side is the product of the eigenvalue and a positive value, hence the positive eigenvalue. 

Self-adjoint (Hermitian) operators, which are identical to their adjoint operators, $\mathbf{Z}=\mathbf{Z}^{\dag}$ (\cite{korn-korn-1968}, paragraph 14.4), are most frequently used in different  physical problems.
Eigenvalues of a self-adjoined operator are real values (\cite{korn-korn-1968}, paragraph 14.8). 

In turn, self-adjoined operators are the particular case of normal operators.
An operator $\mathbf{Z}$ is called normal if it commutes with its adjoint operator: $\mathbf{ZZ}^{\dag}=\mathbf{Z}^{\dag}\mathbf{Z}$ (\cite{korn-korn-1968}, paragraph 14.4).
All eigenvalues of a normal operator are complex conjugate of its adjoint operator's eigenvalues.
Eigenfunctions of the operators $\mathbf{Z}$ and $\mathbf{Z}^{\dag}$ coincide.
Additionally, eigenvectors of a normal operator corresponding to different eigenvalues are orthogonal (\cite{korn-korn-1968}, paragraph 14.8).
Therefore, to calculate {\it operator norm} of these operators, it is sufficient to find their eigenvalues.
We remind that the norm of an operator $\mathbf{Z}$ mapping a Banach space $H$ into itself is the number $||\mathbf{Z}||=\sup\limits_{\mathbf{x}\in H} \frac{||\mathbf{Zx}||}{\mathbf{||x||}}$ (\cite{vilenkin-1972}, Ch. 1).
The norm of the governing operator is very useful, because it allows us to calculate the limit of the vector's norm growth under the action of this operator.

For a normal operator this problem is solved quite easily.
To illustrate this, we (following \cite{schmid-2007}) consider an important particular case in which the operator $\mathbf{Z}$ can be represented as an operator exponent: $\mathbf{Z}=\mathrm{e}^{\mathbf{A}t}$ (see Section \ref{linear_autonomous} for more detail).
The operator $\mathbf{A}$ is time-independent, and its eigenvalues are traditionally denoted as $\{-\mathrm{i}\omega_1, -\mathrm{i}\omega_2, ...-\mathrm{i}\omega_N\}$; here $\omega$ can take both real and complex values.
In this case, eigenvalues of the operator $\mathbf{Z}$ are $\{\mathrm{e}^{-\mathrm{i}\omega_1 t}, \mathrm{e}^{-\mathrm{i}\omega_2 t}, ...\mathrm{e}^{-\mathrm{i}\omega_N t}\}$.
Now let us use the definition of operator's eigenvectors and eigenvalues by writing it in the matrix form:
\begin{equation}
\label{spectral_decomp}
\mathbf{ZX}=\mathbf{X P},
\end{equation}
where $\mathbf{P}$ is a diagonal matrix with eigenvalues of $\mathbf{Z}$, columns of the matrix $\mathbf{X}$ are eigenvectors of $\mathbf{Z}$ standing in correspondence with its eigenvalues in $\mathbf{P}$. 

From (\ref{spectral_decomp}) we find the decomposition $\mathbf{Z}=\mathbf{X P X}^{-1}$.
Next let us use the submultiplicativity of the operator norm (\cite{korn-korn-1968}, paragraph 14.2): $||\mathbf{Z}||\le||\mathbf{X}|| ||\mathbf{P}|| ||\mathbf{X}^{-1}||$.
For orthonormal eigenvectors, the matrix $\mathbf{X}$ is unitary, $\mathbf{XX^{\dag}}=\mathbf{I}$, therefore its norm in this case is unity, $||\mathbf{X}||=1$, and the norm $||\mathbf{Z}||\le ||\mathbf{P}||=\mathrm{e}^{\omega_{max}t}$, where $\omega_{max}=\max\limits_{j\le N} \left(\Im~[\omega_j]\right)$. 

Finally, the most general are {\it non-normal} operators, i.e. those that do not commute with their adjoint operator: $\mathbf{ZZ}^{\dag}\ne\mathbf{Z}^{\dag}\mathbf{Z}$.
Eigenvalues of these operators can be both purely real and complex, and eigenvectors are non-orthogonal to each other.
The non-orthogonality of the eigenvectors complicates the calculation of the operator's norm, since the matrix $\mathbf{X}$ introduced above is not unitary any longer.
For this reason, the energy of a combination of modes is not equal to the sum of energy of each mode, i.e. the Parceval rule is not valid and non-zero cross terms appear.
In other words, due to interference in time between non-orthogonal modes, perturbations described by such an operator can increase even if there are no growing modes.
This energy growth of perturbations, which is mathematically related to the non-normality of the dynamical operator, was dubbed the transient growth of perturbations.
In the context of stability of hydrodynamical flows, non-normal operators and examples were discussed in \cite{farrell-ioannou-1996a}, as well as in Section 3 and 4 of book \cite{schmid-henningson-2001}. 

\subsubsection{Simple geometrical example on the non-orthogonality of eigenvectors}

\begin{figure}[h!]
\includegraphics[width=0.9\linewidth]{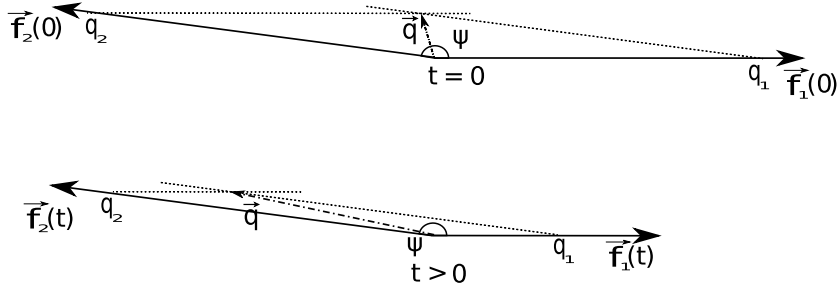}
\caption{\footnotesize{
The increase in the sum of two non-orthogonal vectors, $\mathbf{q}=\mathbf{f_1}+\mathbf{f_2}$, with decreasing lengths but conserving the angle between them.
It is assumed that $q_1=q_2=1$. 
}}
\label{modes_1}
\end{figure}

A simple geometrical example can illustrate the transient growth mechanism.
On the plane $(x,y)$ introduce two vectors symbolizing two perturbation modes.
Write them in the form of two complex numbers, ${\bf f}_1 = f_0 {\rm e}^{-{\rm i}\omega_1 t}$, ${\bf f}_2 =  f_0 {\rm e}^{-{\rm i}\omega_2 t + {\rm i}\psi}$, and numbers $\omega_{1,2}$ can be complex as well.
In this form the analogy between ${\bf f}_{1,2}$ and perturbation modes will be the most clear.
The real and imaginary part of each of vectors ${\bf f}_{1,2}$ yields the $x$- and $y$- vector components, respectively.
Clearly, $\Re[\omega_{1,2}]$ corresponds to the angular velocity with which both vectors rotate on the plane, and $\Im[\omega]_{1,2}$ corresponds to the rate of change of their lengths.
Below we will assume that imaginary parts of $\omega_{1,2}$ are negative, which corresponds to the length shortening of ${\bf f}_{1,2}$.
We remind that in the case of modes, real parts give angular velocities of the solid-body rotation of the spiral pattern in the flow (see Fig. {\ref{pic_modes}}), and imaginary parts give their decay rate, in analogy with a spectrally stable flow.
In addition, we will assume that at the time $t=0$ the vectors had the same length $f_0$ and the angle between them is  $\psi$.

\begin{figure}[h!]
\includegraphics[width=0.7\linewidth]{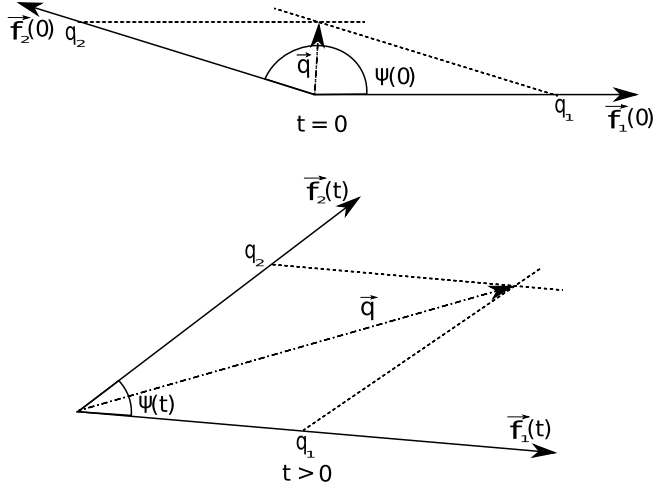}
\caption{\footnotesize{
The increase in the sum of two non-orthogonal vectors, $\mathbf{q}=\mathbf{f_1}+\mathbf{f_2}$, with conserving lengths but changing the angle between them.
It is assumed that $q_1=q_2=1$.
}}
\label{modes}
\end{figure}

Now take the vector ${\bf q} = {\bf f_1} + {\bf f_2}$ and calculate the quantity similar to (\ref{g_t}), which gives the rate of change of the length square ${\bf q}$ with time:
\begin{equation}
\label{g_vector}
g = \frac{{\rm e}^{2\Im [\omega_1] t} + {\rm e}^{2\Im [\omega_2] t} + 2{\rm e}^{\Im[\omega_1+\omega_2] t} \cos(\Re [\omega_1-\omega_2] t + \psi)}{2(1+\cos \psi)}.
\end{equation}
This shows that for angles close to $\pi$ the denominator in (\ref{g_vector}) is small, and any insignificant increase in the nominator will lead to a large increase in $g$.
Consider two particular examples.
In the first case assume that $\Re[\omega_{1,2}] =0$, and in the second case that $\Im[\omega_{1,2}] =0$.
For simplicity, assume $\cos \psi \approx -1 +\epsilon$, where $\epsilon \ll 1$. 

Then for the case $\Re[\omega_{1,2}] =0$ we see that if we additionally admit a large difference in decrements, $|\Im[\omega_1]|\gg |\Im[\omega_2]|$, after some large time $g$ will be 
\begin{equation}
\label{g_tg}
g \approx \frac{e^{2\Im[\omega_2] t}}{2\epsilon},
\end{equation}
which corresponds to $g\gg 1$ on time intervals such that $|\Im[\omega_1] t|\gg 1$ but simultaneously $|\Im[\omega_2] t|\ll 1$.
This means that despite the decrease in length of each particular vector, in the case of strong non-orthogonality (which is characterized by strong difference of $\epsilon$ from 1) their sum exhibits a transient growth up to values $\sim\epsilon^{-1}$ (Fig. \ref{modes_1}).
And only at later times $g$ decreases again at a rate determined by the most slowly decreasing vector.
The similar effect takes place for transient perturbations which can be represented as a sum of decaying modes with zero phase velocity. 

In the opposite case $\Im[\omega_{1,2}] =0$, from (\ref{g_vector}) the following approximate formula can be derived:
\begin{equation}
\label{g_oscill_1}
g \approx \frac{1- \cos (\Re[\omega_1-\omega_2] t)}{\epsilon},
\end{equation}
which is valid when the value of cosine in the nominator is not too close to unity.
Apparently, unlike the example with the sum of non-orthogonal vectors with decreasing length (when the length ${\bf q}$ first increases to maximum and then monotonically decreases down to zero at $t\to \infty$), the length of the sum of rotating vectors exhibits an oscillating growth, by returning many times to ever increasing values $\sim\epsilon^{-1}$ in equal time intervals $\sim |\Re\omega_1-\Re\omega_2|^{-1}$, as is evident from illustration in Fig. \ref{modes}.
Unlike the first case, it would be inappropriate to refer to this second possible variant of the mode superposition growth as 'transient growth', as we did, for example when analyzing local SFH in Section \ref{sect_TG}.
Therefore, it is more appropriate to call it 'non-modal growth'.
One example of such a non-modal growth of a superposition of neutral modes with non-zero phase velocities is considered in Section \ref{sect_martix} and was studied in paper \cite{razdoburdin-zhuravlev-2012}.

\subsubsection{Singular vectors}
Thus, we have just demonstrated how non-orthogonality of the modes leads to transient growth of perturbations.
In many physical and astrophysical problems, the evolution of linear perturbations is determined exactly by non-normal operators with non-orthogonal eigenvectors.
Here the non-normality of ${\bf Z}$ is provided by a shear in the background flow.
We can justify this by deriving the system of {\it adjoint} dynamical equations corresponding to the action of the adjoint operator ${\bf Z}^\dag$ (see Section \ref{sect_adjoint}.1).

Thus, the only knowledge of eigenvalues of a non-normal operator is insufficient to fully describe the possible (transient) growth of perturbations in the system.
In addition, the  pair inner products ('angles') between the eigenvectors on the chosen norm of perturbations should be known.
One more potential complication of the problem with a non-normal dynamical operator is that it is impossible any more to guarantee the completeness of the set of its eigenvectors, and hence, to guarantee the adequacy of the solution of the problem when using the eigenvectors as a basis for decomposition of arbitrary perturbation.

For all these reasons, in order to compute the maximal transient growth rate of perturbations, below we will use the technique of {\it singular} values and vectors.
As will be shown below, the singular vectors form the complete orthonormal set, which allows us to employ them as a basis to describe the evolution of perturbations.
Moreover, the singular values, unlike eigenvalues, enable us to calculate the perturbation energy growth by any given time even for non-normal operators. 

\begin{defenition}[singular values and vectors]
\label{singular_vector_defenition}
A non-negative real number $\sigma$ is called the singular number of a linear operator $\mathbf{Z}$ if there are such vectors $\mathbf{u}$ and $\mathbf{v}$ of unit length that
\begin{equation}
\begin{aligned}
&\mathbf{Zv}=\sigma \mathbf{u}\\
&\mathbf{Z^{\dag}u}=\sigma \mathbf{v}\\
\end{aligned}
\end{equation}
The vectors $\mathbf{u}$ and $\mathbf{v}$ are called the left and right singular vectors, respectively, corresponding to the singular value $\sigma$.
\end{defenition}

Note that the singular values and vectors are related to the eigenvalues and eigenvectors of the composed self-adjoint operators $\mathbf{ZZ^{\dag}}$ and $\mathbf{Z^{\dag}Z}$.
To see this, act by the operator $\mathbf{Z}^{\dag}$ on vector $\mathbf{Zv}$ and by the operator $\mathbf{Z}$ on vector $\mathbf{Z}^{\dag}\mathbf{u}$ and then use the definition \ref{singular_vector_defenition}:
\begin{equation}
\mathbf{Z}^{\dag}\left(\mathbf{Zv}\right)=\mathbf{Z}^{\dag}\left(\sigma \mathbf{u}\right)=\sigma \mathbf{Z}^{\dag}\mathbf{u}=\sigma^2 \mathbf{v}
\end{equation}
\begin{equation}
\mathbf{Z}\left(\mathbf{Z}^{\dag}\mathbf{u}\right)=\mathbf{Z}\left(\sigma \mathbf{v}\right)=\sigma \mathbf{Z}\mathbf{v}=\sigma^2 \mathbf{u}
\end{equation}

Thus, vectors $\mathbf{v}$ and $\mathbf{u}$ are eigenvectors of the operator $\mathbf{Z^{\dag}Z}$ and $\mathbf{ZZ^{\dag}}$, respectively.
The singular value squares are eigenvalues of the composite operators.

The operators $\mathbf{ZZ^{\dag}}$ and $\mathbf{Z^{\dag}Z}$ are positive define, since for any vector $\mathbf{f}$ the inequalities $\left(\mathbf{f},\mathbf{ZZ^{\dag}}\mathbf{f}\right)=\left(\mathbf{Z^{\dag}}\mathbf{f},\mathbf{Z^{\dag}}\mathbf{f}\right)>0$ and $\left(\mathbf{f},\mathbf{Z^{\dag}Z}\mathbf{f}\right)=\left(\mathbf{Z}\mathbf{f},\mathbf{Z}\mathbf{f}\right)>0$ hold.
As all eigenvalues of a positive definite operator are positive, the singular values are real. 

As the operators $\mathbf{ZZ}^{\dag}$ and $\mathbf{Z}^{\dag}\mathbf{Z}$ are self-adjoint and completely continuous, their limit spectrum consists of one point equal to zero (\cite{vilenkin-1972}, Ch. 4).
Next, as the limit spectrum of an operator includes all points of the continuous spectrum, limit points of the discrete spectrum, as well as infinite-fold eigenvalues, the complete continuity of the composite operators  implies that for any small $\epsilon>0$ the set of eigenvalues exceeding $\epsilon$ is discrete. 

Thus, the set of singular values is bounded from above due to the boundedness of the operator ($\ref{U_view}$), is discrete and has the limit point $\sigma=0$.
The singular vectors are usually numbered in the order of their decrease \cite{golub-reinsch-1970}, the perturbation growth by the time $t$ is limited by the first singular value by that time, and the first right singular vector is the perturbation exhibiting this growth. 

The above considerations imply that to calculate the maximum possible perturbation growth rate it is sufficient to calculate the first singular value, called the optimal growth in the literature, and the right singular vector corresponding to this value will be the sought for (optimal) perturbation demonstrating the maximum possible growth rate.
Below we present two methods of calculation of singular values and corresponding singular vectors. 

Another important consequence of the complete continuity of the dynamical operator $\mathbf{Z}$ is the validity of the Hilbert-Schmidt theorem for the operators $\mathbf{ZZ}^{\dag}$ and $\mathbf{Z}^{\dag}\mathbf{Z}$.
The theorem states that for any self-adjoint linear operator, there is an orthonormal sequence $\{\varphi_n\}$ of eigenvectors corresponding to eigenvalues $\{\lambda_n\}$, such that each element $\xi$ can be uniquely written in the form
$$
\xi=\sum c_k \varphi_k+\xi^{'},
$$
where the vector $\xi^{'}$ satisfies the condition $\mathbf{U}\xi^{'}=0$; here
$$
\mathbf{U}\xi=\sum \lambda_k c_k \varphi_k
$$
и
$$
\lim_{n\to\infty}{\lambda_n}=0
$$
It follows from here that the set of singular functions is orthogonal and complete, as a sequence of eigenvectors of a self-adjoint operator, and can be used as a basis for decomposition of any perturbation. 

\subsection{Matrix method for optimal solutions}
\label{sect_martix}
The first method to calculate singular vectors will be preferably referred to as {\it matrix} method.
It is based on the singular value decomposition of the matrix of a dynamical operator.
As a rule, the set of eigenvectors is used as the basis for the matrix calculation. 

Note that there can be another possibility, which was used, for example, in paper \cite{ioannou-kakouris-2001}, when the space is covered by a grid of points, and each perturbation is given by a column of numbers corresponding to the values of the perturbation at these points.
A dynamical operator corresponds to a matrix obtained by the difference approximation of derivatives in the dynamical equations.
The singular value decomposition of this matrix enables one to calculate the singular vectors at the grid points.
A large size of the operator matrix is a shortcoming of this approach, which requires a lot of time to calculate the singular value decomposition; an advantage is that it is not necessary to calculate operator's eigenvectors.
In this Section, we describe the matrix method in the eigenvector basis. 

The problem is to find a linear combination of the dynamical operator modes whose norm exhibits the largest growth by the given time.
Assume that the sequence of eigenvectors $\{\mathbf{f}_1,\mathbf{f}_2,\mathbf{f}_3 ... \mathbf{f}_N\}$ and the corresponding eigenvalues $\{{\rm e}^{-{\rm i}\omega_1 t},{\rm e}^{-{\rm i}\omega_2 t},{\rm e}^{-{\rm i}\omega_3 t} ... {\rm e}^{-{\rm i}\omega_N t)}\}$ of the operator $\mathbf{Z}$ are known.
In the space of linear combinations of eigenvectors, the representation of arbitrary perturbation vector has the form (see paragraph 4.3.2 and Section 4.4. in \cite{schmid-henningson-2001} for more detail)
\begin{equation}
\label{decomposition}
\mathbf{q}=\sum\limits_{j=1}^{N}\kappa^j\hat{f_j},
\end{equation}
where the numbers $\{\kappa^1,\kappa^2,\kappa^3 ... \kappa^N\}$ are coordinates of the vector $\mathbf{q}$ in the eigenvector basis.
Note that the time dependence of $\mathbf{q}$ is essentially in its coordinates. 

The inner product of two vectors $\mathbf{q}$ and $\mathbf{g}$ in this representation  can be calculated from the known coordinates using the metric matrix $\mathbf{M}$:
\begin{equation}
\label{modal_scalar_production}
\left(\mathbf{q},\mathbf{g}\right)=\left(\mathbf{q}^{\dag}\right)^{i} M_{ij} \mathbf{g}^j,
\end{equation}
where the elements of the metric matrix are equal to the inner product of eigenvectors:
\begin{equation}
M_{ij}=\left(\mathbf{f}_i,\mathbf{f}_j\right)
\end{equation}
Thus, the matrix $\mathbf{M}$ is positive definite, since the norm of a non-zero vector is always positive.

Now the problem of calculation of maximum possible perturbation growth is reduced to finding such values $\kappa^j$ at which the growth of the perturbation norm, determined using these values according to formula (\ref{decomposition}), is maximal by the given time moment. 

The representation of an operator $\mathbf{Z}$ in the eigenvector basis can be easily calculated by letting this operator act on the basis element:
\begin{equation}
\label{Z_f_j}
\mathbf{Z} \mathbf{f}_j=\mathbf{f}_j(\tau)=\mathrm{e}^{-\mathrm{i}\omega_j\tau}\mathbf{f}_j,
\end{equation}
Therefore, in the set of basis eigenvectors an operator can be represented by a diagonal matrix $\mathbf{P}$ with complex exponents on the main diagonal: $\mathbf{P}=diag\{\mathrm{e}^{-\mathrm{i}\omega_1\tau},\mathrm{e}^{-\mathrm{i}\omega_2\tau},\mathrm{e}^{-\mathrm{i}\omega_3\tau} ... \mathrm{e}^{-\mathrm{i}\omega_N\tau}\}$

Next, let us use the first equality from definition \ref{singular_vector_defenition}, $\mathbf{Zv}=\sigma \mathbf{u}$, and rewrite it in the matrix form:
\begin{equation}
\label{matrix_view}
\mathbf{P}=\mathbf{U\Sigma V}^{-1}
\end{equation}
Matrix $\mathbf{\Sigma}$ is diagonal with the singular values on the diagonal, $\mathbf{\Sigma}=diag\{\sigma_1,\sigma_2,\sigma_3 ... \sigma_N\}$.
Columns of matrices $\mathbf{U}$ and $\mathbf{V}$ represent right and left singular vectors, respectively.

Now let us write the inner product for two arbitrary singular vectors $\mathbf{q}$ and $\mathbf{g}$ as
\begin{equation}
(\mathbf{q},\mathbf{g})=\left(\mathbf{q}^{\dag}\right)^i M_{ij} \mathbf{g}^j=\left((\mathbf{Fq})^{\dag}\right)^i(\mathbf{Fg})^j,
\end{equation}
where matrix $\mathbf{F}$ is the {\it Cholesky decomposition} of the metric matrix $\mathbf{M}=\mathbf{F}^T\mathbf{F}$.
As the matrix $\mathbf{M}$ is positive define, its Cholesky decomposition always exists and is unique.

Sets of singular vectors are orthonormal, therefore the following relations for matrices $\mathbf{V}$ and $\mathbf{U}$ hold:
\begin{equation}
\mathbf{V}^{\dag}\mathbf{F}^T\mathbf{FV}=\mathbf{I},
\end{equation}
\begin{equation}
\mathbf{U}^{\dag}\mathbf{F}^T\mathbf{FU}=\mathbf{I},
\end{equation}
where $\mathbf{I}$ is the identity matrix. 

Therefore matrices inverse to $\mathbf{V}$ and $\mathbf{U}$ are expressed through Hermitian-conjugate as follows:
\begin{equation}
\mathbf{V}^{-1}=\mathbf{V}^{\dag}\mathbf{F}^T\mathbf{F},
\end{equation}
\begin{equation}
\mathbf{U}^{-1}=\mathbf{U}^{\dag}\mathbf{F}^T\mathbf{F}.
\end{equation}
Making use of these relations in (\ref{matrix_view}) yields
\begin{equation}
\mathbf{P}=\mathbf{U\Sigma}\mathbf{V}^{\dag}\mathbf{F}^T\mathbf{F}=\mathbf{F}^{-1}\mathbf{F}\mathbf{U\Sigma}\mathbf{V}^{\dag}\mathbf{F}^T\mathbf{F}.
\end{equation}
Rewrite this in the form:
\begin{equation}
\mathbf{FPF}^{-1}=\left(\mathbf{FU}\right)\mathbf{\Sigma}\left(\mathbf{FV}\right)^{\dag}\equiv\tilde{\mathbf{U}}\mathbf{\Sigma}\tilde{\mathbf{V}}^{\dag}.
\end{equation}

Now it is clear that the right-hand side of this equality coincides with the so-called {\it singular value decomposition} of matrix $\mathbf{FPF}^{-1}$.
We remind the reader that the singular value decomposition is a factorization of a matrix in the form $\tilde{\mathbf{U}}\mathbf{\Sigma}\tilde{\mathbf{V}}^{\dag}$ where $\tilde{\mathbf{U}}$ and $\tilde{\mathbf{V}}$ are orthogonal matrices and $\mathbf{\Sigma}$ is a diagonal matrix with positive numbers on the main diagonal \cite{golub-van-loan-1996}.
This factorization exists for any real matrix and is unique.
It can be easily convinced that matrices $\tilde{\mathbf{U}}$, $\tilde{\mathbf{V}}$ and $\mathbf{\Sigma}$ satisfy the singular value decomposition conditions, and therefore to calculate singular values and vectors it is sufficient to perform this decomposition for matrix $\mathbf{FPF}^{-1}$.
The singular value decomposition procedure is a standard tool in many linear algebra software packages. 

The original matrices $\mathbf{U}$ and $\mathbf{V}$ are calculated using $\mathbf{F}^{-1}$: $\mathbf{U}=\mathbf{F}^{-1}\tilde{\mathbf{U}}$, $\mathbf{V}=\mathbf{F}^{-1}\tilde{\mathbf{V}}$.
The maximal of the numbers on the diagonal of matrix $\mathbf{\Sigma}$ is the first singular value by the time $t$, and the corresponding column of matrix $\mathbf{V}$ is the first singular vector in the eigenvector basis. 

\subsubsection{Illustration of the matrix method}
The matrix method to search for optimal perturbations has been used in many studies on stability of laboratory flows (see, for example, \cite{butler-farrell-1992}, \cite{reddy-henningson-1993}, \cite{hanifi-1996}, \cite{meseguer-2002},
\cite{malik-2006}, \cite{maretzke-2014}) and in astrophysical papers \cite{yecko-2004}, \cite{mukhopadhyay-2005}, \cite{zhuravlev-shakura-2009}.
Here, we elucidate it by a simple semi-analytical study \cite{razdoburdin-zhuravlev-2012}, where the eigenvector basis
\footnote{Henceforth, the eigenvectors of an operator $\mathbf{Z}$ multiplied by the eigenvalues, i.e. by the time dependence ${\rm e}^{-{\rm i}\omega t}$, will be referred to as perturbation modes.}
is calculated in the WKB approximation in a geometrically thin and barotropic quasi-Keplerian torus with free boundaries.
For simplicity, only the modes whose corotation radius is outside the outer boundary of the torus are considered.
(See Section \ref{sect_adiabat_perts} for the discussion of mechanism of energy exchange between the modes and the background flow at the corotation radius in the context of the spectral problem corresponding to equations (\ref{sys_A_1})-(\ref{sys_A_3})).
When the corotation radius is outside the flow, the energy of modes is conserved.
This means that they do not show exponential growth or decay, i.e. their frequencies $\omega$ are real values (see expression (\ref{E_mode_dot})).
These perturbations are referred to as {\it neutral} modes.
Nevertheless, due to their mutual non-orthogonality, in other words, due to non-orthogonality of eigenvectors of the dynamical operator acting on perturbations, we expect a non-modal growth of their linear combinations (see the analogy in Fig. \ref{modes} and comment to it in text). 

The modes we wish to obtain below physically correspond to inertial-acoustic waves that form a solid-body rotating pattern in the disk, i.e. that have constant in time and space azimuthal projection of the wave vector.
Here, as will be seen from the WKB analysis, their characteristic radial wavelength is close to the disk thickness $H$.
As for their characteristic azimuthal scale, $\lambda_\varphi$, it can be both larger and smaller than $H$, which is determined by the azimuthal wave number $m$ entering system of equations (\ref{sys_A_1})-(\ref{sys_A_3}).
Results concerning the optimal perturbation growth will be presented for the case $\lambda_\varphi\gg H$ (see Fig. \ref{Growth}). 

We will see that in that case the optimal perturbation does not have the form of a spiral unwound by the flow, which we discussed in the context of the transient growth of vortices (see Fig. \ref{pic_TG}), but is a wave packet initially located at the outer boundary of the torus and further propagating towards its inner boundary.
At the moment of reflection from the inner boundary, its total acoustic energy reaches maximum and then decreases while the packet goes back to the flow periphery.
After reflection from the outer boundary the process repeats. Thus, the non-modal growth in this case is oscillating rather than transient, as must be the case according to the analogy shown in Fig. \ref{modes}. 

\subsubsection{Background flow}
Consider a toroidal flow of finite radial size as a model background flow.
The azimuthal velocity component will correspond to the power-law angular velocity radial profile:
\begin{equation}
\label{Om_q}
\Omega=\Omega_0\left(\frac{r}{r_0}\right)^{-q},
\end{equation}
where $r_0$ is the distance to the gravity center in the equatorial plane of the torus at which the rotation occurs with the Keplerian frequency $\Omega_0$, $2>q>3/2$.
Assume that matter moves in the external Newtonian gravitational potential produced by the central point-like mass:
$$
\Phi = -\frac{\Omega_0^2r_0^3}{(r^2+z^2)^{1/2}}.
$$
As will be clear below, in this case the parameter $q$ characterizes the torus thickness which tends to zero by approaching the angular velocity profile to the Keplerian one.
As in Section \ref{sect_adiabat_perts}, we use here the polytropic equation of state and write the force balance using the enthalpy $h$:
\begin{equation}
\label{stat_system}
\begin{aligned}
&\frac{\partial h}{\partial r} = \Omega^2 r - \frac{\partial \Phi}{\partial r},\\
&\frac{\partial h}{\partial z} =  - \frac{\partial \Phi}{\partial z}, \\
\end{aligned}
\end{equation}
where the first and the second equations correspond to the projection of the Euler equation on the radial and vertical direction, respectively. The joint integration of (\ref{stat_system}) yields
$$
h(r,z) = \frac{\Omega_0^2r_0^3}{(r^2+z^2)^{1/2}} + \frac{\Omega_0^2 r_0^{2q}}{2(1-q)} r^{2(1-q)}  + C,
$$
where the integration constant $C$ is determined from the condition that $h(r_1,0)=0$ at the inner boundary of the torus $r_1<r_0$. 

Then, in dimensionless coordinates $\hat x \equiv r/r_0$, $\hat y \equiv z/r_0$ we obtain
\begin{equation}
\label{enth_stat}
h = (\Omega_0 r_0)^2  \left [ (\hat x^2 + \hat y^2)^{-1/2} - \hat x_1 ^{-1} + 
\frac{1}{2(q-1)} \left ( \hat x_1^{-2(q-1)} - \hat x^{-2(q-1)} \right )  \right ].
\end{equation}
Here $\hat x_1\equiv r_1/r_0$.
The enthalpy distribution (\ref{enth_stat}) gives also the outer radial boundary of the torus $\hat x_2>1$, where $h(\hat x_2,0)=0$.
The quantity $\hat x_d=\hat x_2-\hat x_1$ will be called the radial extension of the flow. 

Now it is not difficult to pass to the case of quasi-Keplerian, geometrically thin torus of interest here: $q=\frac{3}{2}+\frac{\epsilon^2}{2}$, $\epsilon \ll 1$.
Using this assumption, the enthalpy profile can be simplified to 
\begin{equation}
\label{h}
\frac{h}{\Omega_0^2 r_0^2} = \frac{\hat H^2}{2 \hat x^3} \left [ 1 - \left( \frac{\hat y}{\hat H} \right )^2 \right ],
\end{equation}
where $\hat H(x)$ is the dimensionless thickness of the torus in units of $r_0$:
\begin{equation}
\label{H}
\hat H = \delta \, \hat x \left [  \frac{ \hat x_1 (1+\ln \hat x) - \hat x (1+\ln \hat x_1) } {\hat x_1 - 1 - \ln \hat x_1 } \right ]^{1/2}
\end{equation}
Here, we have introduced a descriptive small parameter
$$
\delta \equiv \hat H(\hat x=1) = 2^{1/2}\epsilon \left ( 1 - \frac{1+\ln \hat x_1}{\hat x_1} \right )^{1/2}\ll 1,
$$ 
that defines the characteristic aspect ratio of the disk-like torus with $\delta \ll \hat x_d$.
It is not difficult to make sure that expression (\ref{h}) coincides with (\ref{vert_enthaply}). 

Equations (\ref{h}), (\ref{H}) fully determine the quasi-Keplerian background flow which will be used to illustrate the matrix method of determination the non-modal growth of the modes superposition.
In the next Section we will solve the spectral problem for such a flow, i.e. will find the perturbation mode profiles. 

\subsubsection{The modes}

The modes are non-stationary perturbations with exponential time dependence $\propto\exp(-{\rm i}\omega t)$.
They are also solutions of the operator equation (\ref{Z_eq}) determining evolution of a linear perturbation in the flow.
This means that the modes are state vectors which we obtain by the operator $\mathbf{Z}$ acting on its eigenvectors $\mathbf{f}_i$:
$$
\mathbf{f}_i(t) =\mathbf{Z f}_i = {\rm e}^{-{\rm i}\omega t}{\bf f}_i.
$$ 
Again, the numbers $\exp(-{\rm i}\omega t)$ are eigenvalues of $\mathbf{Z}$ that we will need to find along with its eigenvectors.

In practice, we will not use the equation exactly in the form (\ref{Z_eq}), but instead derive an equivalent ordinary differential equation of the second order in the radial coordinate for an Eulerian enthalpy perturbation.
As everywhere in this paper, we will assume that the hydrostatic equilibrium  always holds, i.e. that $\delta v_z=0$.
As we deal with thin torus, $\delta\ll 1$, our perturbations taken originally in the form of azimuthal Fourier harmonics $\propto \exp({\rm i}m\varphi)$ satisfy the system of equations (\ref{sys_A_1})-(\ref{sys_A_3}), which contains the background variables integrated over $z$ (see Section \ref{sect_adiabat_perts}).
The modal analysis implies the substitution $\partial /\partial t \to {\rm i}\omega$, after which from (\ref{sys_A_1}) and (\ref{sys_A_2}) we find that complex Fourier harmonics of the Eulerian velocity perturbations, which are denoted here as ${\bf v}_r$ and ${\bf v}_\varphi$, are expressed through the Fourier harmonics of the enthalpy perturbation, which is denoted here as ${\bf W}$, as follows:
\begin{equation}
\label{v_r}
{\bf v}_r=\frac{i}{D}\left[\bar\omega\frac{d {\bf W}}{d\hat x}-\frac{2m\Omega {\bf W}}{\hat x}\right],
\end{equation}

\begin{equation}
\label{v_phi}
{\bf v}_{\varphi}=\frac{1}{D}\left[\frac{\kappa^2}{2\Omega}\frac{d{\bf W}}{d\hat x}-\frac{m\bar\omega {\bf W}}{\hat x}\right],
\end{equation}
where $D\equiv\kappa^2-\bar\omega^2$, $\kappa^2=\frac{2\Omega}{x}\frac{d}{d\hat x}\left(\Omega \hat x^2\right)$ is, as usually, the epicyclic frequency squared, and $\bar\omega\equiv\omega-m\Omega$ is the shifted frequency.
Below in this Section we assume that all frequencies are taken in units of the frequency $\Omega_0$ and time is taken in units of $\Omega_0^{-1}$. 

Plugging (\ref{v_r}) and (\ref{v_phi}) into the continuity equation (\ref{sys_A_3}), we obtain the following equation for ${\bf W}$:
\begin{equation}
\label{eq_3D}
\frac{D}{\hat x\Sigma}\frac{d}{d\hat x}\left(\frac{\hat x\Sigma}{D}\frac{d{\bf W}}{d\hat x}\right)
-\left[\frac{2m}{\bar \omega}\frac{D}{\hat x\Sigma}\frac{d}{d\hat x}\left(\frac{\Omega\Sigma}{D}\right)+(n+1/2)\frac{D}{h_*}+\frac{m^2}{\hat x^2}\right]{\bf W}=0,
\end{equation}
where
\begin{equation}
\label{sur_a}
\Sigma\left(r\right)=\int\limits_{-H}^{H}\rho dz \propto \hat H \left ( \frac{\hat H^2}{\hat x^3}\right )^n
\quad \mbox{и} \quad
h_*=\frac{\hat H^2}{2\hat x^3}.
\end{equation}
Here $h_*$ is the dimensionless background enthalpy in the equatorial disk plane (cf. (\ref{h})).
To reproduce the surface density dependence on $r$ given above, $\Sigma(r)$, it is enough to recall that $\Sigma \sim H \rho|_{z=0}$, and $\rho \sim h^n$ for the polytorpic equation of state.
Equation (\ref{eq_3D}), as well as its more general analogue for three-dimensional perturbation modes, is often used in the literature.
Their derivation and analysis can be found, for example, in papers \cite{goldreich-1986}, \cite{kato-1987}, \cite{sekiya-miyama-1988}, \cite{kojima-1989}, \cite{kato-2001}. 

As we have already said, the integration of equation (\ref{eq_3D}) is complicated by resonances: the corotational one, where $\bar\omega=0$, and the Lindblad's resonances, where $D=0$.
These points are singular for (\ref{eq_3D}).
However, in order to illustrate the matrix method of optimization, we will restrict ourselves by calculation of only part of the modes with resonances lying outside the outer boundary of the flow, $\hat x_2$.
The condition that the inner Lindblad resonance lies at $\hat x > \hat x_2$ implies
\begin{equation}
\label{neutral}
\omega<(m-1)\Omega(\hat x_2),
\end{equation}
where in the condition $D=0$ we have set $\kappa\approx\Omega$ because of a nearly Keplerian angular velocity profile in a thin torus.
We remind also that $\omega$ is a real value.
Note that for $m=1$ the inner Lindblad resonance is at $\hat x=0$, and hence there are no modes with $m=1$ satisfying the condition (\ref{neutral}).
For this reason, we will consider only modes with $m>1$.
Thus, under the restrictions made, the term $\propto D / h_* \sim \delta^{-2}$ will be large everywhere in the flow, and therefore the solution of equation can be looked for in the WKB approximation.

A WKB solution of equation (\ref{eq_3D}) can be written as 
\begin{equation}
\label{wkb}
{\bf W} = {\bf C}_0 S_1 \cos (S_0 + \varphi_0),
\end{equation}
in which $S_0\sim \delta^{-1}$, and $S_1\sim \delta^{0}$. 

Plugging (\ref{wkb}) into (\ref{eq_3D}) yields its decomposition in $\delta$.
By collecting terms with similar powers of $\delta$, namely, $\delta^{-2}$ and $\delta^{-1}$, we find the explicit form of functions $S_0$ and $S_1$:
$$
S_0 = \int\limits_{\hat x_1}^{\hat x} \left ( (n+1/2) \frac{-D}{h_*} - \frac{m^2}{\hat x^2} \right )^{1/2} d\hat x,
$$
$$
S_1 = \left ( \frac{-D}{\hat x\Sigma} \right )^{1/2}
\left ( (n+1/2)\frac{-D}{h_*} - \frac{m^2}{\hat x^2} \right )^{-1/4}.
$$
The phase $\varphi_0$ is fixed by the boundary conditions.

The WKB solution (\ref{wkb}) is irregular at the boundary points $\hat x_1$ and $\hat x_2$ at which $h_*\to 0$.
It is possible to find a WKB-solution that is regular at the boundaries (see \cite{heading-2013}), but here, let us use another common way of matching the (\ref{wkb}) with an approximate {\it regular} solution of the original equation (\ref{eq_3D}) near $\hat x_1$ and $\hat x_2$.
This matching should yield a discrete set of eigenfrequencies $\omega$, as well as the value of $\varphi_0$.

In order to find the regular solution near $\hat x_1$ and $\hat x_2$, we change to the new radial coordinate $\tilde x\equiv |\hat x-\hat x_{1,2}|$ and expand equation (\ref{eq_3D}) in the main order of the variable $\tilde x \ll 1$.
Technically, this means that all variables from (\ref{eq_3D}) that are non-zero at $\hat x_{1,2}$ are set to its exact values at $\hat x_{1,2}$.
The disk semi-thickness, vanishing at the boundaries, is approximated as $\hat H=\hat H_{1,2}\tilde x^{1/2}$.
Here the constant $\hat H_{1,2}$ is as follows
$$
\hat H_{1,2} = \delta \hat x_{1,2} \left | {\frac{\ln{\hat x_{1,2}}}{1+\ln{\hat x_{1,2}}-\hat x_{1,2}}} \right |^{1/2}
$$ 

We obtain the following near-boundary equation:
\begin{equation}
\label{eq_border}
\tilde x\, \frac{d^2 {\bf W}}{d\tilde x^2} + (n+1/2)\,\frac{d{\bf W}}{d\tilde x} + 
E_{1,2}{\bf W}=0,
\end{equation}

$$
\mbox{where} \quad E_{1,2} = \frac{(2n+1) (-D_{1,2})\, \hat x_{1,2}^3}{H_{1,2}^2}, \quad D_{1,2} \,\, 
\mbox{- are the values of } D \mbox{ at points} \,\, \hat x_{1,2}.
$$

The regular at $\tilde x=0$ solution of (\ref{eq_border}) has the form:

\begin{equation}
\label{solve_border}
{\bf W}={\bf C}_{1,2}\,\tilde x^{-(2n-1)/4} \, J_{n-1/2} ( \tilde z ),
\end{equation}
where $\tilde z = 2 E_{1,2}^{1/2}\,\tilde x^{1/2}$.

Note that equation (\ref{eq_border}) at $\tilde x \to 0$ is equivalent to the boundary condition for the enthalpy perturbation at the free boundary of the flow which states that the Lagrangian enthalpy perturbation vanishes at the boundary points $\hat x_{1,2}$, $\Delta h|_{x_{1,2}}=0$ (see, for example \cite{glatzel-1987a}). 

As the denominator $\tilde z$ contains the small $\delta$, at some distance from the boundary points $\tilde z\gg 1$ yet under the condition $\tilde x \ll 1$.
In this region, ${\bf W}$ is given by the asymptotic of (\ref{solve_border}) for large argument:
\begin{equation}
\label{bessel_approx}
{\bf W} \approx {\bf C}_{1,2} \,\tilde x^{-n/2} ({4\pi^2E_{1,2}})^{-1/4} 
\cos{\left(2 E_{1,2}^{1/2}\, \tilde x^{1/2} - n \,\pi/2\right )}
\end{equation}

The matching of (\ref{bessel_approx}) with the WKB decomposition of the solution near $\hat x_1$ and $\hat x_2$ yields the zero phase $\varphi_0 = -n \pi/2$ in equation (\ref{wkb}) and the following dispersion equation:

\begin{equation}
\label{dispersion}
\int\limits_{\hat x_1}^{\hat x_2} \left ( (2n+1) \frac{-D \hat x^3}{\hat H^2} - \frac{m^2}{\hat x^2} \right )^{1/2} d\hat x = \pi (n + p),
\end{equation}
where $p$ is an integer number.
Solving (\ref{dispersion}) for different $p$ yields a discrete set of $\omega$ that enters $D$.
This is the sequence of eigenfrequencies of neutral modes we are interested in. 

The modes profiles are given by equations (\ref{wkb}) and (\ref{solve_border}) with account for the relations between the corresponding constants:
\begin{equation} 
\frac{{\bf C}_0}{{\bf C}_1} = \left ( \frac{\hat H_{1}^{2n+1}}{2\pi \hat x_{1}^{3n-1} (-D_{1})} \right )^{1/2}, \, \quad
\frac{{\bf C}_2}{{\bf C}_1}=\left(-1\right)^p \left [ \left( \frac{\hat x_2}{\hat x_1}\right)^{3n-1}
\frac{D_2}{D_1}\left(\frac{\hat H_1}{\hat H_2}\right)^{2n+1} \right ]^{1/2}
\end{equation}

After obtaining the profile ${\bf W}(\hat x)$ for a given $\omega_i$, the corresponding complex Fourier harmonics of the Eulerian velocity perturbations  ${\bf v}_r(\hat x)$ and ${\bf v}_{\varphi}(\hat x)$ can be calculated from (\ref{v_r}) and (\ref{v_phi}).
Thus, we find the whole eigenvector ${\bf f}_i\equiv \{{\bf v}_r,\, {\bf v}_\varphi,\, {\bf W}\}$ of operator ${\bf Z}$ corresponding to its eigenvalue $\exp(-{\rm i}\omega_i t)$. 

\subsubsection{Optimal growth}
\label{matrix_growth}
\begin{figure}[h!]
\includegraphics[width=1\linewidth]{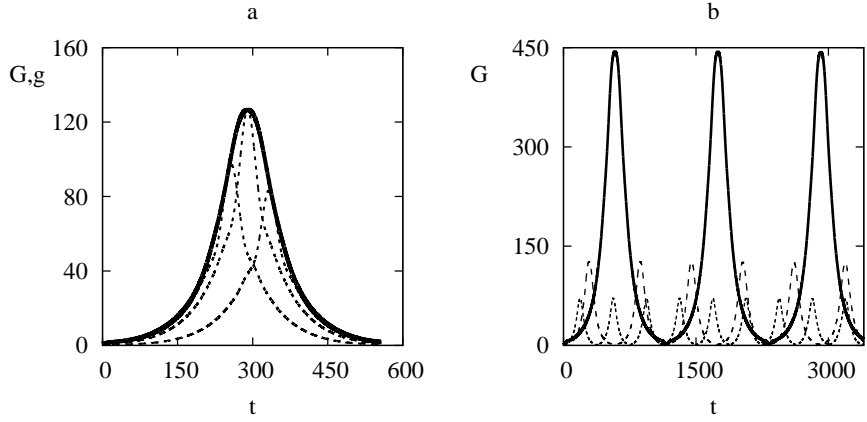}
\caption{\footnotesize{
Panel 'a' shows the optimal growth curve $G(t)$ (the solid curve) of a linear combination of slow modes in a thin disk for $\delta=0.002$.
The dashed lines show the growth factor of the total acoustic energy $g(t)$ of the particular optimal perturbations as a function of time.
These perturbations are optimal for time intervals $t=250, 290, 390$ expressed in units of the characteristic Keplerian period $2\pi\Omega_0^{-1}$.
Panel 'b' shows only curves of $G(t)$.
The solid, dashed and dotted lines correspond to $\delta=0.001,0.002,0.003$, respectively.
The linear combination shown has the dimensionality $N=20$, parameters are $x_d=1.0$, $m=25$, $n=3/2$ (Figure is quoted from \cite{razdoburdin-zhuravlev-2012}). 
}}
\label{Growth}
\end{figure}

The explicit form of eigenvectors of a dynamical operator allows us to calculate {\it the optimal growth}, i.e. to find such a linear combination of these vectors that demonstrates the maximum increase of the norm by the given time. 
The optimal growth by the time $t$ is 
\begin{equation}
\label{G_global}
G(t)=\max\limits_{\mathbf{q}(0)}\frac{||\mathbf{q}(t)||^2}{||\mathbf{q}(0)||^2}.
\end{equation}
This is the generalization of (\ref{G_t}) for the spatially global case.

The inner product of two vectors from the linear span of $N$ eigenvectors of $\mathbf{Z}$ is introduced such that the square of the corresponding norm recovers the acoustic energy of perturbation (\ref{ac_norm}):

\begin{equation}
\label{scalar_production}
(\mathbf{f},\mathbf{g})=\pi\int\limits_{r_1}^{r_2}\Sigma\left((\delta v_r)_f(\delta v_r)^*_g + (\delta v_{\varphi})_f(\delta v_{\varphi})^*_g + (n+1/2)\frac{(\delta h)_f(\delta h)^*_g}{h_*}\right)r dr,
\end{equation}
where the indices $'f'$ or $'g'$ indicate the relation of some physical variable to the vector ${\bf f}$ or ${\bf g}$, respectively.
We remind that by $\delta v_r$, $\delta v_\varphi$ and $\delta h$ here we mean azimuthal Fourier harmonics of the Eulerian perturbations of the velocity and enthalpy components, respectively. 

Now, let us apply the procedure of calculation of the optimal combination of eigenvectors described above.
As we have the eigenvectors in the analytical form, the matrix $\mathbf{M}$ can be obtained by simple numerical integration of the combination of elementary functions using the inner product formula (\ref{scalar_production}):
\begin{equation}
M_{ij}=(\mathbf{f}_i,\mathbf{f}_j)
\end{equation}
Next, we perform the Cholesky decomposition $\mathbf{M}=\mathbf{F}^T\mathbf{F}$ and then the singular value decomposition of the matrix $\mathbf{FPF}^{-1}$.
Both these procedures are standard in numerical methods of matrix algebra.

In Fig. \ref{Growth}a and \ref{Growth}b we show an example of the dependence of the maximum possible energy growth, $G(t)$, among all superpositions of 20 neutral modes by the time $t$ on time scale of the order of the sonic time $t_s \sim (\delta\Omega_0)^{-1}$ and $\sim 10 t_s$, respectively.
Fig. \ref{Growth}a also shows the energy growth of the optimal mode combinations $g(t)$.
Clearly, the curves $g(t)$ touch the general optimal growth curve $G(t)$, as must be the case, each at its own optimization time.
The optimal growth itself in this model has a quasi-periodic form by reaching maxima at times $\sim t_s$, and the thinner the torus, the higher values $g$ the mode superposition can reach. 

\subsubsection{The angular momentum flux}
\begin{figure}[h!]
\includegraphics[width=1\linewidth]{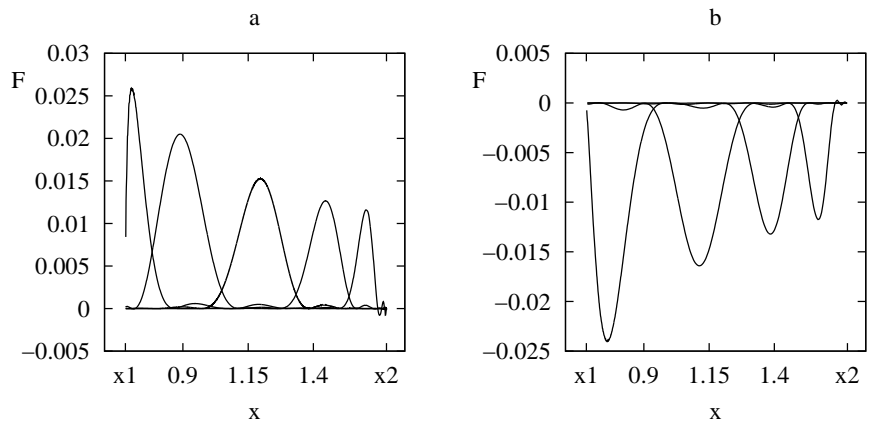}
\caption{\footnotesize{
Radial profiles of the azimuthally averaged angular momentum flux density $F$ of the optimal perturbation, which corresponds to $g(t)$ shown in Fig. \ref{Growth}a for the optimization time $t=290$.
(a) Profiles of $F$ at the instants $t=50, 100, 150, 200, 240$ before the $g(t)$ maximum.
Each profile has one large maximum shifting from the outer disk boundary $x_2$ to the inner disk boundary $x_1$ as $t$ increases.
(b) Profiles of $F$ at the instants $t=290, 350, 400, 450$ after the $g(t)$ maximum. Correspondingly, each profile has one large minimum shifting from the inner disk boundary $x_1$ 
to the outer disk boundary $x_2$ as $t$ increases further. The linear combination has the dimensionality $N=20$, 
the parameters are $\delta=0.002$, $x_d=1.0$, $m=25$, $n=3/2$. (Figure from paper \cite{razdoburdin-zhuravlev-2012}). 
}}
\label{Flux}
\end{figure}

In Section \ref{matrix_growth} we have shown that some combinations of modes can demonstrate a significant growth in the acoustic energy.
Consider in more detail what is this optimal perturbation.
The perturbation amplitude growth suggests that the main flow transfers energy to perturbations.
The first term in the right-hand side of (\ref{ac_en_der}) is responsible for this, and its integrand sometimes is referred to as the Reynolds stress (which we denote as $F_R$, see \cite{kojima-1989}).
It turns out that $F_R$ is simply related to the density of the specific angular momentum flux excited by perturbations, $F$: $F_R = -\frac{d\Omega}{d\hat x}F$ (see Sections 2.3 and 4 of \cite{savonije-heemskerk-1990}).
Clearly, for the Keplerian rotation $F_R$ and $F$ has the same sign: if the perturbation energy increases, $F>0$, the angular momentum flux to the torus periphery occurs, and vice versa. 

In order to calculate the evolution of the profile $F$ for the optimal superposition of modes represented by the curve $g(t)$ for $t=290$ in Fig. \ref{Growth}a, .
let us use the following expression for $F$
\begin{equation}
F = \hat x\Sigma <\delta v_r \delta v_\varphi>.
\end{equation}
Fig. \ref{Flux} shows how the radial distribution of $F$ changes on the interval $(\hat x_1,\hat x_2)$.
At first, we see that $F$ is radially localized, and its localization region changes with time: during the perturbation growth phase it shifts towards the inner torus boundary, whereas during the perturbation decay phase --- back to the outer boundary.
Therefore, in this case the non-modal growing perturbation is represented by a wave packet containing a set of neutral modes (each of the modes, as we remind, rotates like a solid body with an angular velocity somewhat smaller than the angular velocity of the flow).
Initially, this wave packet is localized near the outer boundary of the torus and moves towards the inner boundary.
This causes the angular momentum outflow to the disk periphery, since $F>0$, and its acoustic energy increases.
At the moment of reflection from the inner boundary the sign of $F$ and the direction of motion of the wave packet reverse, which later leads to decrease in its acoustic energy, and the angular momentum inflows back to the inner parts of the torus.
As there is no viscous dissipation and the background flow is stationary, obviously, the evolution the optimal perturbation will further repeat: the wave packet, after reflecting from the outer boundary, will go back towards the inner boundary.
Note also that the shape of $G(t)$, which has been obtained, suggests that during the evolution of this particular type of perturbations there are epochs (time intervals counted from the conventional start of the perturbation evolution) when there is
no combination of modes be amplified.
These epochs correspond to minima on the curve of $G(t)$ (see Fig. \ref{Growth}b). This is because only the wave packets localized near the outer disk boundary can exhibit significant growth. At the same time, the velocity of their radial motion is determined by the sound velocity in the flow, and hence the time intervals 'favorable' for the non-modal growth always take the value $\sim \hat x_d / \delta$. 

If we plot the lines of constant phase of perturbations corresponding to the optimal wave packet on the plane $(r,\varphi)$, it turns out that at the growth stage it corresponds to a {\it trailing} spiral.
It is opened at the initial time, but while propagating towards the inner boundary it winds up stronger and stronger.
Oppositely, after the reflection from the inner boundary, it transforms into a tightly wound {\it leading} spiral, and while moving back towards the outer boundary the degree of winding gradually decreases.
This behavior of the optimal perturbation is similar to the process of enhancement/weakening of the shear density waves by a shear flow we discussed in Section \ref{sect_local_appr} in the context of the spatially local problem. 

\subsection{Alternative: a variational approach}
Singular vectors can be found alternatively by a variational method.
This method represents a generalization of the {\it power iterations} procedure of looking for matrix eigenvalues and eigenvectors in finite dimensional framework (see, for example, monograph \cite{golub-van-loan-1996}).
The variational method requires less computational power than the matrix method \cite{luchini-2000}, and, importantly, it can be applied to non-stationary background flows, as well as used to solve the non-linear problem of transient dynamics of finite-amplitude perturbations.
Unlike the matrix method, it does not require discrete representation of the dynamical operator, i.e., for example the decomposition of perturbations by eigenvectors, whose computation in a shear flow faces the known difficulty while bypassing the corotation and Lindblad resonances (see  \cite{lin-1955}). 

As for linear dynamics, the variational method turns out to be {\it equivalent} to solving a more simple problem of seeking the maximum eigenvalue of the operator $\mathbf{Z^\dag Z}$ (see  Section \ref{sect_singular} and, for example, \cite{andersson-1999} as well). That is why we start with solving exactly that problem, whereas the derivation of the variational method itself directly from the variational principle will be given below during the generalization to the non-linear case.

\subsubsection{Linear autonomous operators}
\label{linear_autonomous}
In Section \ref{sect_singular} after the singular values were introduced, we discussed that the first singular value is simultaneously the maximum eigenvalue of the composite operator $\mathbf{Z}^{\dag}\mathbf{Z}$, and the first right singular vector is the corresponding eigenvector of this operator.
First, let us try to understand what the action of $\mathbf{Z}^{\dag}\mathbf{Z}$ on the initial state vector ${\bf q}(0)$ is equivalent to.
Here, the action of the first (right) part of the composite operator is known from its definition (\ref{Z_eq}): this is the integration of equations of perturbation dynamics, for example, the system (\ref{sys_A_1})-(\ref{sys_A_3}), until the time $t$ starting from the initial condition ${\bf q}(0)$. We symbolically rewrite this as
\begin{equation}
\label{dynamic_equation}
\frac{\partial \mathbf{q}}{\partial t}=\mathbf{Aq}.
\end{equation}

Note that due to linearity of the problem, the operator $\mathbf{A}$ in (\ref{dynamic_equation}) does not depend on $\mathbf{q}$ itself.

The subsequent action of the operator ${\bf Z}^\dag$ on ${\bf q}(t)$ is not difficult to understand if the operator $\mathbf{A}$ is {\it  autonomous}, i.e. time-independent \cite{farrell-ioannou-1996a}. 

Then, the integration of (\ref{dynamic_equation}) can be written in the operator form: $\mathbf{q}(t)=\mathrm{e}^{\mathbf{A}t}\mathbf{q}(0)$, i.e. $\mathbf{A}$ and $\mathbf{Z}$ are related as 
\begin{equation}
\label{oper_exp}
\mathbf{Z}=\mathrm{e}^{\mathbf{A}t}.
\end{equation}
The right-hand side of (\ref{oper_exp}) is called {\it the operator exponent} and should be understood as infinite series ${\bf I} + {\bf A}t + ({\bf A}t)^2/2 + ...$.

The operator adjoint to $\mathbf{Z}$ can also be written through the operator exponent $\mathbf{Z}^{\dag}=\mathrm{e}^{\mathbf{A^{\dag}t}}$, where $\mathbf{A}^{\dag}$ is the operator adjoint to $\mathbf{A}$. 
$\mathbf{A}^{\dag}$ is defined by the Lagrange relation $\left(\mathbf{Aq},\mathbf{\tilde q}\right)=\left(\mathbf{q},\mathbf{A}^{\dag}\mathbf{\tilde q}\right)$, where ${\bf q}$ and $\tilde {\bf q}$ are arbitrary vectors.
This expression for $\mathbf{Z}^\dag$ follows from the application of the conjugation operation to the infinite operator series given above.
Now consider the inner product:
\begin{equation}
\label{scalar1}
\left(\frac{\partial \mathbf{q}}{\partial t}, \tilde{\mathbf{q}}\right)=\left(\mathbf{Aq}, \tilde{\mathbf{q}}\right)=\left(\mathbf{q}, \mathbf{A}^{\dag}\tilde{\mathbf{q}}\right).
\end{equation}
On the other hand,
\begin{equation}
\label{scalar2}
\begin{aligned}
\left(\frac{\partial \mathbf{q}}{\partial t}, \tilde{\mathbf{q}}\right)=
\frac{\partial}{\partial t}\left(\mathbf{q}, \tilde{\mathbf{q}}\right)-\left(\mathbf{q}, \frac{\partial\tilde{\mathbf{q}}}{\partial t}\right)=
\frac{\partial}{\partial t}\left(\mathrm{e}^{\mathbf{A}t}\mathbf{q}(0), \tilde{\mathbf{q}}\right)-\left(\mathbf{q}, \frac{\partial\tilde{\mathbf{q}}}{\partial t}\right)= \\
\left(\mathbf{q}(0), \frac{\partial}{\partial t} \left(\mathrm{e}^{\mathbf{A}^{\dag}t}\tilde{\mathbf{q}}\right)\right)-\left(\mathbf{q}, \frac{\partial\tilde{\mathbf{q}}}{\partial t}\right).
\end{aligned}
\end{equation}
Combining (\ref{scalar1}) and (\ref{scalar2}) yields the identity:
\begin{equation}
\label{result_scalar}
\left(\mathbf{q}(0), \frac{\partial}{\partial t} \left(\mathrm{e}^{\mathbf{A}^{\dag}t}\tilde{\mathbf{q}}\right)\right)-\left(\mathbf{q}, \frac{\partial\tilde{\mathbf{q}}}{\partial t}\right)=
\left(\mathbf{q}, \mathbf{A}^{\dag}\tilde{\mathbf{q}}\right).
\end{equation}
It is easy to note that if $\tilde{\mathbf{q}}$ and $\frac{\partial \mathbf{\tilde q}}{\partial t}$ are related as
\begin{equation}
\label{conjugated_equation}
\frac{\partial \mathbf{\tilde q}}{\partial t}=-\mathbf{A^{\dag}\tilde q},
\end{equation}
then $\mathbf{\tilde q}(t)=\mathrm{e}^{\mathbf{-A^{\dag}t}}\mathbf{\tilde q}(0)$ and the identity (\ref{result_scalar}) is fulfilled for arbitrary $\mathbf{q}$.

Thus, the action of operator $\mathbf{Z}^{\dag}=\mathrm{e}^{\mathbf{A}^{\dag}t}$ is equivalent to integration of equation (\ref{conjugated_equation}) {\it backward} in time from the instant $t$ with initial condition $\mathbf{q}(t)$ down to the instant $t=0$. Equation (\ref{conjugated_equation}) is called {\it the adjoint equation}.

Additionally, note that although the operator $\mathbf{Z}$ can be represented as $\mathbf{Z}=\mathrm{e}^{\mathbf{A}t}$ and $\mathbf{Z}^{\dag}$ as $\mathbf{Z}^{\dag}=\mathrm{e}^{\mathbf{A^{\dag}t}}$, the composite operator {\it cannot} be represented as $\mathbf{Z^{\dag}Z} = \mathrm{e}^{(\mathbf{A} + \mathbf{A^{\dag})t}}$.
In order to see this, it is sufficient to employ the series expansion of those operators.

Thus, the action of the composite operator $\mathbf{Z^\dag Z}$ on the initial vector $\mathbf{q}(0)$ is equivalent to integration of the original equation (\ref{dynamic_equation}) with the initial condition $\mathbf{q}(0)$ forward in time up to the instant $t$, and to subsequent integration of the adjoint equation (\ref{conjugated_equation}) with the initial condition in the form of the vector $\mathbf{q}(t)$ we have just obtained by integrating (\ref{dynamic_equation}) --- backward in time up to $t=0$.

If the action of the composite operator $\mathbf{Z^{\dag}Z}$ on some vector is equivalent to its multiplication by a constant, this vector is a right singular vector of $\mathbf{Z}$, and the constant is the square of the corresponding singular value: $\mathbf{Z^{\dag}Zv}=\sigma^2 \mathbf{v}$.
However, we need only the first, i.e. the largest, right singular vector.
In order to obtain it, consider an iteration procedure with one step consisting of action by the composite operator $\mathbf{Z}^{\dag}\mathbf{Z}$  with subsequent normalization of the result to unity.
To show convergence of iterations to the first singular vector, consider the decomposition of arbitrary state vector over the singular vectors $\mathbf{q}(0)=\sum\limits_{k=1}^{\infty}q_k\mathbf{v}_k(0)$ and act on it by the iteration operator: $\mathbf{Z}^{\dag}\mathbf{Z}\mathbf{q}(0)=\sum\limits_{k=1}^{\infty}\sigma^2_kq_k\mathbf{v}_k(0)$.

Obviously, the iteration operator increases the weight of each singular vector in proportion to the square of its singular value.
Thus, the limit $\left(\mathbf{Z}^{\dag}\mathbf{Z}\right)^{p\to\infty}\mathbf{q}(0)$, where $p$ is a natural number, for arbitrary initial state vector $\mathbf{q}(0)$ is equal to the first right singular vector, since it corresponds to the maximum singular value. The rate of divergence depends on the difference between the singular vectors.

Note that in order to converge exactly to the first singular vector, the initial approximation should not be orthogonal to it, so the in the decomposition of the vector $\mathbf{q}(0)$ the weight of the first singular vector is non-zero, $q_1\neq0$, otherwise, the action of the iteration operator will not increase this weight: $\sigma_1^2q_1=0$.
In the latter case, the iteration scheme will converge to the singular vector with the largest singular value among all vectors that have non-zero weight in the initial decomposition.

After all remarks, we would like to fix once again that in order to find the first right singular vector it is necessary to apply the iteration procedure, which includes the integration of the original equation (\ref{dynamic_equation}) forward in time and of the adjoint equation (\ref{conjugated_equation}) --- backward in time with the subsequent normalization to unity on each iteration step.

\subsubsection{Linear non-autonomous operators}

In the case of a time-dependent operator $\mathbf{A}$ (so-called {\it non-autonomous} operator, see \cite{farrell-ioannou-1996b}), the action of operator $Z^\dag$ also corresponds to integration of equation (\ref{conjugated_equation}) backward in time, what can be verified as follows.

For the non-autonomous operator $\mathbf{A}$, the action of operator $\mathbf{Z}$ can be factorized as a product of actions of infinitesimal operators:

\begin{equation}
\mathbf{Z}(\tau)=\lim_{n\to\infty}\prod\limits_{j=1}^n \mathrm{e}^{\mathbf{A}(t_j)\delta t},
\end{equation}
where $\delta t=\tau/n$; $(j-1)\delta t<t_j<j\delta t$, see \cite{farrell-ioannou-1996b}. 

The conjugation of the product of operators yields
\begin{equation}
\mathbf{Z}^{\dag}(\tau)=\lim_{n\to\infty}\prod\limits_{j=n}^1 \mathrm{e}^{\mathbf{A}^{\dag}(t_j)\delta t}.
\end{equation}
Clearly, at each time interval $\delta t$ the integration is performed backward in time, and the intervals themselves are ordered with decreasing $j$, therefore the action of $\mathbf{Z}^{\dag}$ is again equivalent to integration of (\ref{conjugated_equation}) backward in time. 

Thus, like in the case of autonomous operators, the action of $\mathbf{Z^\dag Z}$ is equivalent to consecutive integration of (\ref{dynamic_equation}) forward in time and of (\ref{conjugated_equation}) backward in time. 

Correspondingly, the iteration procedure to search for the first singular vector presented above is applicable to non-autonomous operators as well. 

\subsubsection{Calculation of the consecutive singular vectors}
Singular vectors produce an orthogonal set of vectors that can be used as a basis for decomposition of any linear perturbation.
Thus, it could be useful to calculate not only the first but also the consecutive singular vectors. Therefore, below we will briefly describe their calculation by the variational method.

In order that the iterations described above converge not to the first singular vector but to vector by number $N$, it is sufficient that the domain of the iteration operator completes the subset of linear combinations of previous $N-1$ vectors, or, what is equivalent, the initial approximation is orthogonal to the already obtained singular vectors, i.e. the condition $\left(\mathbf{q}(0),\mathbf{v}_j(0)\right)=0$ should be satisfied for $j<N$.
In this case, the action of the iteration operator will be orthogonal to the obtained singular vectors:
\begin{equation}
\left(\mathbf{Z}^{\dag}\mathbf{Z}\mathbf{q}(0),\mathbf{v}_j(0)\right)=\left(\mathbf{Z}^{\dag}\mathbf{Z}\sum\limits_{k=N}^{\infty}q^k\mathbf{v}_k(0),\mathbf{v}_j(0)\right)=\left(\sum\limits_{k=N}^{\infty}\sigma_k^2q^k\mathbf{v}_k(0),\mathbf{v}_j(0)\right)=0
\end{equation}

Thus, if we expand some vector over the singular vectors in the form
\begin{equation}
\mathbf{q}(0)=\sum\limits_{k=1}^{\infty}q^k\mathbf{v}_k(0),
\end{equation}
then the change of the initial condition in the iteration procedure by $\mathbf{q}(0)-\sum\limits_{k=1}^{N-1}q^k\mathbf{v}_k(0)$ provides the convergence of power iterations to the singular vector number $N$.
Thus, having the previous $N-1$  singular vectors it is always possible to calculate the next one. 

\subsubsection{Generalization to non-linear case}

In the case of non-linear dynamics, the justification of iterative computation of optimal growth presented in two previous Sections becomes invalid, however, in somewhat generalized form it can be obtained directly from the variational principle, as we will show below.

The problem is formulated as a search for the initial condition demonstrating the maximum growth of the norm by the given time, i.e. it is required to find such vector $\mathbf{q}(0)$ for which the functional 
\begin{equation}
\label{functional}
\mathcal{G\left(\tau\right)}=\frac{||\mathbf{q}(t)||^2}{||\mathbf{q}(0)||^2}
\end{equation}
reaches maximum provided that the vector $\mathbf{q}$ satisfies the dynamical equations written in the operator form (\ref{dynamic_equation}).
To do this, a technique similar to the well-known Lagrange multipliers method of finding conditional extremum of a function is used. 

The Lagrangian necessary to find the conditional extremum in this case includes two terms: the functional whose maximum is searched for, and the so-called 'penalty' term, which is non-zero only if $\mathbf{q}$ does not satisfy the dynamical equations (\ref{dynamic_equation}) (see also \cite{corbett-bottaro-2001},  \cite{guegan-2006} and the review \cite{schmid-2007}):
\begin{equation}
\label{lagrangian}
\mathcal{L}\left(\mathbf{q},\tilde{\mathbf{q}}\right)=\mathcal{G}\left(\mathbf{q}\right)-\int\limits_0^{t}\left(\tilde{\mathbf{q}},\dot{\mathbf{q}}-\mathbf{A(q)q}\right)d\tau.
\end{equation}
Apparently, the penalty term in (\ref{lagrangian}) is written as inner product of the Lagrange multipliers (entering $\mathbf{\tilde q}$) and equation (\ref{dynamic_equation}), and additionally integrated over the time.
Unlike the well-known problem of finding conditional extremum of a function, the Lagrangian in this case is a functional defined for all possible shapes of $\mathbf{q}$, and the Lagrange multipliers themselves t are functions rather than the numbers. 

The extremum of (\ref{lagrangian}) is reached when variations of the Lagrangian with respect to $\mathbf{q}$ and $\mathbf{\tilde{q}}$ vanish simultaneously.
These variations are defined as (see book \cite{gunzburger-2003})
\begin{equation}
\frac{\partial \mathcal{L}}{\partial \mathbf{q}}\delta \mathbf{q}=\lim\limits_{\epsilon\to 0}\frac{\mathcal{L}\left(\mathbf{q}+\epsilon\delta\mathbf{q},\mathbf{\tilde{q}}\right)-\mathcal{L}\left(\mathbf{q},\mathbf{\tilde{q}}\right)}{\epsilon}
\end{equation}
\begin{equation}
\frac{\partial \mathcal{L}}{\partial \mathbf{\tilde{q}}}\delta \mathbf{\tilde{q}}=
\lim\limits_{\epsilon\to 0}\frac{\mathcal{L}\left(\mathbf{q},\mathbf{\tilde{q}}+\epsilon\delta\mathbf{\tilde{q}}\right)-\mathcal{L}\left(\mathbf{q},\mathbf{\tilde{q}}\right)}{\epsilon},
\end{equation}
where $\delta \mathbf{q}$ and $\delta \mathbf{\tilde{q}}$ are arbitrary functions taken at any time.

Variation with respect to indefinite multipliers clearly reads
\begin{equation}
\label{variation_tilde_q}
\frac{\partial \mathcal{L}}{\partial \mathbf{\tilde{q}}}\delta \mathbf{\tilde{q}}=
-\lim\limits_{\epsilon\to 0}\frac{1}{\epsilon}\int\limits_0^{t}\left(\epsilon\delta \mathbf{\tilde{q}},\dot{\mathbf{q}}-\mathbf{A(q)q}\right)d\tau=-\int\limits_0^{t}\left(\delta \mathbf{\tilde{q}},\dot{\mathbf{q}}-\mathbf{A(q)q}\right)d\tau
\end{equation}
Equating (\ref{variation_tilde_q}) to zero we obtain, by arbitrariness of $\delta \mathbf{\tilde{q}}$, equation (\ref{dynamic_equation}).
To compute variations with respect to the state vectors, use the Lagrange identity: $\left(\tilde{\mathbf{q}},\mathbf{Aq}\right)=\left(\mathbf{A}^{\dag}\tilde{\mathbf{q}},\mathbf{q}\right)$ (see, for example, \cite{marchyk-1998} for more detail about adjoint operators in non-linear problems) and take the penalty term by parts, after which the Lagrangian can be rewritten as
\begin{equation}
\mathcal{L}\left(\mathbf{q},\tilde{\mathbf{q}}\right)=\mathcal{G}\left(\mathbf{q}\right)-\left(\tilde{\mathbf{q}},\mathbf{q}\right)\bigg|^{t}_0+
\int\limits_0^{t}\left(\dot{\tilde{\mathbf{q}}}+\mathbf{A}^{\dag}(\mathbf{\tilde q})\tilde{\mathbf{q}},\mathbf{q}\right)d\tau
\end{equation}

Taking into account the smallness of $\epsilon$ and the real-valued inner product 
\footnote{
Real-valued inner product is additionally required only in this Section to obtain in the simple form the constraints (\ref{coupling_condition_tau}) and (\ref{coupling_condition_zero}), see below.},
calculate variation with respect to the state vectors:
\begin{equation}
\begin{aligned}
&\frac{\partial \mathcal{L}}{\partial \mathbf{q}}\delta \mathbf{q}=\lim\limits_{\epsilon\to 0}\frac{1}{\epsilon}\left[\frac{||\mathbf{q}(t)+\epsilon\delta\mathbf{q}(t)||^2}{||\mathbf{q}(0)+\epsilon\delta\mathbf{q}(0)||^2}-\frac{||\mathbf{q}(t)||^2}{||\mathbf{q}(0)||^2}-\left(\mathbf{\tilde{q}}(t),\epsilon\delta\mathbf{q}(t)\right)+\right.\\
&\left.+\left(\mathbf{\tilde{q}}(0),\epsilon\delta\mathbf{q}(0)\right)+\int\limits_0^{t}\left(\mathbf{\dot{\tilde{q}}}+\mathbf{A}^{\dag}(\mathbf{\tilde{q}})\mathbf{\tilde{q}},\epsilon\delta\mathbf{q}\right)d\tau\right],
\end{aligned}
\end{equation}
Here the first term can be recast to the form
\begin{equation}
\lim\limits_{\epsilon\to 0}\frac{1}{\epsilon}\frac{||\mathbf{q}(t)+\epsilon\delta\mathbf{q}(t)||^2}{||\mathbf{q}(0)+\epsilon\delta\mathbf{q}(0)||^2}=
\lim\limits_{\epsilon\to 0}\frac{1}{\epsilon}\frac{||\mathbf{q}(t)||^2+\epsilon(\delta\mathbf{q}(t),\mathbf{q}(t))+\epsilon(\mathbf{q}(t),\delta\mathbf{q}(t))}{||\mathbf{q}(0)||^2+\epsilon(\delta\mathbf{q}(0),\mathbf{q}(0))+\epsilon(\mathbf{q}(0),\delta\mathbf{q}(0))}
\end{equation}
As the inner product is real-valued, we have $(\delta\mathbf{q}(t),\mathbf{q}(t))=(\mathbf{q}(t),\delta\mathbf{q}(t))$, and so the transformation can be continued:
\begin{equation}
\begin{aligned}
&\lim\limits_{\epsilon\to 0}\frac{1}{\epsilon}\left[\frac{||\mathbf{q}(t)+\epsilon\delta\mathbf{q}(t)||^2}{||\mathbf{q}(0)+\epsilon\delta\mathbf{q}(0)||^2}-\frac{||\mathbf{q}(t)||^2}{||\mathbf{q}(0)||^2}\right]=
\lim\limits_{\epsilon\to 0}\frac{1}{\epsilon}\left[\frac{||\mathbf{q}(t)||^2+2\epsilon(\delta\mathbf{q}(t),\mathbf{q}(t))}{||\mathbf{q}(0)||^2+2\epsilon(\delta\mathbf{q}(0),\mathbf{q}(0))}-\frac{||\mathbf{q}(t)||^2}{||\mathbf{q}(0)||^2}\right]=\\
&=\lim\limits_{\epsilon\to 0}\frac{1}{\epsilon}\left[\frac{2\epsilon(\delta\mathbf{q}(t),\mathbf{q}(t))}{||\mathbf{q}(0)||^2}-\frac{2\epsilon(\delta\mathbf{q}(0),\mathbf{q}(0))||\mathbf{q}(t)||^2}{||\mathbf{q}(0)||^4}\right]=\\
&=\frac{2(\delta\mathbf{q}(t),\mathbf{q}(t))}{||\mathbf{q}(0)||^2}-2(\delta\mathbf{q}(0),\mathbf{q}(0))\frac{||\mathbf{q}(t)||^2}{||\mathbf{q}(0)||^4},
\end{aligned}
\end{equation}
which ultimately gives the variation:
\begin{equation}
\begin{aligned}
\label{variation_q}
&\frac{\partial \mathcal{L}}{\partial \mathbf{q}}\delta \mathbf{q}=\frac{2(\delta\mathbf{q}(t),\mathbf{q}(t))}{||\mathbf{q}(0)||^2}-2(\delta\mathbf{q}(0),\mathbf{q}(0))\frac{||\mathbf{q}(t)||^2}{||\mathbf{q}(0)||^4}-\left(\mathbf{\tilde{q}}(t),\delta\mathbf{q}(t)\right)+\\
+&\left(\mathbf{\tilde{q}}(0),\delta\mathbf{q}(0)\right)+\int\limits_0^{t}\left(\mathbf{\dot{\tilde{q}}}+\mathbf{A}^{\dag}(\mathbf{\tilde{q}})\mathbf{\tilde{q}},\delta\mathbf{q}\right)d\tau
\end{aligned}
\end{equation}

\begin{figure}[h!]
\includegraphics[width=0.8\linewidth]{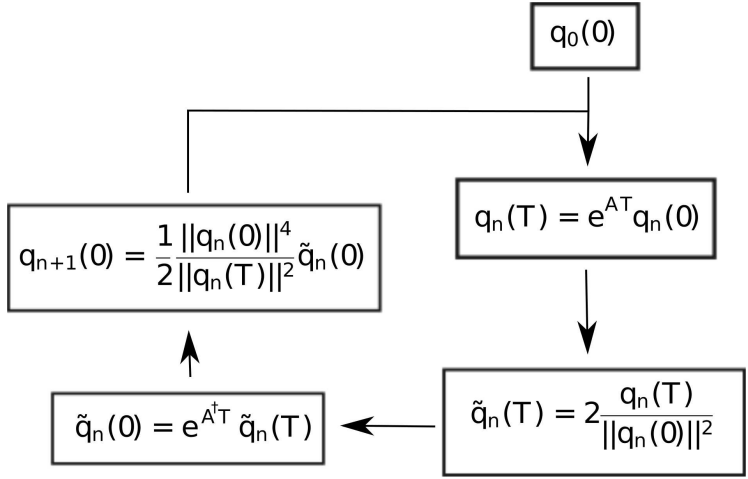}
\caption{\footnotesize{
Schematics of the iteration loop to search for the optimal perturbation for the time $T$ satisfying the general system (\ref{dynamic_equation}) (see the review \cite{schmid-2007}).
}}
\label{scheme}
\end{figure}

Since the variations of $\delta\mathbf{q}$ taken at different instants are independent from each other, equating (\ref{variation_q}) to zero yields the equation for indefinite multipliers (\ref{conjugated_equation}), 
which provides the vanishing of the Lagrangian variation in the interval $0<\tau<t$, as well as the relations between $\mathbf{q}$ and $\mathbf{\tilde q}$, 
which are necessary to vanish the Lagrangian variations at the moments $\tau=0$ and $\tau=t$:

\begin{equation}
\label{coupling_condition_tau}
\mathbf{\tilde q}(t)=\frac{2}{||\mathbf{q}(0)||^2}\mathbf{q}(t)
\end{equation}
\begin{equation}
\label{coupling_condition_zero}
\mathbf{q}(0)=\frac{||\mathbf{q}(0)||^4}{2||\mathbf{q}(t)||^2}\mathbf{\tilde q}(0)
\end{equation}

Vectors $\mathbf{q}$ and $\tilde{\mathbf{q}}$ satisfying equations (\ref{dynamic_equation}), (\ref{conjugated_equation}) and constraints (\ref{coupling_condition_tau}), (\ref{coupling_condition_zero}) turn the Lagrangian variations to zero, and hence exactly for them the functional (\ref{functional}) reaches extremum. 

Like for linear systems, the joint solution of equations can be looked for by power iteration method schematically shown in Fig. \ref{scheme}.
This issue is further discussed in papers \cite{cherubini-2010}, \cite{cherubini-2011}, \cite{cherubini-de-palma-2013}. 
Note once again that for linear perturbations the optimization of the functional (\ref{functional}) is reduced to looking for the maximal eigenvalue of the composite operator $\mathbf{Z}^{\dag}\mathbf{Z}$. 

\subsection{Adjoint equations}
\label{sect_adjoint}
\subsubsection{Derivation of adjoint equations}
In order to obtain explicit form of adjoint equations to the set (\ref{sys_A_1})-(\ref{sys_A_3}), we will use the norm identical to the total acoustic energy of perturbations (\ref{ac_norm}).
The inner product corresponding to this norm is given by equation (\ref{scalar_production}) we have already used.
Using it we represent $(\mathbf{\tilde{q}},\mathbf{Aq})$ as
\begin{equation}
\label{conj_1}
\begin{aligned}
&(\mathbf{\tilde{q}},\mathbf{Aq})=\pi\int\limits_{r_{in}}^{r_{out}}\Sigma\left[\delta\tilde{v}_r\left(im\Omega \delta v_r^*+2\Omega \delta v_{\varphi}^*-\frac{\partial \delta h^*}{\partial r}\right)+\delta\tilde{v}_{\varphi}\left(-\frac{\kappa^2}{2\Omega}\delta v_r^*+im\Omega\delta v_{\varphi}^*+\frac{im\delta h^*}{r}\right)+\right.\\
&\left.+\frac{\delta\tilde{h}}{a_*^2}\left(-\frac{a_*^2}{\Sigma r}\frac{\partial}{\partial r}\left(r\Sigma \delta v_r^*\right)+\frac{ima_*^2}{r}\delta v_{\varphi}^*+im\Omega\delta h^*\right) \right]rdr
\end{aligned}
\end{equation}

Now, using the Lagrange identity $(\mathbf{\tilde{q}},\mathbf{Aq})=(\mathbf{A^{\dag}\tilde{q}},\mathbf{q})$ and equation (\ref{conjugated_equation}) in the left-hand side of this expression, represent the inner product according to  (\ref{scalar_production}).
The right-hand side can be rearranged in a way to get the components of $\delta\mathbf{q}$ in the form of multipliers.
Here, the spatial derivatives are rearranged using integration by parts.
We obtain
\begin{equation}
\label{conj_2}
\begin{aligned}
&\pi\int\limits_{r_{in}}^{r_{out}}\Sigma rdr \left[-\delta v_r^*\frac{\partial\tilde{v}_r}{\partial t}-\delta v_{\varphi}^*\frac{\partial\tilde{v}_{\varphi}}{\partial t}-\delta h^*\frac{\partial\tilde{h}}{\partial t}\right]=\\
&=\pi\int\limits_{r_{in}}^{r_{out}}\Sigma rdr \left[\delta v_r^*\left(im\Omega \delta \tilde{v}_r-\frac{\kappa^2}{2\Omega}\delta \tilde{v}_{\varphi}+\frac{\partial \delta \tilde{h}}{\partial r}\right) +\delta v_{\varphi}^*\left(2\Omega\delta \tilde{v}_r+im\Omega\delta \tilde{v}_{\varphi}+\frac{im}{r}\delta \tilde{h}\right)+\right.\\
&+\left. \delta h^*\left(\frac{1}{r\Sigma}\frac{\partial}{\partial r}\left(r\Sigma\delta\tilde{v}_r\right)+\frac{im}{r}\delta \tilde{v}_{\varphi}+\frac{im\Omega}{a_*^2}\delta \tilde{h}\right) \right]-\pi r\Sigma \delta \tilde{h}\delta v_r^*\bigg|_{r_{in}}^{r_{out}}-\pi r\Sigma\delta\tilde{v}_r\delta h^*\bigg|_{r_{in}}^{r_{out}}.
\end{aligned}
\end{equation}
The substitutions in the right-hand side of (\ref{conj_2}) vanish since $\Sigma\to 0$ at the boundaries.

The components of variation $\delta\mathbf{q}$ are arbitrary and independent, so \ref{conj_2}) is transformed into the three independent equalities each corresponding to certain component of $\delta\mathbf{q}$.
These equalities result in the system of adjoint equations:
\begin{equation}
\label{adj_sys_A_1}
\frac{\partial \delta \tilde{v}_r}{\partial t} = -{\rm i}m\Omega\, \delta \tilde{v}_r + \frac{\kappa^2}{2\Omega} \delta \tilde{v}_\varphi - \frac{\partial \delta \tilde{h}}{\partial r},
\end{equation}

\begin{equation}
\label{adj_sys_A_2}
\frac{\partial \delta \tilde{v}_\varphi}{\partial t} =  -2\Omega \delta \tilde{v}_r  -{\rm i}m\Omega\, \delta \tilde{v}_\varphi -\frac{{\rm i} m}{r} \delta \tilde{h}, 
\end{equation}

\begin{equation}
\label{adj_sys_A_3}
\frac{\partial \delta \tilde{h}}{\partial t} = -\frac{a_*^2}{r\Sigma} \frac{\partial}{\partial r} (r\Sigma \delta \tilde{v}_r) -\frac{{\rm i} m a_*^2}{r} \delta \tilde{v}_\varphi  -{\rm i}m\Omega \,\delta \tilde{h},
\end{equation}

Changing to the local space limit in (\ref{adj_sys_A_1})-(\ref{adj_sys_A_3}) (as we did in Section \ref{sect_local_appr} to obtain the system  (\ref{sonic_sys1})-(\ref{sonic_sys3}) from equations (\ref{sys_A_1})-(\ref{sys_A_3})) we get an explicit form of equations adjoint to the set (\ref{sonic_sys1})-(\ref{sonic_sys3}):
\begin{equation}
\label{adj_sonic_sys1}
\left ( \frac{\partial}{\partial t} - q\Omega_0 x\frac{\partial}{\partial y} \right ) \tilde u_x - (2 - q)\Omega_0\tilde u_y =
-\frac{\partial\tilde W}{\partial x},
\end{equation}
\begin{equation}
\label{adj_sonic_sys2}
\left ( \frac{\partial}{\partial t} - q\Omega_0 x\frac{\partial}{\partial y} \right ) \tilde u_y + 
2\Omega_0 \tilde u_x =
-\frac{\partial \tilde W}{\partial y},
\end{equation}
\begin{equation}
\label{adj_sonic_sys3}
\left ( \frac{\partial}{\partial t} - q\Omega_0 x\frac{\partial}{\partial y} \right ) \tilde W + 
a_*^2 \left ( \frac{\partial \tilde u_x}{\partial x} + \frac{\partial \tilde u_y}{\partial y} \right ) = 0,
\end{equation}
where tildes above $u_x$, $u_y$ and $W$ means that these quantities compose an adjoint state vector. 

Finally, passing to the co-moving reference frame in (\ref{adj_sonic_sys1})-(\ref{adj_sonic_sys3}) yields adjoint equations for particular SFH:
\begin{equation}
\label{adj_sonic_sys1_sh}
\frac{d \hat{\tilde u}_x}{d t} = 
(2 - q) \hat{\tilde u}_y - {\rm i}\, \tilde k_x(t) \hat {\tilde W},
\end{equation}

\begin{equation}
\label{adj_sonic_sys2_sh}
\frac{d \hat {\tilde u}_y}{d t} = 
- 2 \hat {\tilde u}_x - {\rm i}\, k_y \hat {\tilde W},
\end{equation}

\begin{equation}
\label{adj_sonic_sys3_sh}
\frac{d \hat {\tilde W}}{d t} = 
- {\rm i}\, ( \, \tilde k_x(t) \hat {\tilde u}_x + k_y \hat {\tilde u}_y \,).
\end{equation}

Applying the power iteration method jointly to (\ref{sys_A_1})-(\ref{sys_A_3}) and (\ref{adj_sys_A_1})-(\ref{adj_sys_A_3}) for global azimuthal Fourier harmonics of two-dimensional perturbations or to the sets (\ref{sonic_sys1_sh})-(\ref{sonic_sys3_sh}) and (\ref{adj_sonic_sys1_sh})-(\ref{adj_sonic_sys3_sh}) for the local SFH, 
we automatically arrive at the optimal initial profiles of the enthalpy and velocity component perturbations that maximize the total acoustic energy growth by the given time interval.
This problem in application to Keplerian flows was solved by \cite{zhuravlev-razdoburdin-2014}. 

\subsubsection{Non-normality condition for ${\bf Z}$}

Here we show that non-normality of the dynamical operator determined by the set of equations (\ref{sys_A_1})-(\ref{sys_A_3}) is the direct consequence of the angular velocity gradient in the flow.
We already discussed this in Section \ref{sect_singular} where we introduced the notion of singular vectors.
Now we can prove this rigorously in a quite general case, since the explicit form of the operator ${\bf A}^\dag$ defined by the system (\ref{adj_sys_A_1})-(\ref{adj_sys_A_3}) is known.
First, let us calculate the commutator of $\mathbf{A}$ and $\mathbf{A}^{\dag}$:
\begin{equation}
\left[\mathbf{A},\mathbf{A}^{\dag}\right]
=\left ( \begin{array}{ccc}
\frac{16\Omega^4-\kappa^4}{4\Omega^2} & 0 & \frac{\mathrm{i}m}{2r\Omega}\left(4\Omega^2-\kappa^2\right) \\
0 &  \frac{\kappa^4-16\Omega^4}{4\Omega^2} & \frac{4\Omega^2-\kappa^2}{2\Omega}\frac{\partial}{\partial r} \\
\frac{\mathrm{i}ma_*^2}{2r\Omega}\left(\kappa^2-4\Omega^2\right) & \frac{a_{*}^2}{r\Sigma}\frac{\partial}{\partial r}\left(\frac{r\Sigma}{2\Omega}\left(\kappa^2-4\Omega^2\right)\right)+\frac{a_*^2}{2\Omega}\left(\kappa^2-4\Omega^2\right)\frac{\partial}{\partial r}  & 0
\end{array} \right )
\end{equation}

It is not difficult to notice that $\left[\mathbf{A},\mathbf{A}^{\dag}\right]$ vanishes for $\kappa=2\Omega$, what corresponds to solid-body rotation.
In this case the commutator $\left[\mathbf{Z},\mathbf{Z}^{\dag}\right]=\left[\mathrm{e}^{\mathbf{A}t},\mathrm{e}^{\mathbf{A}^{\dag}t}\right]$ can be easily found, 
because for commuting operators the product of their operator exponents is equal to the exponent of their sum, what can be easily verified by writing the operator exponents as the corresponding infinite series
\begin{equation}
\left[\mathrm{e}^{\mathbf{A}t},\mathrm{e}^{\mathbf{A}^{\dag}t}\right]=\mathrm{e}^{\mathbf{A}t}\mathrm{e}^{\mathbf{A}^{\dag}t}-\mathrm{e}^{\mathbf{A}^{\dag}t}\mathrm{e}^{\mathbf{A}t}=\mathrm{e}^{\left(\mathbf{A}+\mathbf{A}^{\dag}\right)t}-\mathrm{e}^{\left(\mathbf{A}^{\dag}+\mathbf{A}\right)t}=0
\end{equation}
Thus, the operator $\mathbf{Z}$ becomes normal for solid-body rotation.

The inverse statement is also valid: if $\mathbf{Z}$ is normal for any time moment $t$, the rotation is solid-body.
To see this, use the Campbell-Baker-Hausdorff formula (\cite{richtmyer-1981}, Ch. 25) to represent the composite operators ${\bf Z}{\bf Z}^\dag$ and ${\bf Z}^\dag {\bf Z}$:
\begin{equation}
\mathrm{e}^{\mathbf{A}t}\mathrm{e}^{\mathbf{A}^{\dag}t}=\exp\left(\left(\mathbf{A}+\mathbf{A}^{\dag}\right)t+\frac{t^2}{2}\left[\mathbf{A},\mathbf{A}^{\dag}\right]+\frac{t^3}{12}\left[\mathbf{A},\left[\mathbf{A},\mathbf{A}^{\dag}\right]\right]-\frac{t^3}{12}\left[\mathbf{A}^{\dag},\left[\mathbf{A},\mathbf{A}^{\dag}\right]\right]+...\right)
\end{equation}

\begin{equation}
\begin{aligned}
&\mathrm{e}^{\mathbf{A}^{\dag}t}\mathrm{e}^{\mathbf{A}t}=\exp\left(\left(\mathbf{A}^{\dag}+\mathbf{A}\right)t+\frac{t^2}{2}\left[\mathbf{A}^{\dag},\mathbf{A}\right]+\frac{t^3}{12}\left[\mathbf{A}^{\dag},\left[\mathbf{A}^{\dag},\mathbf{A}\right]\right]-\frac{t^3}{12}\left[\mathbf{A},\left[\mathbf{A}^{\dag},\mathbf{A}\right]\right]+...\right)=\\
&=\exp\left(\left(\mathbf{A}^{\dag}+\mathbf{A}\right)t-\frac{t^2}{2}\left[\mathbf{A},\mathbf{A}^{\dag}\right]+\frac{t^3}{12}\left[\mathbf{A},\left[\mathbf{A},\mathbf{A}^{\dag}\right]\right]-\frac{t^3}{12}\left[\mathbf{A}^{\dag},\left[\mathbf{A},\mathbf{A}^{\dag}\right]\right]+...\right)
\end{aligned}
\end{equation}

The equality $\left[\mathrm{e}^{\mathbf{A}t},\mathrm{e}^{\mathbf{A}^{\dag}t}\right]=0$ is fulfilled for any $t$, therefore the terms with the same powers of $t$ must be independently equal to zero,  which is possible only if the commutator $\left[\mathbf{A},\mathbf{A}^{\dag}\right]=0$.
The last equality is valid for solid-body rotation only. 

This implies that solid-body rotation is necessary and sufficient for the dynamical operator $\mathbf{Z}$ of the set (\ref{sys_A_1})-(\ref{sys_A_3}) to be normal.
Thus, any deviation from solid-body rotation, i.e. the appearance of angular velocity gradient in astrophysical disks, makes the dynamical operator non-normal and perturbation modes non-orthogonal to each other. 

\section{Optimal perturbations in Keplerian disks}

In the concluding Section of the this paper we would like to briefly discuss the use of the variational method to search for optimal perturbations in astrophysical disks.
We consider geometrically thin disks with an almost Keplerian azimuthal velocity profile in the background flow.
In numerical calculations we are going to consider a radially infinite disk with only the inner (free) boundary and the thin quasi-Keplerian torus with inner and outer radial boundaries. 
The latter configuration was used in Section \ref{sect_martix} for the analysis of superposition of neutral modes to illustrate the matrix method of optimization.
However, for the sake of methodology, we start with the simplest analytically tractable problem of transient growth of local short-wave perturbations with $k_y\gg 1$ which we discussed in detail in Section \ref{sect_TG} 

\subsection{Local approximation}

\label{sect_local_adjoint}

Indeed, let us apply the power iteration method to the system (\ref{sonic_sys1_sh})-(\ref{sonic_sys3_sh}), (\ref{adj_sonic_sys1_sh})-(\ref{adj_sonic_sys3_sh}) in the limit $k_y\gg 1$ corresponding to incompressible fluid.
In this limit, the system (\ref{sonic_sys1_sh})-(\ref{sonic_sys3_sh}) can be reduced to one equation for $\hat u_x$:
\begin{equation}
\label{SFH_u_x}
\frac{d \hat u_x}{d t} +
2qk_y \frac{\tilde k_x}{k_y^2 + \tilde k_x^2} \, \hat u_x = 0,
\end{equation}
giving the analytical solution
\begin{equation}
\label{u_x_sol}
\hat u_x(t) = \hat u_x(0) \, \frac{k_x^2 + k_y^2}{\tilde k_x^2 + k_y^2},
\end{equation}
which, of course, repeat (\ref{vort_u_x}) for $k_y\gg 1$. 

At the same time, the adjoint equations (\ref{adj_sonic_sys1_sh})-(\ref{adj_sonic_sys3_sh}) in the limit of incompressible fluid suggest that the quantity $\hat {\tilde u}_x$ conjugate to $\hat u_x$ is {\it conserved}
\footnote{
It can be checked that the value of $I$, which was conserved in the direct equations (\ref{sonic_sys1_sh})-(\ref{sonic_sys3_sh}), becomes time-dependent in the adjoint equations (\ref{adj_sonic_sys1_sh})-(\ref{adj_sonic_sys3_sh}) 
(see the Appendix in paper \cite{zhuravlev-razdoburdin-2014}).
}:
\begin{equation}
\label{adj_SFH_eq}
\frac{d\hat {\tilde u}_x}{d t} = 0.
\end{equation}

Obviously, after $p$ iterations of arbitrary initial profile $\hat u_x^{in}(k_x, k_y, t=0)$ we obtain that it is multiplied by the factor:
\begin{equation}
\label{opt_factor}
\left [ \frac{k_x^2 + k_y^2}{(\tilde k_x(t)^2 + k_y^2} \right ]^p.
\end{equation}

With account for renormalization of the solution at each iteration, while $p\to\infty$, the factor (\ref{opt_factor}) suppresses all SFH composing $\hat u_x^{in}(k_x, k_y, t=0)$ except the optimal SFH corresponding to a maximum of 
(\ref{opt_factor}) as a function of $k_x$.
For a fixed time interval $t$ this $k_x$ takes the value 
\begin{equation}
\label{max_k_x}
k_x = 1/2 k_y \,( -q t - ( (q t)^2+4)^{1/2} ).
\end{equation}

Plugging (\ref{max_k_x}) into the SFH growth factor (\ref{g_incompr}) yields the sought for optimal growth $G$, which for the local problem is defined as (\ref{G_t}):
\begin{equation}
\label{variational_G}
G(t) = \frac{ (q t)^2 + q t [(q t)^2+4]^{1/2} +4 }
{ (q t)^2 - q t [(q t)^2+4]^{1/2} +4 }.
\end{equation}

Expression (\ref{variational_G}) represents the first singular value which the iteration loop for short-wave local vortices converges to.
Apparently, for large time intervals, $q t\gg 1$, it gives $G\approx (q t)^2$, which reproduces the approximate estimate of $G$ according to formula (\ref{G_high_ky}).

Also note that an exact result (\ref{variational_G}) could be obtained in this simple example directly from expression for the growth factor (\ref{g_incompr}) by calculating maximum of $g$ as a function of $k_x$ at a fixed $t$.

For arbitrary $k_y$ the optimal growth can be obtained by numerical forward-backward integration of the full system of direct and adjoint equations, which are ordinary differential equations for SFH. 

\subsection{Global problem}
\label{sect_global_var}

In the case where the azimuthal scale of perturbations is comparable to the horizontal disk scale it is necessary to numerically solve the system of partial differential equations (\ref{sys_A_1})-(\ref{sys_A_3}) and (\ref{adj_sys_A_1})-(\ref{adj_sys_A_3}), which was done in paper \cite{zhuravlev-razdoburdin-2014} using a second-order explicit difference scheme (leap-frog) (see, for example, \cite{frank-robertson-1988}).

\begin{figure}[h!]
\includegraphics[width=0.8\linewidth]{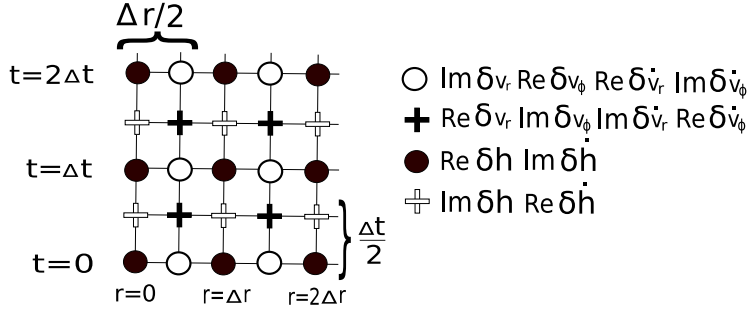}
\caption{\footnotesize{
Illustration to the numerical scheme of integration of equations (\ref{sys_A_1})-(\ref{sys_A_3}) and (\ref{adj_sys_A_1})-(\ref{adj_sys_A_3}).
}}
\label{fig_leap_frog}
\end{figure}

In this difference scheme, each equation is separated into real and imaginary parts and on the plane $(r,t)$ 4 grids are introduced.
Unknown variables are calculated in the nodes of these grids using the corresponding differences (Fig. \ref{fig_leap_frog}).
The nodes are shifted with respect to each other by half time step $\Delta t$ and/or by half radial step $\Delta r$.
This allows one to use the central approximation to calculated derivatives with respect to $r$ and $t$, which provides an accuracy of order of $(\Delta r)^2$ and $(\Delta t)^2$.
The time step is determined using the radial step and the Courant condition that follows from the local dispersion relation which can be obtained from the equations being integrated. 

\subsubsection{Comparison of the transient growth of vortices in global and local space limits}

As a background flow, consider an unlimited Keplerian disk that has only inner boundary at $r=r_1$.
To see how the cylindrical geometry of disk and, mainly, the accurate profile of Keplerian angular velocity, $\Omega = \Omega(r_1) (r/r_1)^{-3/2}$, affect the transient growth, we assume for simplicity that all other values in the equations for perturbations are constant:
\begin{equation}
\label{homogen_disc}
\Sigma=const, \quad a_{eq} = (\delta / \sqrt{2n}) (\Omega r)|_{r_1}.
\end{equation}

\begin{figure}[h!]
\includegraphics[width=1\linewidth]{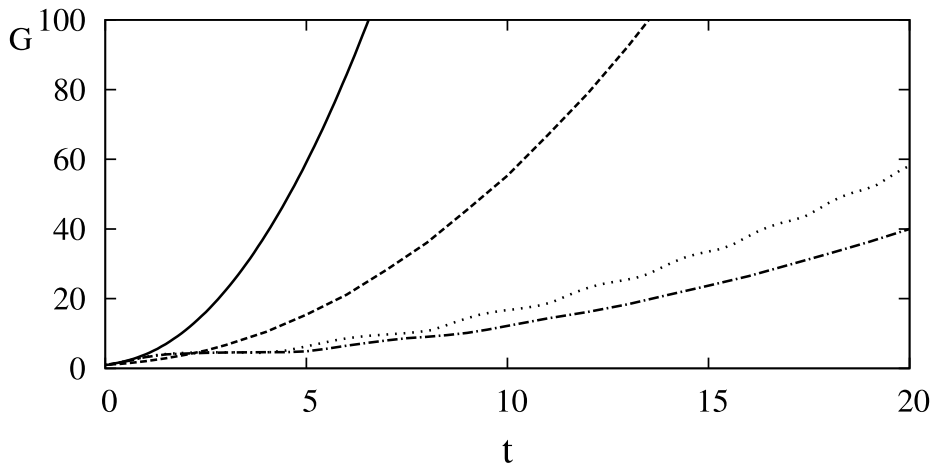}
\caption{\footnotesize{
Comparison of the optimal growth for small-scale and large-scale vortices (see Section \ref{sect_TG}) in the global and local space limits.
The solid and dotted curves are calculated for local SFH using formula (\ref{G_t}) by the iteration loop for equations (\ref{sonic_sys1_sh})-(\ref{sonic_sys3_sh}) and (\ref{adj_sonic_sys1_sh})-(\ref{adj_sonic_sys3_sh}) for $k_y=12.5$ and $k_y=0.125$, respectively.
Harmonics with $m=5$ are taken as global perturbations.
The optimal perturbations are calculated using formula (\ref{G_global}) by iteration loop for equations (\ref{sys_A_1})-(\ref{sys_A_3}) and (\ref{adj_sys_A_1})-(\ref{adj_sys_A_3}) with the polytropic index $n=3/2$.
Using the relation $(m/r)H \sim k_y$, for similar large-scale and small-scales vortices, a disk with $\delta=0.05$ (the dashed-dotted line) and a formally thick disk with $\delta=5$ (the dashed line) were considered, respectively.
In both cases the time is expressed in units of $\Omega(r_1)^{-1}$.
}}
\label{fig_10}
\end{figure}

As shown in paper \cite{zhuravlev-razdoburdin-2014}, the account for a more realistic distributions of $\Sigma$ and $a_{eq}$ (for example, as in the standard accretion disks) does not change the qualitative conclusions presented below.
The results of local and global calculations of optimal perturbations by the variational method are shown in Fig. \ref{fig_10}.

Here we compare the transient growth of vortices with azimuthal scale both smaller and larger than the disk thickness.
The main qualitative conclusion is that the growth rate of small-scale vortices ($\lambda_\varphi<H$) decreases much faster as one proceeds from the shearing sheet approximation to the scales of order of the disk radial size 
(i.e. from $m=\infty$ to  $m\sim 1$) than that of large-scale vortices ($\lambda_\varphi>H$).
It can be verified that in the limiting case of global perturbations with $m=1$, the value of $G$ for small-scale and large-scale vortices differ to within a factor of 1.5-2 only, for the given parameters and on time intervals up to $t\sim 20$.
At the same time, for local vortices, the value of $G$ for $\lambda_\varphi<H$ and $\lambda_\varphi>H$ differs by several orders of magnitude.
This suggests that global large-scale vortices in thin Keplerian disks can also exhibit the growth of dozens of times on quite short time intervals of order of several Keplerian periods at the inner disk boundary.
In turn, this may imply the importance of the perturbations transient growth for the angular momentum transfer on scales much larger than the disk thickness.

\subsubsection{Transient spirals and modes in a quasi-Keplerian torus}

Finally, let us return to the disk model considered above for illustration of the matrix method (see Section \ref{sect_martix}).
As is well known (see, for example, \cite{glatzel-1987a}, \cite{glatzel-1987b}, \cite{glatzel-1988}, as well as \cite{zhuravlev-shakura-2007b}), this flow demonstrates a weak spectral instability, since there are exponentially growing inertial-acoustic modes.
As we have already mentioned in Section \ref{sect_adiabat_perts}, their increments rapidly decrease with decreasing relative geometrical thickness of the torus, i.e. with approaching $\Omega$ the Keplerian profile.
Then, perturbations can grow only due to the transient mechanism of shortening of leading spirals by the shear flow (see the discussion in Section \ref{sect_TG}), which occurs on short time scales of order of several Keplerian periods in the flow.
However, in the intermediate case, where the pressure gradient in the torus is sufficiently high, both non-modal and modal perturbation growth can occur simultaneously but on different time scales.
The exponential growth of modes will always dominate over the transient growth starting from some large time intervals.
Interestingly, essentially this means that as calculating the first singular value of the dynamical operator employing the variational method, starting from some $t$ the curve $G(t)$ should become exponential corresponding to the most unstable mode.
At the same time, the iteration loop, which always converges to the optimal initial perturbation vector ${\bf q}(t=0)$, must then give not a leading spiral, but a {\it mode}.
Whereas the spiral starts being shrunk by the shear flow and enhanced due to the perimeter shortening (see the discussion in Section \ref{sect_adiabat_perts}) at the time $t>0$, the mode rotates like a solid body with angular velocity equal to $\Omega$ at the corotation radius inside the flow, since its amplitude increases due to the resonance energy exchange with the flow at this radius.
Thus, method of optimization of perturbations can be applied both to study the transient growth of perturbations and to find the profiles and increments of the mostly unstable modes in arbitrary complex shear flows, i.e. to solve the spectral problem as well.

An example of the calculation of a transient spiral and of an unstable mode in the same toroidal flow by joint solution of the systems (\ref{sys_A_1})-(\ref{sys_A_3}) and (\ref{adj_sys_A_1})-(\ref{adj_sys_A_3}) employing the variational method was presented in Fig. \ref{pic_TG} and \ref{pic_modes} in the Introduction.
As we see, even for $\delta=0.3$ the maximal increment is very low, and it takes $\sim 10^3$ Keplerian periods for the most unstable mode to get at least twice amplitude.
At the same time, the transient growing spiral increases by a factor of 6 already after several rotational periods at the inner  disk boundary. 

\section{Conclusion}
This review is devoted to the transient dynamics of perturbations, which is of special interest in theory of astrophysical disks, in particular accretion disks.
Exponentially growing perturbations do not exist in a homogeneous inviscid Keplerian flow provided that there are no conditions for the magneto-rotational instability.
Nevertheless, observations suggest that also in this case angular momentum should be somehow transported outwards.
At least, this implies that there should be some mechanism of energy transfer from the regular rotational motion to hydrodynamical perturbations.
In spectrally stable flows the transient growth mechanism is responsible for this.
Here it was introduced by a simple example of two-dimensional vortices and it was discussed that the reason for their growth is the shortening of the length of leading spirals by the differential rotation of the flow (see Fig. \ref{pic_TG} and \ref{contour}). Nonwithstanding their seeming simplicity, those (quasi-)columnar structures exhibit the strongest ability to extract energy from the spectrally stable differentially rotating flows (see \cite{maretzke-2014} about it). 
Physically, the energy growth of vortices takes place due to their own angular momentum conservation, which in the local limit is expressed by the conservation of their potential vorticity and the existence of the invariant $I$ (see Section \ref{sect_local_appr}).
Here we considered both small-scale ($k_y\gg 1$) and large-scale ($k_y\ll 1$) vortices and compared their optimal growth with account for non-zero effective viscosity in the disk (see Fig. \ref{G_max}).
Importantly, the transient growth of large-scale vortices strongly increases for a super-Keplerian rotation, which can be significant in relativistic disks where $q>3/2$.
In this paper, special attention was paid to mathematical aspects of non-modal analysis and to methods of optimal perturbations computation.
We have discussed in detail that the transient growth is a consequence of non-normality of the governing dynamical operator of the problem and non-orthogonality of its eigenvectors, i.e. modes of perturbations (see Fig. \ref{modes_1} and \ref{modes}).
Therefore, the growth of arbitrary perturbations can be adequately studied by calculating not eigenvectors but singular vectors of this operator.
We have considered two methods: a matrix and variational one and applied them to the particular problems (see the corresponding results in Fig. \ref{Growth} and \ref{fig_10}).
The matrix method requires a discrete representation of the dynamical operator, for example, in the  basis of its eigenvectors.
The variational method is reduced to iterative integration of the system of direct and adjoint equations forward and backward in time, respectively.
We have emphasized that the variational method is more universal and can be applied to study of non-modal dynamics of perturbations in non-stationary flows, as well as to non-linear problems.

As was discussed, the transient growth of perturbations is used in the concept of bypass transition to turbulence in laminar flows.
It can be also important as a mechanism of enhanced angular momentum transfer and stimulation of the accretion rate in weakly turbulized disks.
Note that turbulence emerging due to bypass mechanism is fundamentally different from the 'classical' turbulence, in which the energy transfer from the background flow is mediated by modes exponentially growing on large spatial scales, whereas the non-linear interactions nothing but redistribute this energy between modes with other wave vectors $k$ (the so-called direct or inverse cascade).
This means that in the phase space the energy flux $\epsilon_T(k)$ arises which brings (in the case of direct cascade) the kinetic energy of perturbations to small scales where viscous dissipation occurs.
In this picture, the mode distribution over the directions of ${\bf k}$ in the phase space is of minor importance, and $\epsilon_T$ can be non-zero only along the direction of change of the module ${\bf k}$.
A completely different situation should take place when the transient growth of perturbations is responsible for the energy transfer from the background flow.
This linear mechanism  appears as leading spirals in the disk, i.e. spatial Fourier harmonics corresponding to only such values ${\bf k}$ that $k_x/k_y<0$.
In a spectrally stable flow, where there are no energy supply to the leading spirals, initial perturbations inevitably decay because the leading spirals turn into trailing ones.
Thus, the turbulent state here is possible only due to a positive non-linear feedback, which can exist only in the appearance of non-zero $\epsilon_T$ also in the direction of positional change of vector ${\bf k}$, i.e. in the phase space angles, when the trailing spirals give back to the leading ones a part of their energy sufficient to sustain the transient growth.
Simultaneously, the other part of energy stored in the trailing spirals dissipates into the heat just due to their ultimate transition to higher $k$.
Here, the heat dissipation can be not due to the direct cascade, but due to a purely linear winding up of the trailing spiral by the flow, i.e. due to the increase in time of the ratio $k_x/k_y >0$ at $k_y=const$.
As we see, the {\it transverse} cascade is an essential part of the alternative picture of turbulence in a shear flow, which is the angular redistribution of spatial Fourier harmonics of perturbations (see, for example, the appendix in \cite{chagelishvili-2003}).
The maintainance of the transient growth of small perturbations by the transverse cascade was studied in detail in \cite{horton-2010} for a two-dimensional Couette flow.
Such profound changes in the concept of the possible structure of turbulent flows should affect both analytical estimates of turbulent viscosity coefficient (see, for example, \cite{canuto-1984}) and numerical simulations of turbulence in astrophysical disks (see, for example, \cite{simon-hawley-beckwith-2009}, \cite{davis-stone-pessah-2010}, where spectral properties of turbulence averaged over the directions of ${\bf k}$ were mostly studied).
Note that we deliberately cited here numerical simulations in disks with magnetic field, in which the modal growth of perturbations due to the magneto-rotational instability takes place.
The point is that recent studies \cite{squire-bhattacharjee-2014} and \cite{squire-bhattacharjee-2014b} show that even in Keplerian flows, where the magneto-rotational instability operates, the optimal transiently growing perturbations dominate over exponentially growing modes on short time-scales.
Like in an unmagnetized flow, these transient perturbations are locally represented by shear harmonics.
Thus, the non-modal dynamics of perturbations can be essential tool in taking energy from the background flow in MHD-turbulent accretion disks as well.
Another hint of this is the recent paper \cite{mamatsashvili-2014}, which studied numerically (similar to \cite{horton-2010}) the transverse cascade of shear harmonics in a spectrally stable plane-parallel magnetized flow and demonstrated that two-dimensional turbulence arises due to a positive feedback with linear transient growth of shear harmonics.
The plane Poiseuille flow provides another example of shear flow in which the bypass transition to turbulence turns out to be more preferable than the 'classical' mechanism despite the presence of growing modes.
Here we mention \cite{schmid-1996} and \cite{reddy-1998}, which numerically studied not the developed turbulence (as is usually done in most of papers on MHD-turbulence in Keplerian flows), but some scenarios of the transition to turbulence from regular initial small perturbations of different types (see also Ch. 9 of book \cite{schmid-henningson-2001}).
It turned out that the previously accepted scenario of the transition due to the secondary instability of saturated modes requires much more time and/or significantly higher initial perturbation amplitudes than the transition due to the secondary instability of the so-called streaks grown due to the transient mechanism. For the sake of clarity, we give a remark that streaks in 3D model of plane Poiselle flow grow from the so called vortex {\it rolls} due to lift-up mechanism, which 
is also a variant of transient growth, but differs from the (swing) amplification of 2D vortices considered throughout this work.
Anyhow, as follows from Fig. 1 of \cite{schmid-1996}, the time of turbulence development from regular initial perturbations strongly depends on their amplitudes.
This is not surprising, since vortex rolls (just like 2D spatial Fourier harmonics studied above) of smaller amplitude require more time to saturate, after what the secondary instability comes into play leading directly to the breakdown to turbulence.
Clearly, the time of such a transition can be as long as hundreds of characteristic shear times, and nevertheless this does not affect later the properties and power of turbulent motions.
Although, presently we have only the results of such studies in the case of plane-parallel flows, in future they can be obtained for quasi-Keplerian flows with high Reynolds numbers, since locally such flows differ from plane-parallel flows only by the presence of the Coriolis force stabilizing the flow.
At last, an additional useful evidence here are the simplified finite-dimensional dynamical models of non-normal systems with positive feedback that recover basic properties of transition to turbulence in spectrally stable shear flows (see \cite{trefethen-1993} and \cite{waleffe-1995}).
For example, in Fig. 10 of \cite{trefethen-1993} it can be seen that the time for a such simplified model to reach one and the same 'turbulent' state increases with decreasing of intial perturbation amplitude and ultimately becomes infinite. 

To conclude, we note once again that here we have not discussed the aspects of three-dimensional perturbation dynamics.
Meantime, there are indications that the account of the natural inhomogeneity of the disk due to vertical density and pressure gradients gives a qualitatively new picture of both the transient growth of perturbations and the subsequent transition to turbulence (see \cite{lominadze-2011}).
Here, the perturbation dynamics is essentially three-dimensional, and it can be shown that for three-dimensional transient vortices there is a time-conserved analogue of the invariant of motion $I$ (see \cite{tevzadze-2003} и \cite{tevzadze-2008}).
The new numerical calculations carried out in \cite{marcus-2014} also point out that taking into account of the disk vertical inhomogeneity can result in its destabilization in the subcritical regime at high Reynolds numbers, unlike the case observed in a homogeneous flow (see \cite{shen-2006}).

The work by D.N. Razdoburdin was supported by RSF grant 14-12-00146 for writing Section 3 of this review. 
The work by V.V. Zhuravlev was partially supported by RFBR grants 14-02-91172 and 15-02-08476, and also by Program 9 of the Praesidium of RAS.

\end{document}